\documentclass{aastex62}

\submitjournal{ApJ}

\shorttitle{Sample article}
\shortauthors{Katsianis et al.}

\begin{document}

\title{An evolving and mass dependent ${\rm \sigma_{\text{sSFR}}}$-${\rm M_{\star}}$ relation for galaxies.}

\correspondingauthor{Antonios Katsianis}
\email{kataunichile@gmail.com, kata@sjtu.edu.cn}

\author{Antonios Katsianis}
\affiliation{Tsung-Dao Lee Institute, Shanghai Jiao Tong University, Shanghai 200240, China}
\affiliation{Department of Astronomy, Shanghai Key Laboratory for Particle Physics and Cosmology, Shanghai Jiao Tong University, Shanghai 200240, China}
\affiliation{Department of Astronomy, Universitad de Chile, Camino El Observatorio 1515, Las Condes, Santiago, Chile}

\author{Xianzhong Zheng}
\affiliation{Purple Mountain Observatory, CAS, 8 Yuanhua Road, Nanjing, China}
\affiliation{Chinese Academy of Sciences South America Center for Astronomy, China-Chile Joint Center for Astronomy, Camino del Observatorio 1515, Las Condes, Chile }

\author{Valentino Gonzalez}
\affiliation{Chinese Academy of Sciences South America Center for Astronomy, China-Chile Joint Center for Astronomy, Camino del Observatorio 1515, Las Condes, Chile }
\affiliation{Centro de Astrofísica y Tecnologías Afines (CATA), Camino del Observatorio 1515, Las Condes, Santiago, Chile}

\author{Guillermo Blanc}
\affiliation{Observatories  of  the  Carnegie  Institution  for  Science, Pasadena, CA, USA}

\author{Claudia del P. Lagos}
\affiliation{International Centre for Radio Astronomy Research, University of Western Australia, 35 Stirling Hwy, Crawley, WA 6009, Australia}
\affiliation{Cosmic Dawn Center (DAWN), Denmark, Norregade 10, 1165 Kobenhavn, Denmark}

\author{Luke J. M. Davies}
\affiliation{International Centre for Radio Astronomy Research (ICRAR), M468, University of Western Australia, 35 Stirling Hwy, Crawley, WA 6009, Australia}

\author{Peter Camps}
\affiliation{Sterrenkundig Observatorium, Universiteit Gent, Krijgslaan 281, B-9000 Gent, Belgium}

\author{Ana Tr\v{c}ka}
\affiliation{Sterrenkundig Observatorium, Universiteit Gent, Krijgslaan 281, B-9000 Gent, Belgium}

\author{Maarten Baes}
\affiliation{Sterrenkundig Observatorium, Universiteit Gent, Krijgslaan 281, B-9000 Gent, Belgium}

\author{Joop Schaye}
\affiliation{Leiden Observatory, Leiden University, PO Box 9513, NL-230
0 RA Leiden, the Netherlands.}

\author{James W. Trayford}
\affiliation{Leiden Observatory, Leiden University, PO Box 9513, NL-230
  0 RA Leiden, the Netherlands.}

\author{Tom Theuns}
\affiliation{Institute for Computational Cosmology, Department of Physics, University of Durham, South Road, Durham, DH1 3LE, UK.}

\author{Marko Stalevski}
\affiliation{Astronomical Observatory, Volgina 7, 11060 Belgrade, Serbia}
\affiliation{Department of Astronomy, Universitad de Chile, Camino El Observatorio 1515, Las Condes, Santiago, Chile}
\affiliation{Sterrenkundig Observatorium, Universiteit Gent, Krijgslaan 281, B-9000 Gent, Belgium}




\begin{abstract}

  The scatter (${\rm\sigma_{\text{sSFR}}}$) of the specific star formation rates (sSFRs) of galaxies is a measure of the diversity in their star formation histories (SFHs) at a given mass. In this paper we employ the EAGLE simulations to study the dependence of the  ${\rm \sigma_{\text{sSFR}}}$ of galaxies on stellar mass (${\rm M_{\star}}$) through the  ${\rm \sigma_{\text{sSFR}}}$-${\rm M_{\star}}$ relation in $ {\rm z \sim 0-4}$. We find that the relation evolves with time, with the dispersion depending on both stellar mass and redshift. The models point to an evolving U-shape form for the  ${\rm \sigma_{\text{sSFR}}}$-${\rm M_{\star}}$ relation with the scatter being minimal at a characteristic mass $M^{\star}$ of ${\rm 10^{9.5}}$ ${\rm M_{\odot}}$ and increasing both at lower and higher masses. This implication is that the diversity of SFHs increases towards both at the low- and high-mass ends. We find that active galactic nuclei feedback is important for increasing the ${\rm \sigma_{\text{sSFR}}}$ for high mass objects. On the other hand, we suggest that SNe feedback increases the ${\rm \sigma_{\text{sSFR}}}$ of galaxies at the low-mass end. We also find that excluding galaxies that have experienced recent mergers does not significantly affect the ${\rm \sigma_{\text{sSFR}}}$-${\rm M_{\star}}$ relation. Furthermore, we employ the combination of the EAGLE simulations with the radiative transfer code SKIRT to evaluate the effect of SFR/stellar mass diagnostics in the ${\rm \sigma_{\text{sSFR}}}$-${\rm M_{\star}}$ relation and find that the ${\rm SFR/M_{\star}}$ methodologies (e.g. SED fitting, UV+IR, UV+IRX-$\beta$) widely used in the literature to obtain intrinsic properties of galaxies have a large effect on the derived shape and normalization of the ${\rm \sigma_{\text{sSFR}}}$-${\rm M_{\star}}$ relation.

\end{abstract}



\section{Introduction}
\label{intro}

The scatter ( ${\rm \sigma_{\text{sSFR}}}$) of the specific star formation rate (the ratio of star formation rate and stellar mass)-stellar mass (${\rm M_{\star}}$) relation provides a measurement for the variation of star formation across galaxies of similar masses with physical mechanisms important for galaxy evolution make their imprint to it. These processes include gas accretion, minor mergers,  disc dynamics, halo heating, stellar feedback and AGN feedback. The above are typically dependent on galaxy stellar mass and cosmic epoch \citep{Cano2016,Abbot2017,Wang2017,Chiosi2017,Garcia2017,Eales2018,Qin2018,Sanchez2018,Canas2018,Blanc2019} and possibly affect the shape of the ${\rm \sigma_{\text{sSFR}}}$-${\rm M_{\star}}$ differently. However, it is difficult to study and especially quantify the effect of the different prescriptions important for galaxy evolution to the scatter solely by using insights from observations.

In addition, the shape of the  ${\rm \sigma_{\text{sSFR}}}$-${\rm M_{\star}}$ relation and the value of the dispersion is a matter of debate in the literature. The scatter is usually reported to be constant ($ {\rm \sim 0.3}$ dex) with stellar mass in most studies, especially those which address the high and intermediate redshift Universe (${\rm z>1}$). For example, \citet{Rodighiero10} and \citet{Schreiber2015}, using mostly UV-derived SFRs, suggest that the dispersion is independent on galaxy mass and constant (${\rm \sim 0.3}$ dex) over a wide $M_{\star}$ range for ${\rm z \sim2}$ star forming galaxies (${\rm 10^{9}-10^{11}}$ ${\rm M_{\odot}}$). \citet{Whitaker2012} reported a variation of 0.34 dex from Spitzer MIPS observations. Similarly, \citet{Noeske2007} and \citet{Elbaz2007} reported a 1${\rm \sigma}$ dispersion in ${\rm \log(SFR)}$ of around ${\rm 0.3}$ dex at ${\rm z \sim1}$ for their flux-limited sample. However, other studies suggest that the dispersion tends to be larger for more massive objects and in the lower-redshift Universe \citep{Guo2013,Ilbert2015} implying that mechanisms important for galaxy evolution are prominent and contribute to a variety of star formation histories for massive galaxies. On the other hand, \citet{Santini2017} suggested that the scatter decreases with increasing mass and this implies that mechanisms important for galaxy formation are giving a larger diversity of SFHs to low mass objects. In addition, \citet{Boogaard2018} using the MUSE Hubble Ultra Deep Field Survey suggest that the intrinsic scatter of the relation at the low mass end is $\sim 0.44$ dex and larger than what is typically found at higher masses. In disagreement with all the above, \citet{Willet2015} using the Galaxy Zoo survey for ${\rm z < 0.085}$ find a dispersion that actually decreases with mass at ${\rm 10^{8}-10^{10}}$ ${\rm M_{\odot}}$ from ${\rm \sigma = 0.45}$ dex to ${\rm 0.35}$ dex and increases again at ${\rm 10^{10}-10^{11.5}}$ ${\rm M_{\odot}}$ to reach a scatter of ${\rm \sim 0.5}$ dex. All the above observational studies have conflicting results and this is possibly because they are affected by selection effects, uncertainties originating from star formation rate diagnostics \citep{Katsianis2015,Davies2017}, different separation criteria for passive/star forming galaxies \citep{Renzini2015} and usually focus on different redshifts and masses. In addition, the observed scatter can be different than the intrinsic value. More specifically, at the high-mass end an increased scatter can be inferred due to the uncertainties in removing passive objects, while at the low-mass end an increased scatter can be due to poor signal-to-noise ratio \footnote{According to \citet{Kurczynski2017} the intrinsic scatter is $\sim 0.10$/$\sim 0.15$ dex lower than the observed value at $z \sim 0.5-1.0 $/$\sim 2.5-3.0$.}. Due to the above conflicting results and limitations it is almost impossible to decipher if there is an evolution in the scatter of the relation and if it is mass/redshift dependent or not, solely by relying on observations.

Cosmological simulations are able to reproduce realistic star formation rates and stellar masses of galaxies \citep{TescariKaW2013,Katsianis2014,Furlong2014,Katsianis2017,Pillepich2018} and thus are a valuable tool to address the questions related to the shape of the  ${\rm \sigma_{\text{sSFR}}}$-${\rm M_{\star}}$, the value of the dispersion, and the way mechanisms important to galaxy evolution affect it. Simulations have limitations in resolution and box size.  Thus they suffer from small number statistics of galaxies at a given mass, especially at the high-stellar mass end and cosmic variance due to finite box-size. However, despite their limitations the retrieved properties of galaxies do not suffer from poor S/N at the low mass end or uncertainties brought by different methodologies employed in observational studies \citep{Katsianis2015,Katsianis2016} and thus can provide a useful guide to future surveys or address controversies in galaxy formation Physics. \citet{Dekel2009} pointed out that the scatter of the specific star formation rate - stellar mass relation in cosmological simulations is about 0.3 dex and driven mostly by the galaxies' gas accretion rates. \citet{Hopkins2014} using the FIRE zoom-in cosmological simulations studied the dispersion in the SFR smoothed over various time intervals  and pointed out that the star  formation  main  sequence  and  distribution of specific SFRs emerge naturally from the shape of the galaxies’ star formation histories, from $M_{\star} \sim 10^8 - 10^{11}$ $M_{\odot}$ at $z \sim 0-6$. The authors suggested that the scatter is larger on small timescales and masses, while dwarf galaxies ($< 10^8$) exhibit much more bursty SFHs (and therefore larger scatter) due to stochastic processes like star cluster formation, and their associated feedback. \citet{Matthee2018} argued that the scatter of the main sequence ${\rm \sigma_{\text{sSFR}}}$-${\rm M_{\star}}$ relation, defined by a sSFR cut in galaxies, is mass dependent and decreasing with mass at $z \sim0$, while presenting a comparison between the EAGLE reference model and SDSS data. The authors suggested that the scatter of the relation originates from a combination of fluctuations on short time-scales (ranging from 0.2-2 Gyr) that are presumably associated with self-regulation from cooling, star formation and outflows (which nature is stochastic), but is dominated by long time-scale variations \citep{Hopkins2014,Torrey2017}, which is governed by the SFHs of galaxies, especially at high masses ($\log(M_{\star}/M_{\odot}) > 10.0)$\footnote{The authors pointed out that the total scatter of $\sim 0.4$ is driven by a combination of short and long time-scale variations, while for massive galaxies ($\log(M_{\star}/M_{\odot}) = 10.0-11.0$) the contribution of stochastic fluctuations \citep{Kelson2014} is not significant  ($<$ 0.1 dex). For objects with $\log(M_{\star}/M_{\odot}) <  10.0$ the contribution of fluctuations on short time scales, which nature is more stochastic, becomes relatively more important ($\sim$ 0.2 dex).}. \citet{Dutton2010} using a semi-analytic model suggested that the scatter of the SFR sequence appears to be invariant with redshift and with a small value of ${\rm  < 0.2}$ dex. 

The Virgo project Evolution and Assembly of GaLaxies  and  their  Environments  simulations \citep[EAGLE,][]{Schaye2015,Crain2015} is a suite of cosmological hydrodynamical simulations in cubic, periodic volumes ranging from 25 to 100 comoving Mpc per side. The reference model reproduces the observed star formation rates of ${\rm z \sim0-8}$ galaxies \citep{Katsianis2017} and the evolution of the stellar mass function \citep{Furlong2014}. In addition, EAGLE allows us to investigate this problem with superb statistics (several thousands of galaxies at each redshift) and investigate different configurations that include different subgrid Physics. All the above provide a powerful resource for understanding the ${\rm \sigma_{\text{sSFR}}}$-${\rm M_{\star}}$ relation, address the shortcomings of observations, study its evolution across cosmic time and decipher its shape.

In this paper, we examine the dependence of the sSFR dispersion on ${\rm M_{\star}}$ using the EAGLE simulations \citep{Schaye2015,Crain2015,Katsianis2017}. In section \ref{EAGLE} we present the simulations used for this work. In section \ref{specificSFR-Ma} we discuss the evolution of the  ${\rm \sigma_{\text{sSFR}}}$-${\rm M_{\star}}$ relation (subsection \ref{specificSFR-M} presents the reference model) and how different feedback prescriptions (subsection \ref{specificSFR-MFeedback}) and ongoing mergers (subsection \ref{specificSFR-Mmergers}) affect its shape. In addition, we employ the EAGLE+SKIRT data \citep{Camps2018}, which represent a post-process of the simulations with the radiative transfer code SKIRT, in order to decipher how star formation rate and stellar mass diagnostics affect the relation in section \ref{specificSFR-Mdiagnostics}. Finally in section \ref{ConDis} we draw our conclusions. Studies of the dispersion that rely solely on 2d scatter plots (i.e. displays of the location of the individual sources in the plane) are not able to provide a quantitative information of how galaxies are distributed around the mean sSFR and cannot account for galaxies that could be under-sampled or missed by selection effects so we extend our analysis of the dispersion of the sSFRs at different mass intervals on their distribution/histogram, namely the specific Star Formation Rate Function (sSFRF) by comparing the results of EAGLE with the observations present in \citet{Ilbert2015}. In appendix \ref{SFRFEAGLE} we present the evolution of the simulated specific star formation rate function in order to present how the sSFRs are distributed.

\section{The EAGLE simulations used for this work}
\label{EAGLE}

The EAGLE simulations track the evolution of baryonic gas, stars, massive  black  holes  and  non-baryonic dark  matter particles from a starting  redshift  of ${\rm z = 127}$  down to ${\rm z = 0}$. The different runs were performed to investigate the effects of resolution, box size and various physical prescriptions (e.g. feedback and metal cooling). For this work we employ the reference model (L100N1504-Ref), a configuration with smaller boxsize (50 Mpc) but same resolution and physical prescriptions (L50N752-Ref), a run without AGN feedback (L50N752-NoAGN) and a simulation without SN feedback but AGN included (L50N752-OnlyAGN). We outline a summary of the different configurations in Table \ref{tab:sim_runs}.

\startlongtable
\begin{deluxetable}{cccc}
\tablecaption{The EAGLE cosmological simulations used for this work \label{tab:sim_runs}}
\tablehead{
\colhead{Run} & \colhead{L [Mpc]} & \colhead{N$_{\rm TOT}$} & \colhead{Feedback}  \\
}
\colnumbers
\startdata
  L100N1504-Ref & 100 & 2 $\times$ $1504^3$ & AGN $+$ SN \\ \hline
  L100N1504-Ref+SKIRT & 100 & 2 $\times$ $1504^3$ & AGN $+$ SN \\ \hline
  L50N752-Ref & 50 & 2 $\times$ $752^3$ & AGN $+$ SN  \\ \hline
  L50N752-NoAGN & 50 & 2 $\times$ $752^3$ & No AGN $+$ SN\\ \hline
  L50N752-OnlyAGN & 50 & 2 $\times$ $752^3$ & AGN $+$ No SN\\ \hline
\enddata
\tablecomments{Summary of the different EAGLE simulations used in this work. Column 1, run name; column 2, Box size of the simulation in comoving Mpc; column 3, total number of particles (N$_{\rm TOT} =$ N$_{\rm GAS}$ $+$ N$_{\rm DM}$ with N$_{\rm GAS}$ $=$ N$_{\rm DM}$); column 4, combination of feedback implemented; The mass of the dark matter particle m$_{\rm DM}$ is 9.70$\times10^{6}$ [$\rm M_{\odot}$], the mass of the initial mass of the gas particle m$_{\rm gas}$ is 1.81$\times10^{6}$ [$\rm M_{\odot}$] and the comoving gravitational softening length ${\rm \epsilon_{com}}$ is 2.66 in KPc in all configurations.}
\end{deluxetable}

The EAGLE reference simulation has $2 \, \times \, 1504^{3}$ particles (an  equal  number of gas and dark matter elements) in an L =  100 comoving Mpc volume box, initial gas particle mass of ${\rm m_{g}} =  1.81 \times 10^6 \, {\rm M_{\odot}}$, and mass of dark matter particles of ${\rm m_{g}} =  9.70 \times 10^6 \, {\rm M_{\odot}}$. The simulations  were run using an improved and updated version  of  the N-body  TreePM  smoothed  particle  hydrodynamics code GADGET-3 \citep{Springel2005} and employ the star formation recipe of \citet{Schaye2008}. In this scheme, gas with densities exceeding the critical density for the onset of the thermo-gravitational instability (${\rm n_H \sim 10^{-2} - 10^{-1} \, cm^{-3}}$) is treated as a multi-phase mixture of cold molecular clouds, warm atomic gas and hot ionized bubbles which are all approximately in pressure equilibrium \citep{Schaye2004}. The above mixture is modeled using a polytropic equation of state  ${\rm P = k \rho ^{\gamma _{eos}}}$, where P is the gas pressure, ${\rm \rho}$ is the gas density and ${\rm k}$ is a constant which is normalized to ${\rm P/k = 10^3 \, cm^{-3} \, K}$ at the density threshold ${\rm n_H^{\star}}$ which marks the onset of star formation. The simulations adopt the stochastic thermal feedback scheme described in \citet{DVecchia2012}. In addition to the effect of re-heating interstellar gas from star formation, which  is  already  accounted for by the equation of state, galactic winds produced by Type II Supernovae are also considered. EAGLE models AGN feedback by seeding galaxies with BHs as described by \citet{Springel2005}, where seed BHs are placed at the center of every halo more massive than $10^{10}$ M$_{\rm\odot}/{\rm h}$ that does not already contain a BH. When a seed is needed to be implemented at a halo, its highest density gas particle is converted into a collisionless BH particle inheriting the particle mass. These BHs grow by accretion of nearby gas particles or through mergers. A radiative efficiency of ${\rm \epsilon_r = 0.1}$ is assumed for the AGN feedback.  Other prescriptions such as inflow-induced starbursts, stripping of gas due to different interactions between galaxies, stochastic IMF sampling or variations to the AGN feedback prescription such as torque-driven accretion models \citep{Angles-Alcazar2017} or kinetic feedback \citep{Weinberger2017} are not currently modeled in EAGLE. The EAGLE reference model and its feedback prescriptions have been calibrated to reproduce key observational constraints, into the present-day stellar mass function of galaxies \citep{LiWhite2009,Baldry2012}, the correlation between the black hole and bulge masses \citep{McConnell2013} and the dependence of galaxy sizes on mass \citep{Baldry2012} at $z \sim 0$. Alongside with these observables the simulation was able to match many other key properties of galaxies in different eras, like molecular hydrogen abundances \citep{Lagos2015}, colors and luminosities at ${\rm z \sim 0.1}$ \citep{Trayford2015}, supermassive black hole mass function \citep{Rosas2016}, angular momentum evolution \citep{Lagos2017}, atomic hydrogen abundances \citep{Crain2017}, sizes \citep{Furlong2017}, SFRs \citep{Katsianis2017}, Large-scale outflows \citep{Tescari2018} and ring galaxies \citep{Elagali2018}. In addition, \citet{Schaller2015} pointed out that there is a good agreement between the normalization and slope  of  the  main  sequence  present in \citet{Chang2015} and the  EAGLE reference model. \citet{Katsianis2015} demonstrated that cosmological hydrodynamic simulations like EAGLE, Illustris \citep{Sparre2014} and ANGUS \citep{TescariKaW2013} produce very similar results for the SFR-$M_{\star}$ relation with a normalization being in agreement with that found in observations at $z \sim 0-4$ \citep{Kajisawa2010,Delosrayes2014,Bauer2013,Salmon2015} and a slope close to unity. In this work, galaxies and their host halos are identified by a friends-of-friends (FoF) algorithm \citep{Davis1985} followed by the SUBFIND algorithm \citep{springel2001,DolagSta2009} which is used to identify substructures or subhalos across the simulation. The star formation rate of each galaxy is defined to be the sum of the star formation rate of all gas particles that belong to the corresponding subhalo and that are within a 3D aperture with radius 30 kpc \citep{Schaye2015,Crain2015,Katsianis2017}.

\begin{deluxetable}{ccccccccc}
  \tablecaption{Fraction of passive galaxies excluded in order to define a main sequence \label{tab:sim_runs2}}
  \tablehead{
\colhead{z} & \colhead{0} & \colhead{0.350} & \colhead{0.615} & \colhead{0.865} & \colhead{1.400} & \colhead{2.000} & \colhead{3.000} & \colhead{4.000} }
\startdata
$\sigma_{\text{sSFR, MS}}$-$M_{\star}$ sSFR Cut ($yr^{-1}$) & $ 10^{-11.0} $ & $ 10^{-10.8}$ & $ 10^{-10.3}$ & $ 10^{-10.2}$ & $ 10^{-9.9}$ & $ 10^{-9.6}$ & $ 10^{-9.4}$  &  $ 10^{-9.1}$ \\
$F_{Passive}$, $\log(M_{\star}/M_{\odot}) = 10^{8.0-9.5}$ & 0.08 & 0.06 & 0.10 & 0.07 & 0.04 & 0.08  & 0.08 & 0.05 \\
$F_{Passive}$, $\log(M_{\star}/M_{\odot}) = 10^{9.5-10.5}$ & 0.12  & 0.10 & 0.09 & 0.07 & 0.04 & 0.08 & 0.12 & 0.09 \\
$F_{Passive}$, $\log(M_{\star}/M_{\odot}) = 10^{10.5-11.0}$ & 0.32 &  0.24 & 0.37 & 0.32 & 0.27 & 0.33 & 0.35 & 0.29 \\
&  & & & & & & & \\ \hline
$\sigma_{\text{sSFR, MS, Moderate}}$-$M_{\star}$ sSFR Cut ($yr^{-1}$) & \citet{Guo2015} & $ 10^{-11.0}$ & $ 10^{-11.0} $ & $ 10^{-11.0} $ & $ 10^{-11.0}$ & $ 10^{-11.0}$ & $ 10^{-11.0} $ &  $ 10^{-11.0}$ \\
$F_{Passive}$, $\log(M_{\star}/M_{\odot}) = 10^{8.0-9.5}$  & 0.02  & 0.03 & 0.02 &  0.03 &  0.02 & 0.02 & 0.01 & $<$ 0.01\\
$F_{Passive}$, $\log(M_{\star}/M_{\odot}) = 10^{9.5-10.5}$  & 0.03  & 0.05  & 0.03  &  0.02 & 0.01 & 0.01 & 0.01 & $<$ 0.01 \\
$F_{Passive}$, $\log(M_{\star}/M_{\odot}) = 10^{10.5-11.0}$  & 0.17  &  0.16 & 0.17  & 0.10 & 0.04 &  0.02 & 0.01 & 0.01 \\
&  & & & & & & & \\
\enddata
\tablecomments{The fraction of passive galaxies ($F_{Passive}$) at each redshift excluded in order to define a main sequence. We adopt the effect of two different sSFR cuts. The first criterion \citep{Furlong2014,Matthee2018} is used to define the  $\sigma_{\text{sSFR, MS}}$-$M_{\star}$ relation, while the second, more moderate criterion \citep{Ilbert2015,Guo2015} is used to define the $\sigma_{\text{sSFR, MS, Moderate}}$-$M_{\star}$ relation.}
\end{deluxetable}

\begin{table}
  \caption{Summary of the different observations used for this work.}
  \resizebox{0.99\textwidth}{!}{%
\begin{tabular}{llcccccccccc}
  \\ \hline & Publication & Redshift range & Technique to obtain  & $ \langle \sigma_{\text{sSFR}} \rangle $ $\pm$ Uncertainty, $\sigma_{\text{sSFR}}$ with $M_{\star}$ $ \uparrow $  [dex] &   \\ & & Stellar mass range &  sSFRs and SFRs & Intrinsic or Observed $\sigma_{\text{sSFR}}$, Shape of $\sigma_{\text{sSFR}}$-$M_{\star}$  & &  \\ \hline \hline

  & \citet{Noeske2007} & z = 0.32, 0.59, 1.0 & EL+UV+$IR_{24 \mu m}$ & 0.3 $\pm 0.05$ [0.3 $\rightarrow $  0.3] &  \\

  &  & $\log(M_{\star}/M_{\odot}) = 9.5 - 11.45$  &  &  Observed, Constant \\ \hline

  & \citet{Rodighiero10} & z = 1.47, 2.2 & UV+$IR_{24 \mu m}$  &  0.3 $\pm 0.05$ [0.3 $\rightarrow $  0.3] &  \\

   &  & $\log(M_{\star}/M_{\odot}) = 9.5 - 11.45$  &  & Observed, Constant \\ \hline

  & \citet{Guo2013} & z = 0.7 &  UV+$IR_{24 \mu m}$  & 0.24 $\pm 0.04$ [0.182 $\nearrow$  0.307] & \\

   &  & $\log(M_{\star}/M_{\odot}) = 9.75 - 11.25$  &  &  Observed, increases with $M_{\star}$ \\ \hline

  & \citet{Schreiber2015} & z = 0.5, 1.0, 1.5, 2.2, 3.0 & $UV+IR_{SED}$ &  0.35 $\pm 0.03$ [0.29  $\nearrow$ 0.37]  & \\

   &  & $\log(M_{\star}/M_{\odot}) = 9.45 - 10.95$  &  & Observed, increases with $M_{\star}$ \\ \hline

  & \citet{Ilbert2015} & z = 0.3, 0.7, 0.9, 1.3,  & $UV+IR_{SED}$   & 0.33 $\pm 0.03$ [0.22 $\nearrow$ 0.481]  & \\

   &  & $\log(M_{\star}/M_{\odot}) = 9.75-11.25$  &  & Observed, increases with $M_{\star}$ \\ \hline

  & \citet{Willet2015} & z $ < 0.085$ & $UV_{SED}+H \alpha$  & 0.33 $\pm 0.03$ [0.52 $ \searrow  $ 0.37 $\nearrow$  0.48]  &  \\

   &  & $\log(M_{\star}/M_{\odot}) = 8.35 - 11.5$  &  &  Observed, U-Shape  \\ \hline

  & \citet{Guo2015} &
  $0.01 < z < 0.03 $ & $H \alpha +IR_{22 \mu m}$  &  0.44 $\pm 0.012$ [0.366 $\nearrow$ 0.557] &  \\

   &  & $\log(M_{\star}/M_{\odot}) = 8.85 - 10.75$  &  & Observed, increases with $M_{\star}$ \\ \hline

  & \citet{Kurczynski2017} & z = 0.75, 1.25, 1.75, 2.25, 2.75 & SED fitting  & 0.40 $\pm 0.02$ [0.404 $ \searrow  $ 0.315 $\nearrow$ 0.435] &  \\

   &  & $\log(M_{\star}/M_{\odot}) = 6.85 - 10.25$  &  & Intrinsic, redshift/mass dependent \\ \hline

  & \citet{Santini2017} & z = 1.65, 2.5, 3.5 & UV + $ \beta $  slope & 0.42 $\pm 0.05$ [0.54 $ \searrow  $  0.31] &  \\

   &  & $\log(M_{\star}/M_{\odot}) = 6.85-10.25 $  &  & Observed, decreases with $M_{\star}$ \\ \hline

  & \citet{Boogaard2018} & z = 0.1 - 0.9 & H$ \alpha $+H$ \beta $ &  0.44 $\pm ^{0.05}_{0.04}$  [0.44 $\rightarrow $  0.44] &   \\

   &  & $\log(M_{\star}/M_{\odot}) = 8.0-10.5 $  &  &  Intrinsic, constant \\ \hline

  & \citet{Davies2019} & z $ < $ 0.1 & $SFR_{H \alpha}$ &  0.66 $\pm 0.02$ [0.74308562 $ \searrow  $ 0.53393775 $\nearrow$ 0.70720613] &  \\

   &  & $\log(M_{\star}/M_{\odot}) = 7.5 - 11.0$  &  & Observed, U-shape \\ \hline
  
  & \citet{Davies2019} & z $ < $ 0.1 & $SFR_{W}$ &  0.44 $\pm 0.02$ [0.42797872 $ \searrow  $   0.3490565 $\nearrow$ 0.53074792] &  \\

   &  & $\log(M_{\star}/M_{\odot}) = 7.5 - 11.0$  &  & Observed, U-Shape  \\ \hline
\end{tabular}%
}
\vspace{-0.1cm}
\medskip\\
\tablecomments{Summary of the different observations used in this work. Column 1, Publication name ; Column 2 (top), Redshift range ; Column 2 (bottom), Stellar mass range ; Column 3, Technique to obtain galaxy sSFRs and SFRs; Column 4 (top), Average $\sigma_{\text{sSFR}}$ $\pm$ uncertainty, behavior of the scatter with increasing mass at the lowest redshift considered by the authors (Column 2, top) ; Column 4 (bottom), Type of scatter used by the authors (Intrinsic or Observed), Shape of the $\sigma_{\text{sSFR}}$-$M_{\star}$.}
\label{Observationspre}
\end{table}

\section{The evolution of the intrinsic $\sigma_{\text{sSFR}}$-$M_{\star}$ relation}
\label{specificSFR-Ma}

\begin{figure*}
  \centering
\includegraphics[scale=0.42]{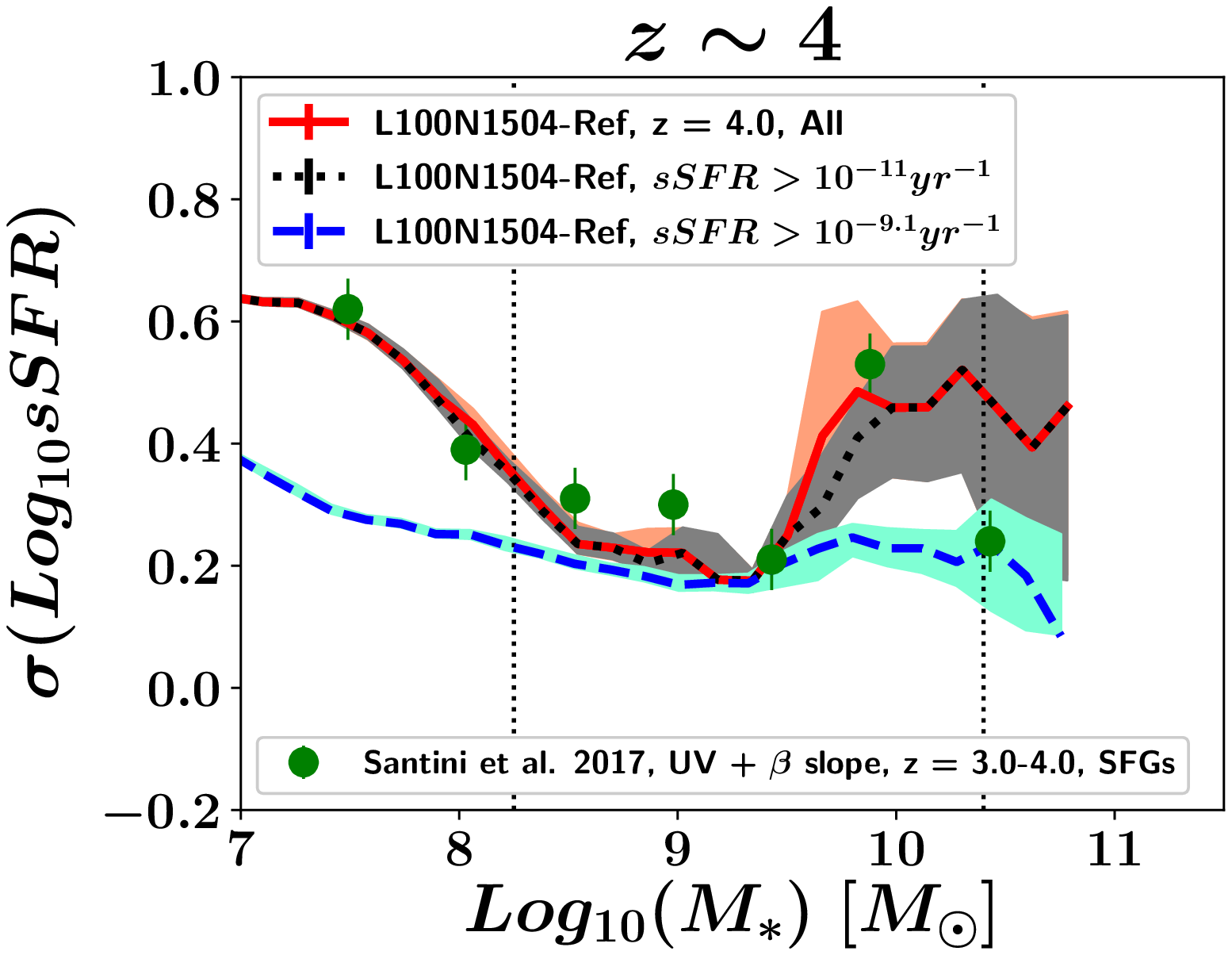} \hspace{-1.0em}  \vspace{0.5em}
\includegraphics[scale=0.42]{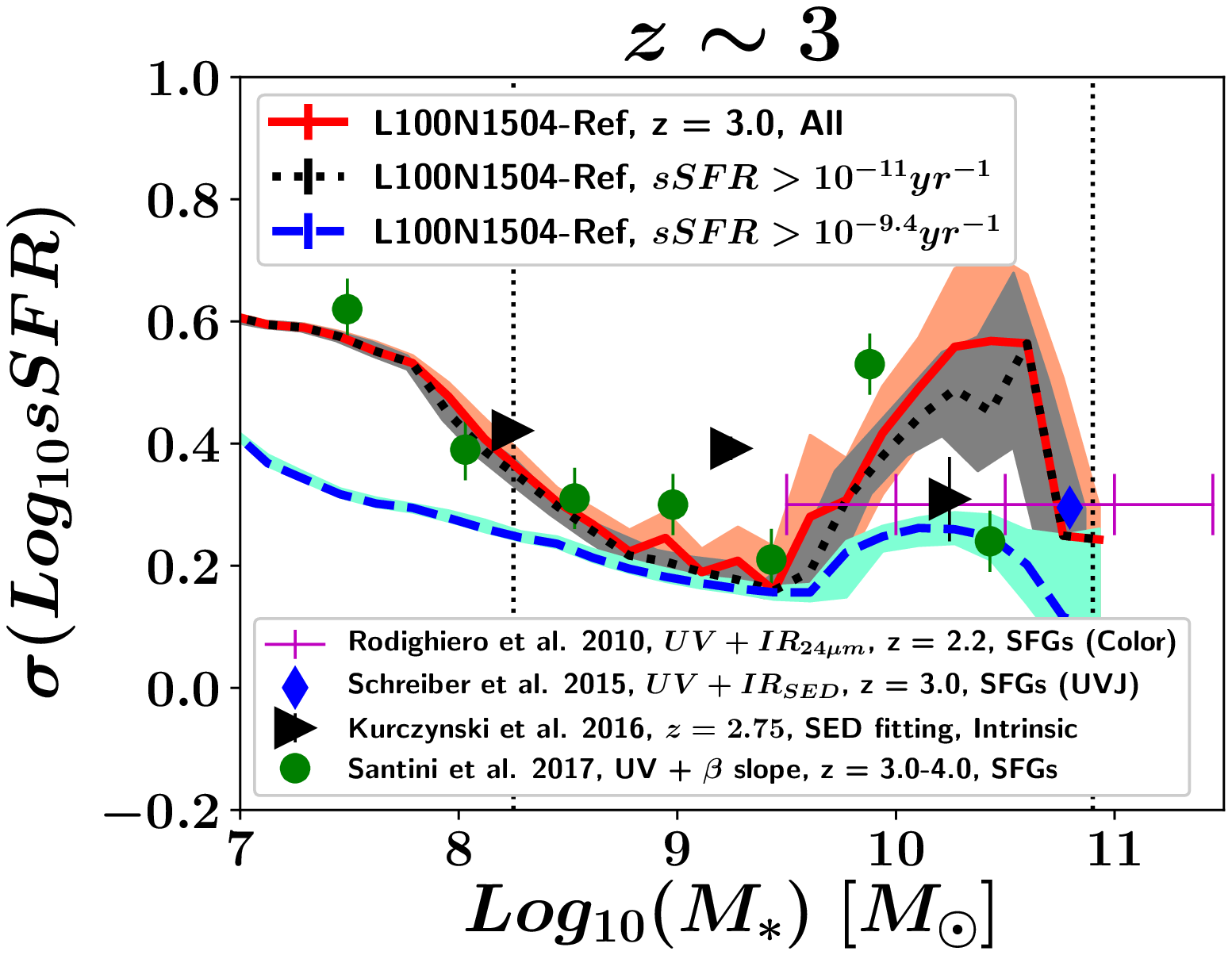} \hspace{-1.0em} 
\includegraphics[scale=0.42]{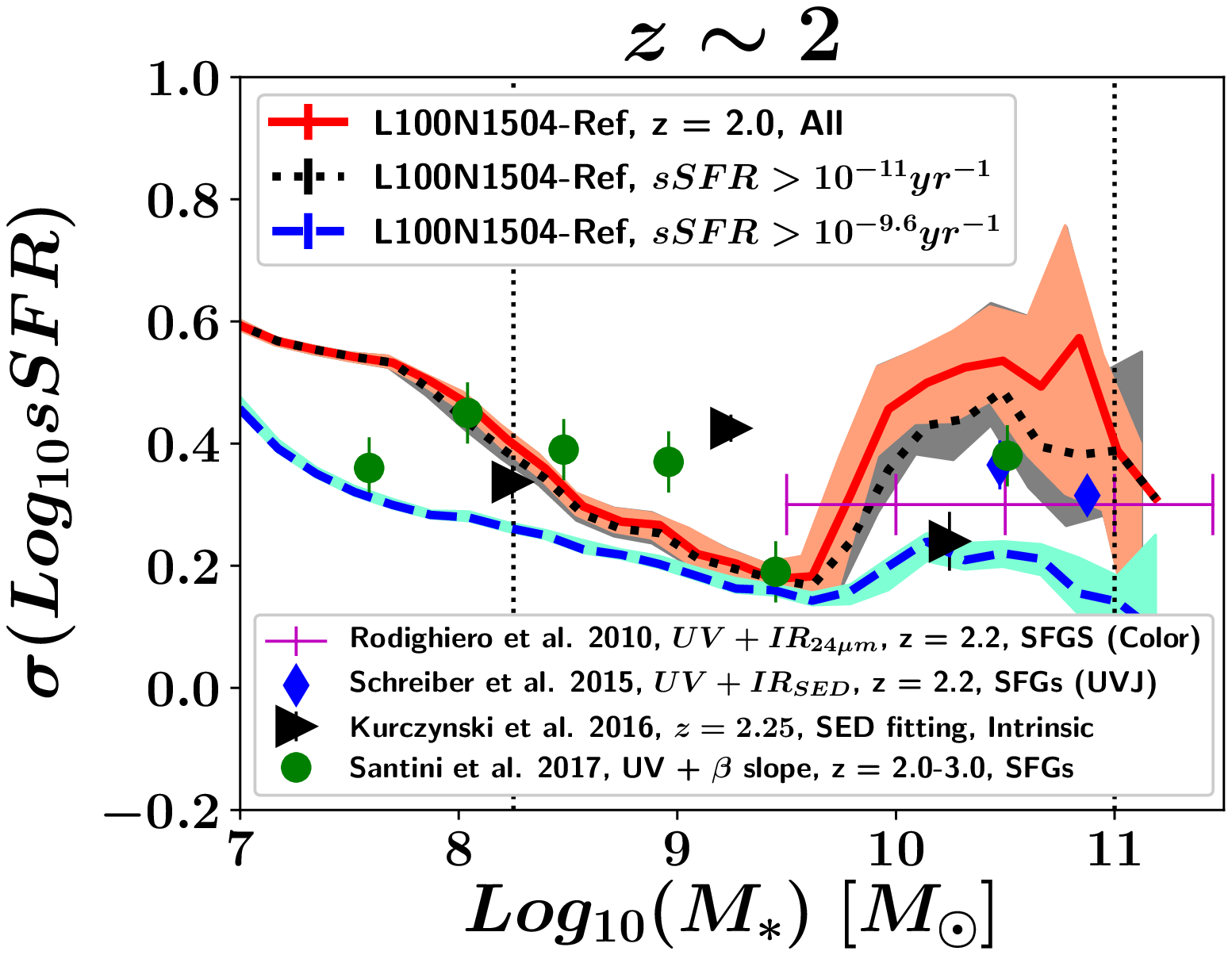}   \hspace{-1.0em} \vspace{0.5em}
\includegraphics[scale=0.42]{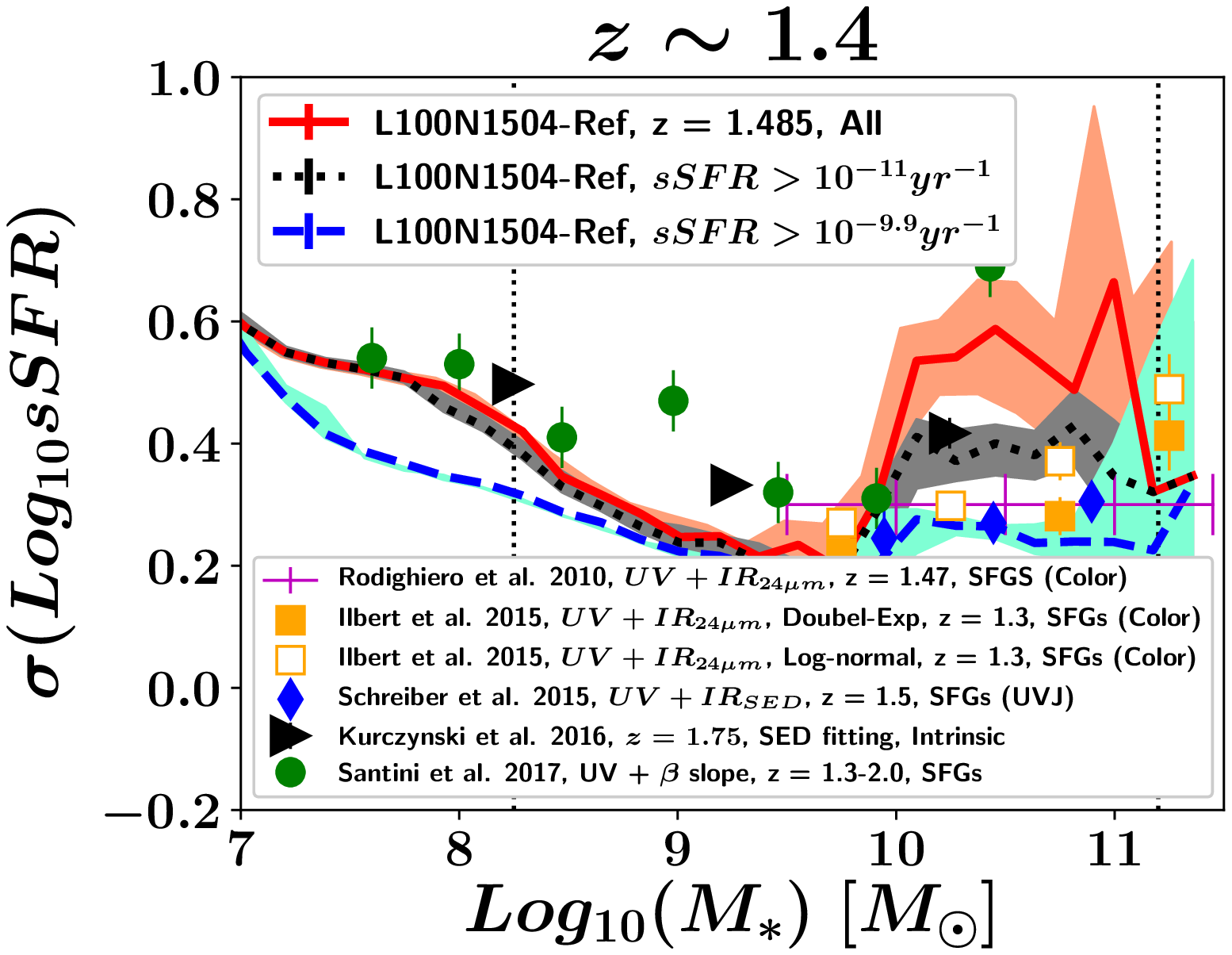} \hspace{-1.0em} 
\includegraphics[scale=0.42]{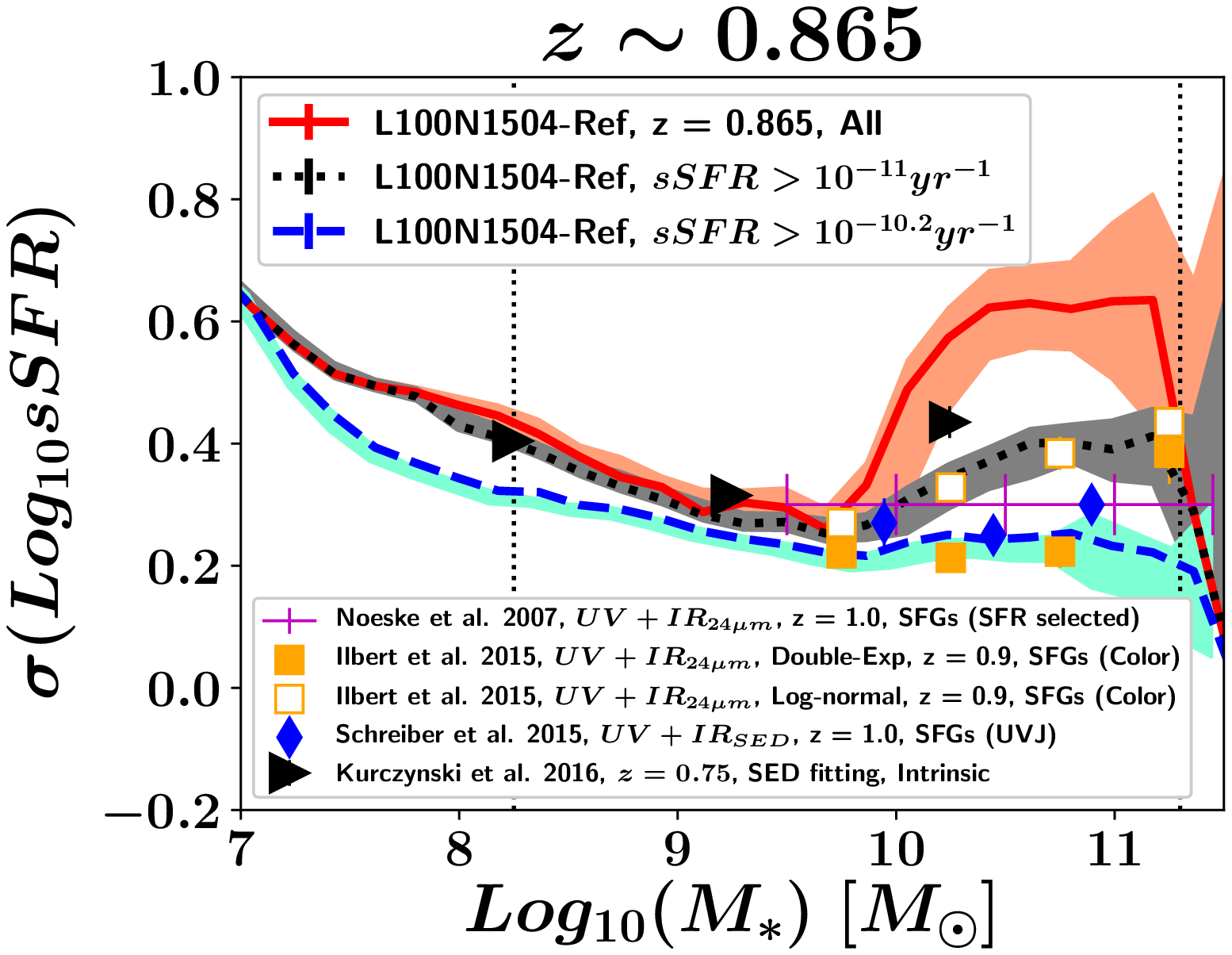} \hspace{-1.0em} 
\includegraphics[scale=0.42]{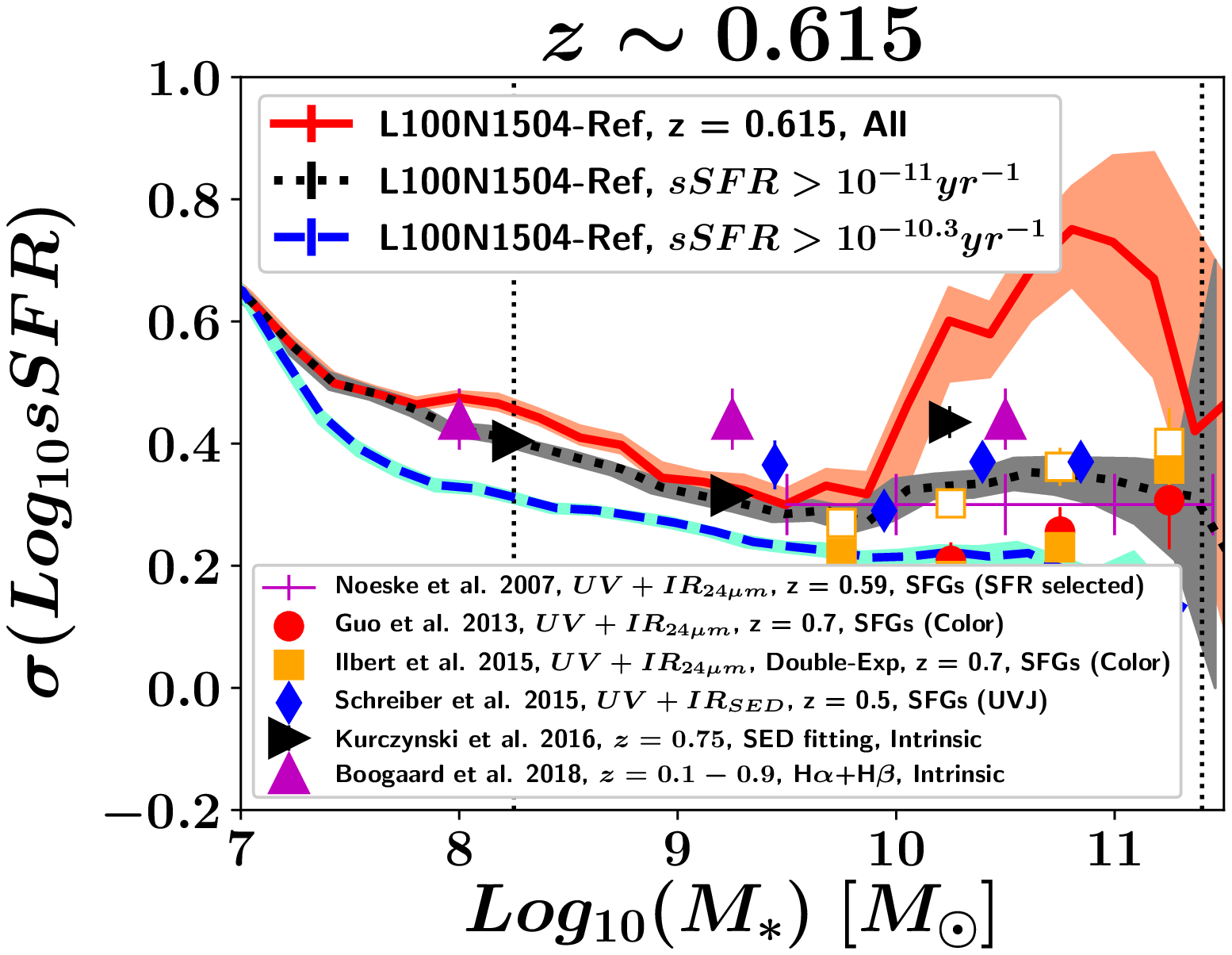} \hspace{-1.0em} 
\includegraphics[scale=0.42]{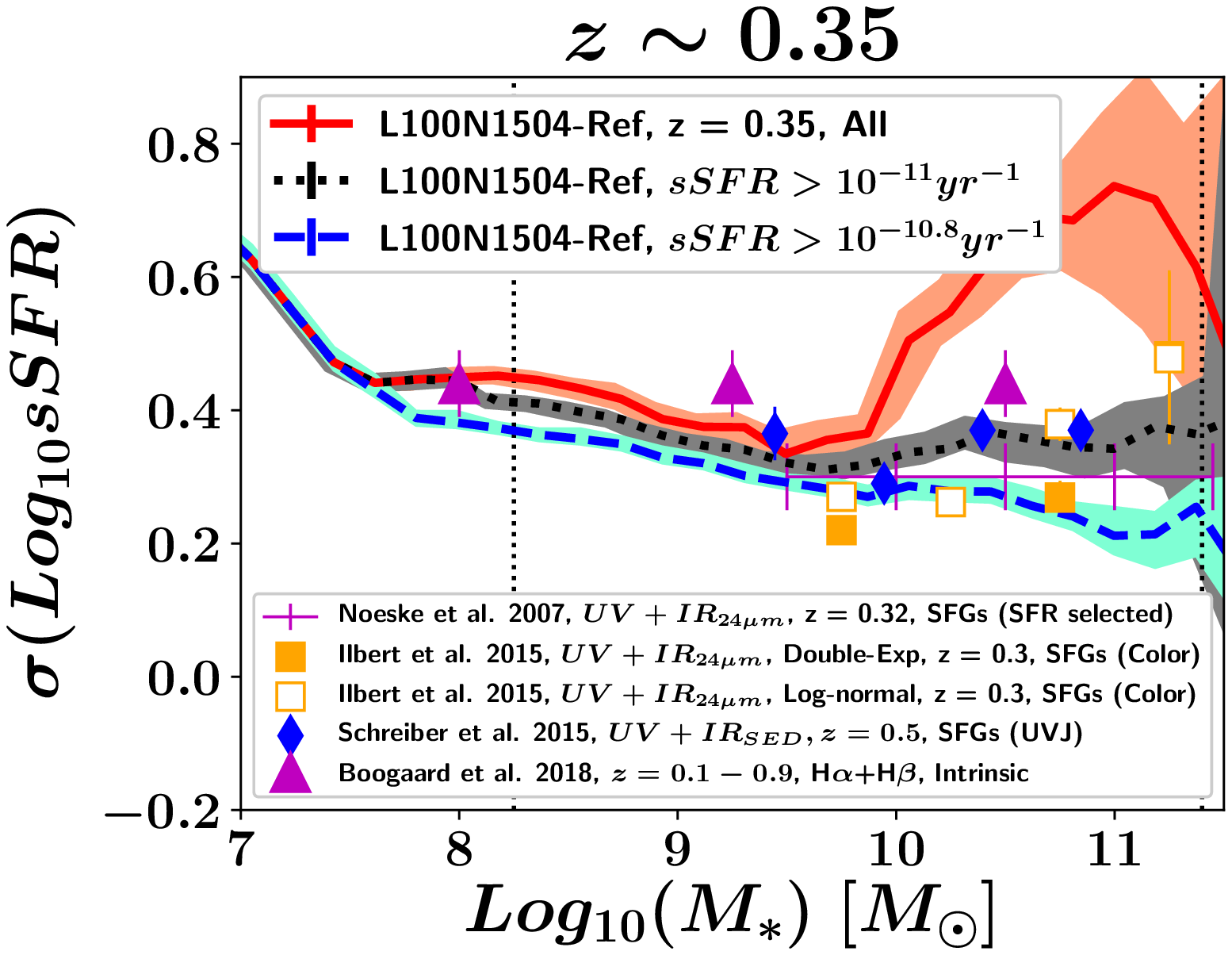} \hspace{-1.0em} \vspace{0.5em}
\includegraphics[scale=0.42]{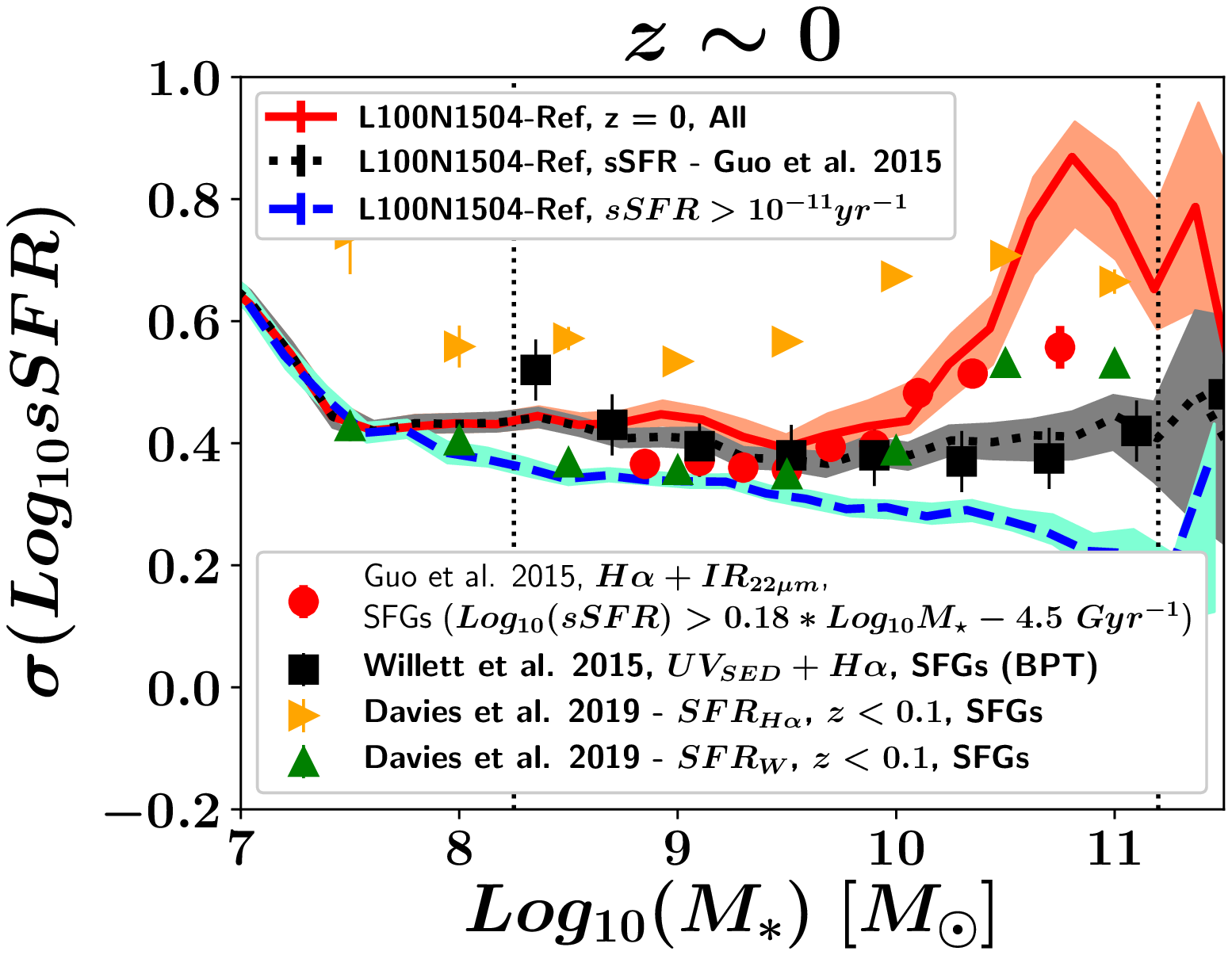}
\vspace{-0.9em}%
\caption{The evolution of the  $\sigma_{\text{sSFR}}$-$M_{\star}$, $\sigma_{\text{sSFR, MS, Moderate}}$-$M_{\star}$ and $\sigma_{\text{sSFR, MS}}$-$M_{\star}$ relations at $z \sim 0-4$ of the EAGLE reference model, L100N1504-Ref (red solid line, black dotted line and blue dashed line, respectively). The vertical dotted lines represent the mass limit of 100 baryonic particles and the statistical limit where there are fewer than 10 galaxies  at the low- and high-mass ends \citep{Furlong2014,Katsianis2017}, respectively. The pink, grey and cyan areas represent the 95 $\%$ bootstrap confidence interval for 5000 re-samples for the $\sigma_{\text{sSFR}}$-$M_{\star}$, $\sigma_{\text{sSFR}, MS, Moderate}$-$M_{\star}$ and $\sigma_{\text{sSFR}, MS}$-$M_{\star}$ relations, respectively. For both $\sigma_{\text{sSFR}}$-$M_{\star}$ and $\sigma_{\text{sSFR, Moderate}}$-$M_{\star}$ the scatter decreases with mass for the  $\log(M_{\star}/M_{\odot}) \sim 8-9.5$ interval but then increases at $\log(M_{\star}/M_{\odot}) \sim 9.5-11.0$. This U-shape behavior is consistent with recent observations \citep{Guo2013,Ilbert2015,Schreiber2015,Willet2015,Santini2017}. On the other hand, the scatter for the $\sigma_{\text{sSFR, MS}}$-$M_{\star}$ relation is constant with mass at the $\log(M_{\star}/M_{\odot}) \sim 9.5-11.0$ interval around $\sim 0.2-0.3$ dex for $z \sim 0.35-0.85$, while at $z \sim0 $ the $\sigma_{\text{sSFR, MS}}$ decreases with mass.}
\label{fig:sSFRFscatter}
\end{figure*}

In this section we present the evolution of the  $\sigma_{\text{sSFR}}$-$M_{\star}$ relation in order to quantify and decipher its evolution and its dependence (or not) upon stellar mass and redshift. In subsection \ref{specificSFR-M} we present the results of the EAGLE reference model and the compilation of observations used in this work, while in subsections \ref{specificSFR-MFeedback} and \ref{specificSFR-Mmergers} we focus on the effect of feedback and mergers on the  $\sigma_{\text{sSFR}}$-$M_{\star}$ relation, respectively. For the simulations,  we  split  the  sample of galaxies  at each redshift in 30 stellar mass bins from $\log(M_{\star}/M_{\odot}) \sim 6.0$ to $\log(M_{\star}/M_{\odot}) \sim 11.5$ (stellar mass bins of 0.18 dex at $z \sim 0$) and measure the 1 $\sigma$ standard deviation $\sigma(\log_{10}sSFR)$ in each bin.

We compare our simulated results with a range of observational studies in which usually different authors employ different techniques to exclude quiescent objects in their samples. In order to select only Star Forming Galaxies (SFGs) the authors may select only blue cloud galaxies \citep{Peng2010}, or use the BzK two-color selection \citep{Daddi2007},  or the standard BPT \citep{Baldwin1981} criterion, or employ the rest-frame UVJ selection \citep{Whitaker2012,Schreiber2015} or an empirical color selection \citep{Rodighiero10,Guo2013,Ilbert2015} or specify a sSFR separation criterion \citep{Guo2015}. All these criteria  {\it should} ideally cut out galaxies  with  low  sSFR,  but the thresholds differ significantly in value from  one study to another, with some being redshift dependent and others not \citep{Renzini2015}. In order to surpass this complication and uncertainty of the effectiveness of excluding ``passive objects'' in observational studies, in the following subsection we present:
\begin{itemize}
\item the $\sigma_{\text{sSFR}}$-$M_{\star}$ of the full (Star forming $+$ Passive) EAGLE sample,
\item the $\sigma_{\text{sSFR, MS}}$-$M_{\star}$ relation of a ``main sequence'' defined by excluding passive objects with sSFRs $ < 10^{-11} $yr$^{-1}$, $ < 10^{-10.8} $yr$^{-1}$, $ < 10^{-10.3}$ yr$^{-1}$, $ < 10^{-10.2}$ yr$^{-1}$, $ < 10^{-9.9}$ yr$^{-1}$, $ < 10^{-9.6} $yr$^{-1}$, $ < 10^{-9.4} $yr$^{-1}$, $ < 10^{-9.1} $yr$^{-1}$ for redshifts $z = 0$, $z = 0.35$, $z = 0.615$, $z = 0.865$, $z = 1.485$, $z = 2.0$, $z = 3.0$, $z = 4.0$, respectively \citep{Furlong2014,Matthee2018} and
\item the $\sigma_{\text{sSFR, MS, Moderate}}$-$M_{\star}$ relation of a ``main sequence'' defined by the more conservative sSFR cuts (that ensures a more complete SF sample in the expense of some possible passive galaxy contamination) of $ < 10^{-11.0} yr^{-1}$ for $z > 0$ and $ \log10(sSFR) < 0.18 \times \log10(M_{\star}) - 4.5 \, Gyr^{-1}$ for $z = 0$ \citep{Ilbert2015,Guo2015}.
\end{itemize}

The differences between observations used in this work in terms of assumed methodology to exclude (or not) quiescent objects is described in the previous paragraph and table \ref{tab:sim_runs2}. The different data sets and methodologies to obtain SFRs for the same studies are described below and table \ref{Observationspre}. \citet{Noeske2007} used 2095 field galaxies from the Wavelength Extended Groth Strip International Survey (AEGIS) and derived SFRs from emission lines, GALEX, and Spitzer MIPS and 24$\mu$m photometry. \citet{Guo2013} used 12,614 objects from the multi-wavelength data set of COSMOS, while SFRs are obtained combining 24 $\mu$m and UV luminosities. \citet{Schreiber2015} used GOODS-North, GOODS-South, UDS, and COSMOS extragalactic fields and derived SFRs using UV+FIR luminosities. \citet{Ilbert2015} based their analysis on a 24 $\mu$m selected catalogue combining the COSMOS and GOODS surveys. The authors estimated SFRs by combining mid- and far-infrared data for 20,500 galaxies. \citet{Willet2015} used optical observations in the SDSS DR7 survey, while stellar  masses and star formation rates are computed from optical diagnostics and taken from the MPA-JHU catalogue \citep{Salim2007}. \citet{Guo2015} used the SDSS data release 7, while in their studies SFRs are estimated from H$\alpha$ in combination with 22 $\mu$m observation from the Wide-field Infrared Survey Explorer. \citet{Santini2017} used the Hubble Space Telescope Frontier Fields, while SFRs were estimated from observed UV rest-frame photometry \citep{meurer1999,Kennicutt2012}. \citet{Davies2019} used 9,005 galaxies from the GAMA survey \citep{Driver2011,Driver2016}. The SFR indicators used are described in length in \citet{Driver2016b} and involve a) the SED fitting code magphys, b) a combination of Ultraviolet and Total Infrared (UV+TIR), c) H$\alpha$ emission line d) the Wide-field  Infrared  Survey  Explorer (WISE)  W3-band \citep{Cluver2017}, and e) extinction-corrected u-band luminosities derived using  the  GAMA  rest-frame u -band  luminosity  and u-g colors.

\begin{figure*}
\centering
\includegraphics[scale=0.9]{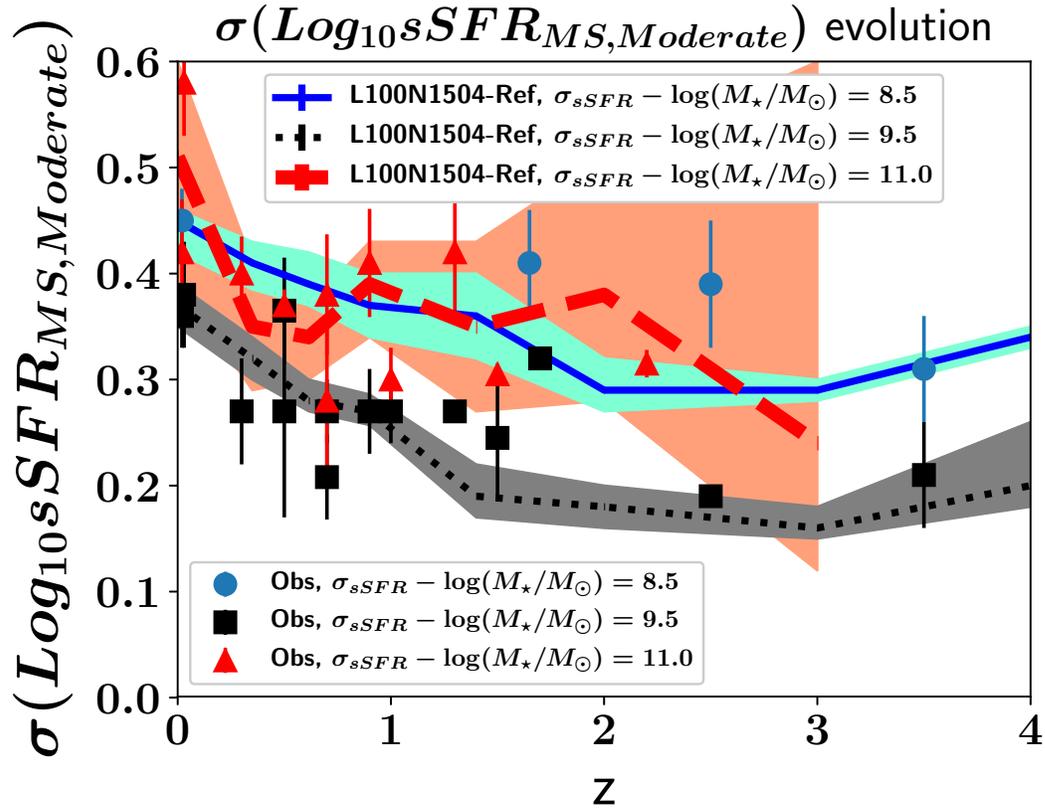} \hspace{-2.5em} 
\vspace{-0.9em}%
\caption{The evolution of the $\sigma_{\text{sSFR}, MS, Moderate}$ relation at $z \sim 0-4$ for the stellar masses of  $\log(M_{\star}/M_{\odot}) \sim 8.5$ (blue solid line), $\log(M_{\star}/M_{\odot}) \sim 9.5$ (black dotted line), $\log(M_{\star}/M_{\odot}) \sim 11.0$ (red dashed line). The blue circles, black squares and red triangles represent the compilation of observations \citep{Guo2013,Ilbert2015,Schreiber2015,Willet2015,Santini2017} at $\log(M_{\star}/M_{\odot}) \sim 8.5$, $\log(M_{\star}/M_{\odot}) \sim 9.5$ and $\log(M_{\star}/M_{\odot}) \sim 11.0$, respectively. Both in simulations and the above observations the $\sigma_{\text{sSFR}}$ at $\log(M_{\star}/M_{\odot}) \sim 8.5$ increases steadily from $\sim$ 0.3 dex at $z \sim 4$ to $\sim$ 0.55 dex. For $\log(M_{\star}/M_{\odot}) \sim 9.5$ the scatter remains almost constant at $\sim0.2$ dex for $z \sim 4.0-1.5$ but increases up to 0.35 at $z \sim 0$ for the redshift interval of $z \sim 1.5-0$. Last, the scatter increases from 0.2 dex to 0.45 dex at $\log(M_{\star}/M_{\odot}) \sim 11.0$. We note that the scatter around the characteristic mass ($\log(M_{\star}/M_{\odot}) \sim 9.5$, black dotted line) is always smaller than that found at the low- and high- mass ends.}
\label{fig:sSFRFscatterevl}
\end{figure*}

\begin{figure*}
\centering
\includegraphics[scale=0.9]{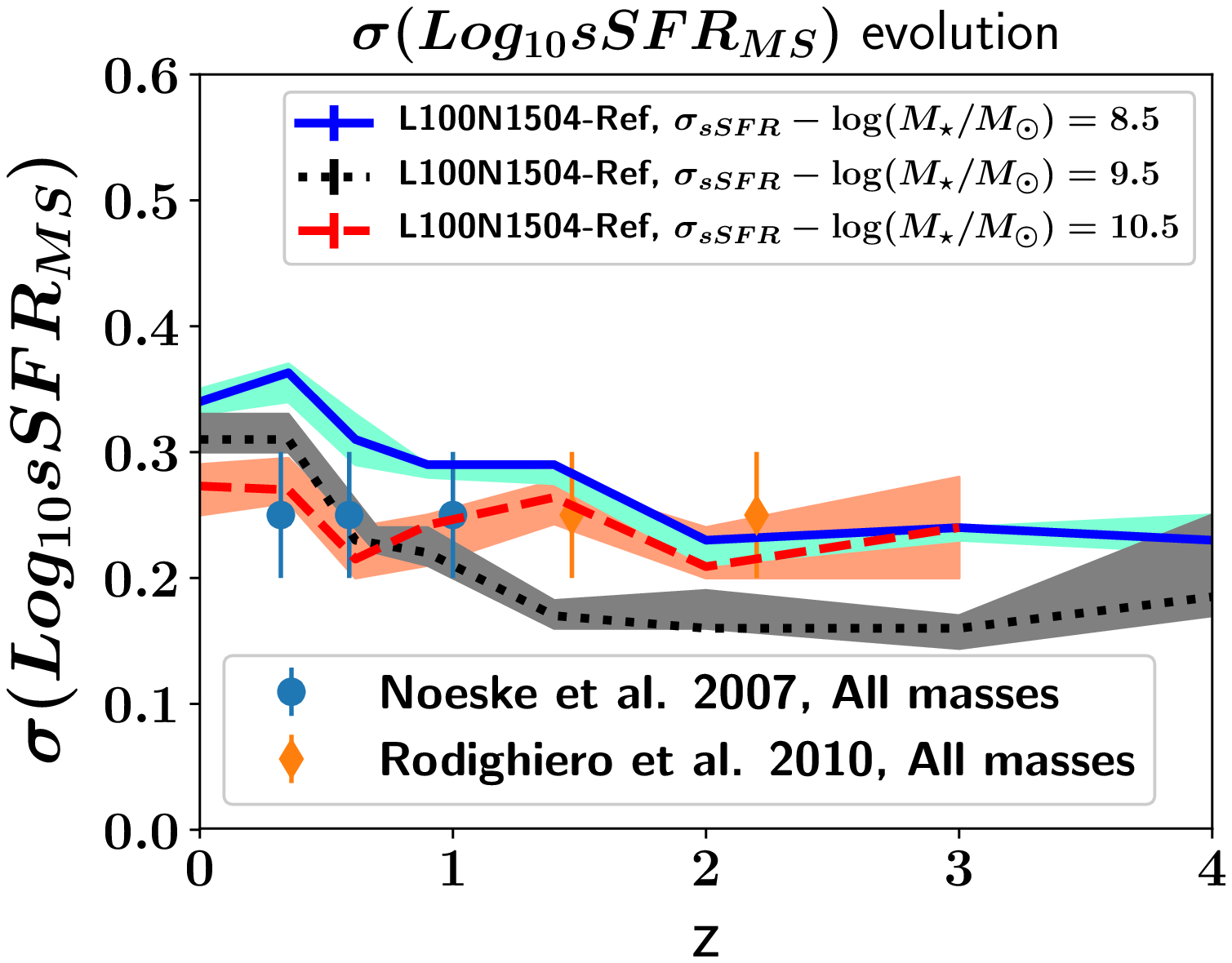} \hspace{-2.5em} 
\vspace{-0.9em}%
\caption{The evolution of the $\sigma_{\text{sSFR}, MS}$ relation at $z \sim 0-4$ for the stellar masses of  $\log(M_{\star}/M_{\odot}) \sim 8.5$ (blue solid line), $\log(M_{\star}/M_{\odot}) \sim 9.5$ (black dotted line), $\log(M_{\star}/M_{\odot}) \sim 11.0$ (red dashed line). Alongside the observations of \citet{Noeske2007} and \citet{Rodighiero10}, which suggest that the scatter is constant and non-evolving at $0.2-0.3$ dex. We note that the $\sigma_{\text{sSFR}, MS}$ in simulations is time-dependent at  $\log(M_{\star}/M_{\odot}) \sim 8.5$ and  $\log(M_{\star}/M_{\odot}) \sim 9.5$.}
\label{fig:sSFRFscatterevlMain}
\end{figure*}

\subsection{The evolution of the $\sigma_{\text{sSFR}}$-$M_{\star}$, $\sigma_{\text{sSFR, MS}}$-$M_{\star}$ and  $\sigma_{\text{sSFR, MS, Moderate}}$-$M_{\star}$ relations in the EAGLE model}
\label{specificSFR-M}

In Fig. \ref{fig:sSFRFscatter} we present the evolution of the  $\sigma_{\text{sSFR}}$-$M_{\star}$, which includes both passive and star forming objects (represented by the red solid line), and the ``main sequence'' $\sigma_{\text{sSFR, MS}}$-$M_{\star}$ (represented by the black dotted line) and $\sigma_{\text{sSFR, MS, Moderate}}$-$M_{\star}$ (represented by the blue dashed line) relations in the EAGLE L100N1504-Ref at $\log(M_{\star}/M_{\odot}) \sim 7.0$ to $\log(M_{\star}/M_{\odot}) \sim 11.5$ and compare them with observations. The two vertical dotted lines represent the mass resolution limit of 100 baryonic particles ($\log(M_{\star}/M_{\odot}) \sim 8.25$) and the statistic limit where there are fewer than 10 galaxies at the low- and high-mass ends \citep{Furlong2014,Katsianis2017}. The shaded regions represent the 95 \% bootstrap confidence interval for 5000 re-samples for the $\sigma_{\text{sSFR}}$-$M_{\star}$ relation.

Starting from redshift $z = 4.0$ (top left panel of Fig. \ref{fig:sSFRFscatter}) we see that the $\sigma_{\text{sSFR}}$-$M_{\star}$ of the reference model has a U-shape form. The dispersion decreases with mass at the $\log(M_{\star}/M_{\odot}) \sim 8.5-9.5$ interval from $\sigma_{\text{sSFR}} =0.4$ to $0.2$ dex while it increases with mass at the $\log(M_{\star}/M_{\odot}) \sim 9.5-10.5$ interval from $\sigma_{\text{sSFR}} \sim 0.2$ to $0.6$ dex. For the ``main sequence'',  $\sigma_{\text{sSFR, MS}}$-$M_{\star}$ relation, defined by the exclusion of objects with $ < 10^{-9.1} yr^{-1}$ \citep{Furlong2014,Matthee2018}, the scatter increases more moderately at the high mass end (from 0.2 dex to 0.3 dex), since passive objects which would increase the dispersion are excluded using a sSFR cut. We note that at this era the fraction of quiescent galaxies is expected to be small, thus the exclusion of quiescent objects should not affect significantly the relation (especially at the low mass end), and it is very possible that the above selection criterion is too strict. However, a more moderate cut of  $ < 10^{-11.0} yr^{-1}$ \citep{Ilbert2015} results in a relationship that is closer to that of the full EAGLE sample since the exclusion of quenched objects is less severe. We note that the observations of \citet{Santini2017} are broadly consistent with the $\sigma_{\text{sSFR, MS, moderate}}$-$M_{\star}$ (represented by the black dotted line)  and $\sigma_{\text{sSFR}}$-$M_{\star}$ (represented by the red solid line) relations (green filled circles representing the observations of \citet{Santini2017} within 0.1 dex with respect the simulated results). A similar behavior is found for lower redshifts up to $z \sim2.0$. This is possibly due to the fact that the moderate sSFR cut of $ < 10^{-11.0} yr^{-1}$ \citep{Ilbert2015} resembles more closely the selection performed by \citet{Schreiber2015} and \citet{Santini2017}.

\begin{figure}
\centering
\includegraphics[scale=0.45]{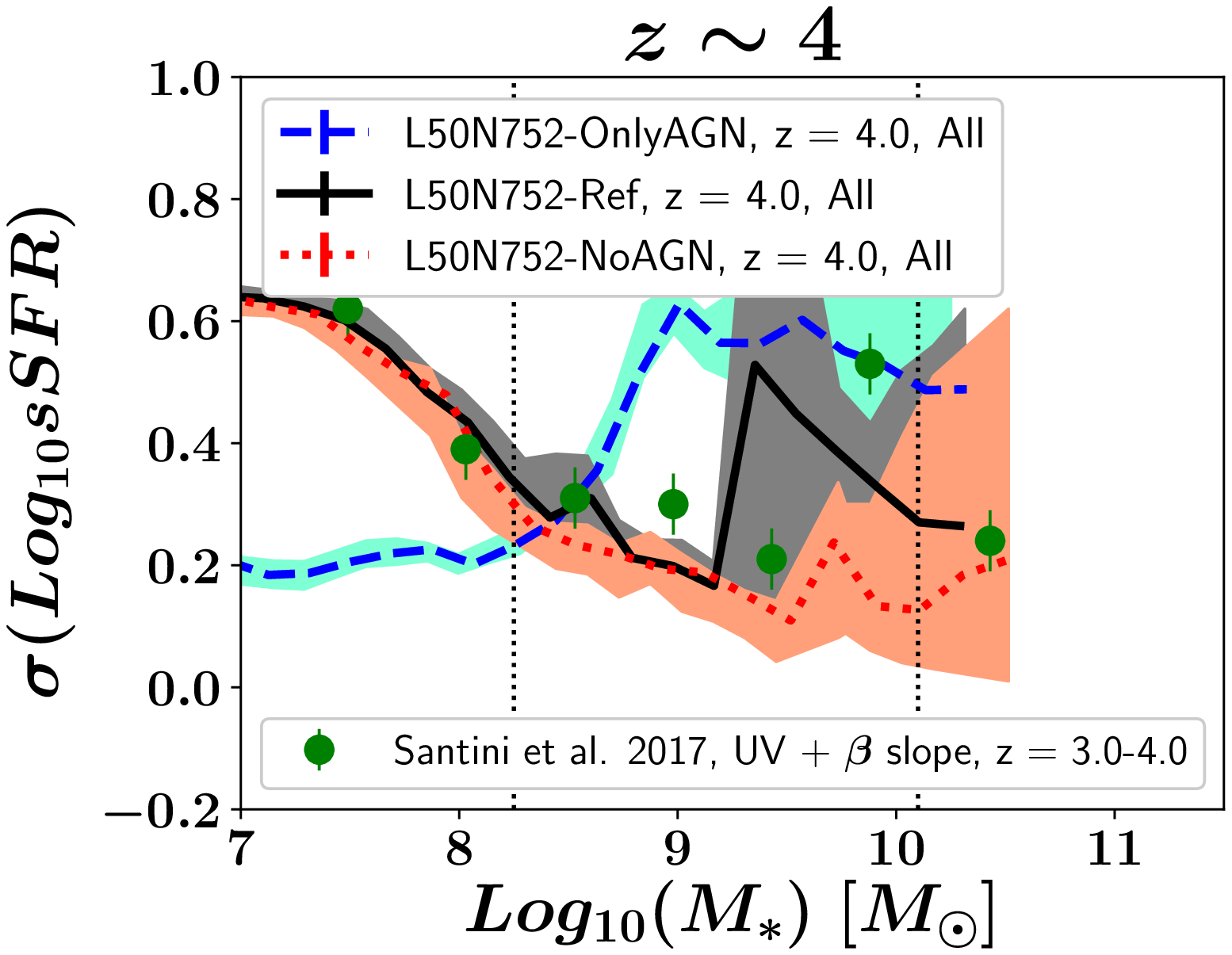}  \hspace{-0.2em} \hspace{-2.5em} \vspace{1em}
\includegraphics[scale=0.45]{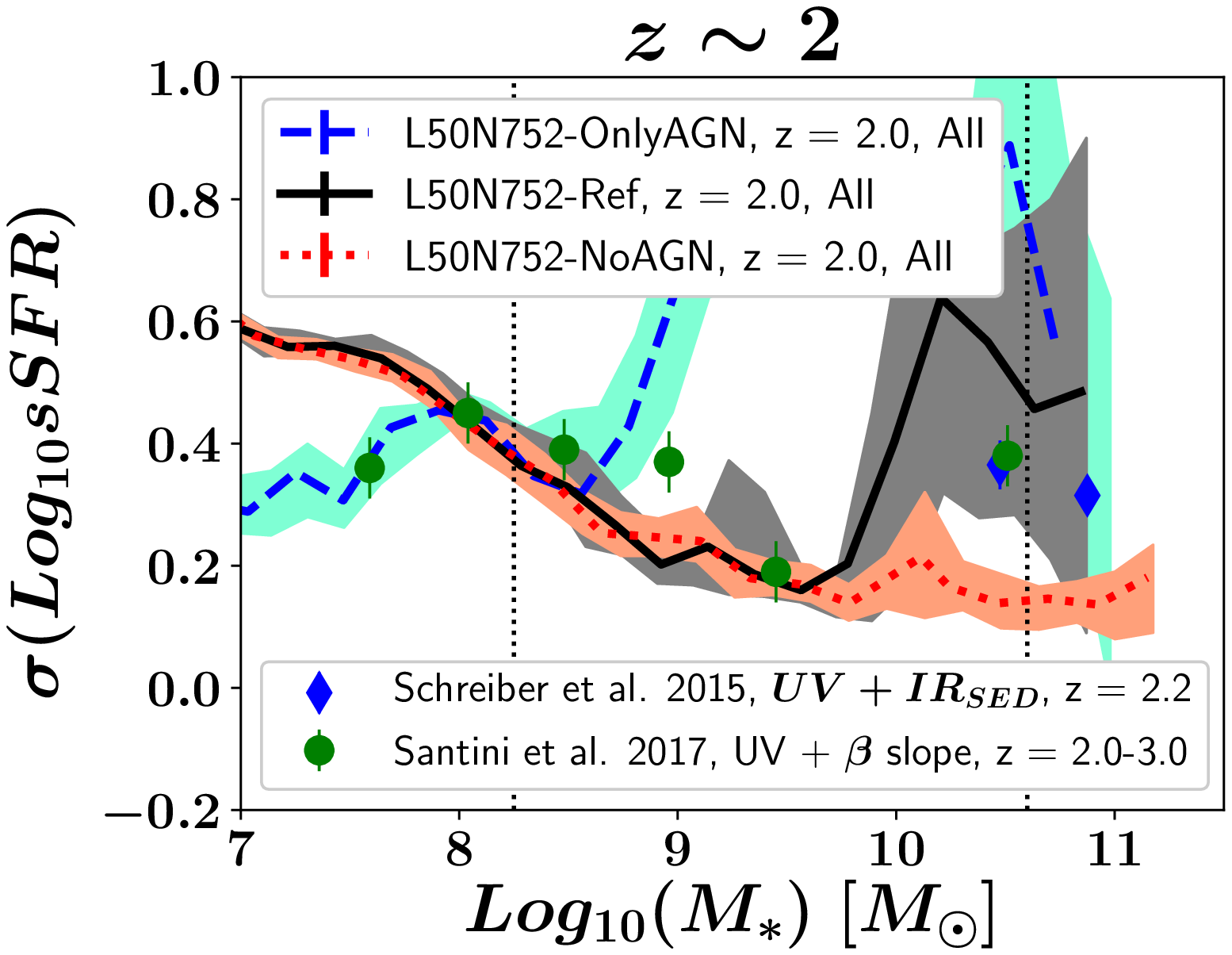}  \hspace{-0.2em} \hspace{-2.5em} \vspace{1em}
\includegraphics[scale=0.45]{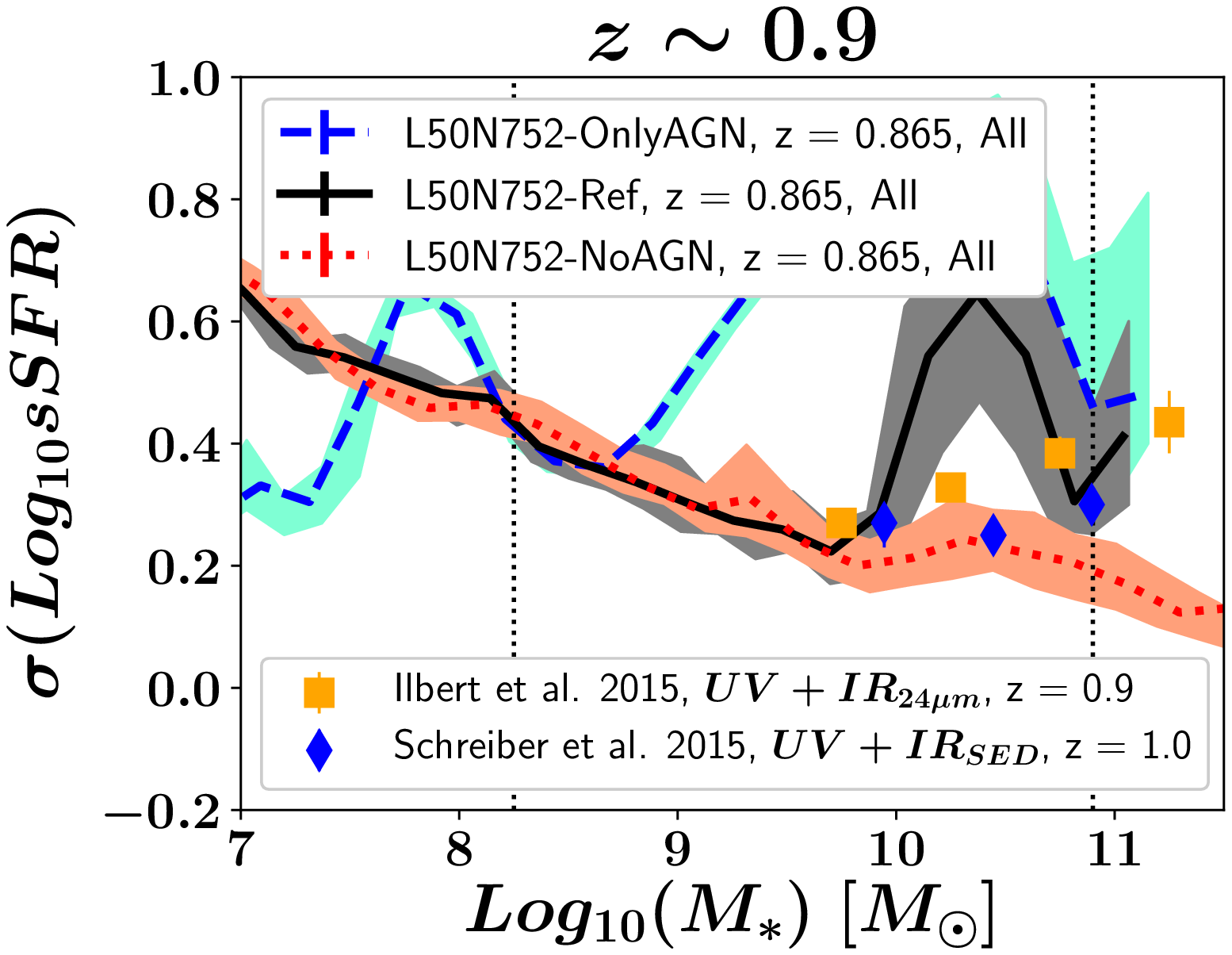}  \hspace{-0.2em} \hspace{-2.5em} \vspace{1em}
\includegraphics[scale=0.45]{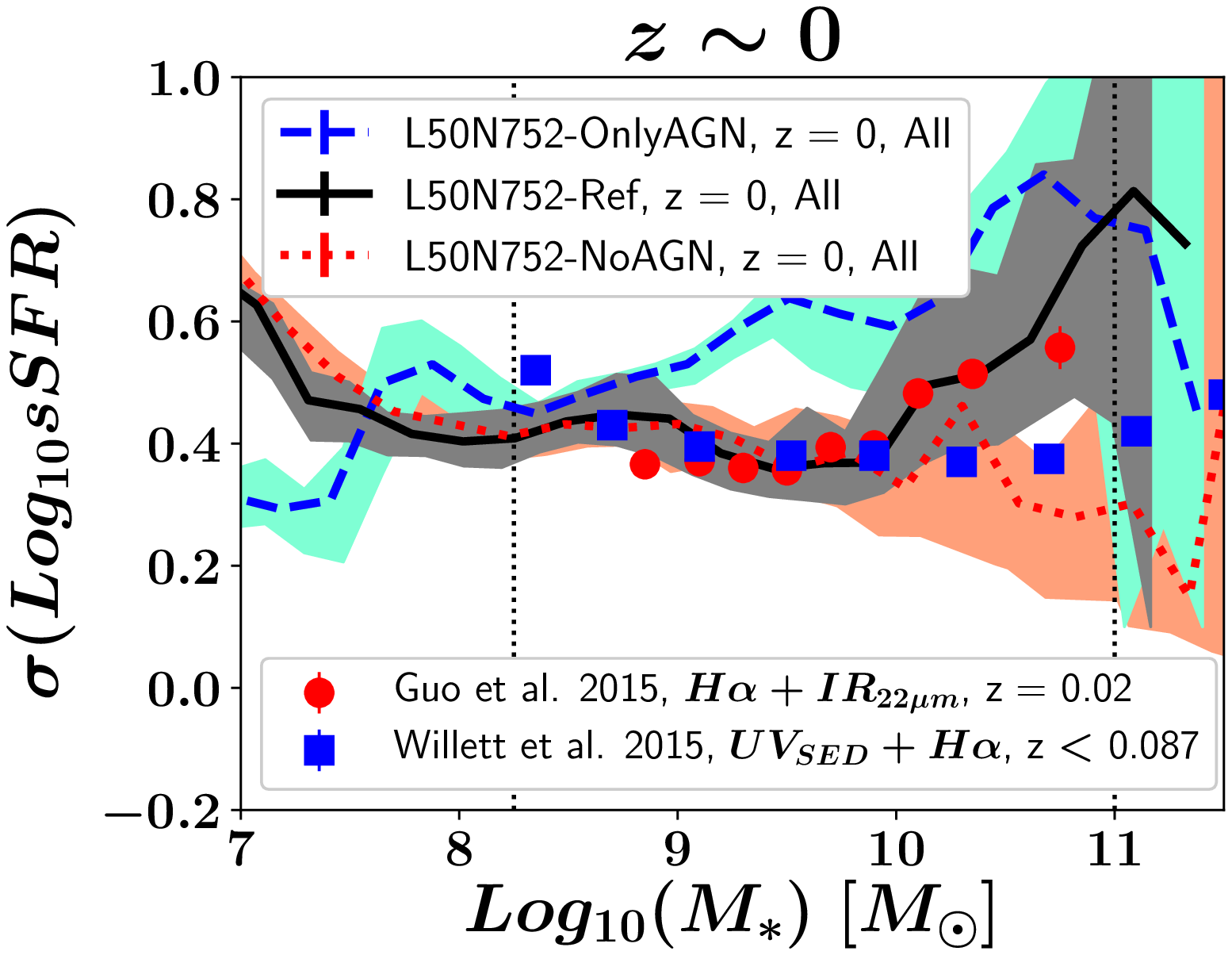}
\caption{The evolution of the  $\sigma_{\text{sSFR}}$-$M_{\star}$ relation at $z \sim 0-4$ for simulations that employ different feedback prescriptions. For a simulation without AGN feedback (L50N752-NoAGN, red dotted line) the scatter decreases with stellar mass at all redshifts. Comparing the configuration with L50N752-Ref which includes the AGN prescription (black solid line) reveals that the effect of the mechanism is to increase the dispersion and is more severe for objects with high stellar masses. On the contrary the simulation which includes only the AGN feedback prescription but not SN (L50N752-OnlyAGN, blue dashed line) has a lower scatter of sSFRs for low mass objects ($\log(M_{\star}/M_{\odot}) \sim 8.5$). This implies that SN play a crucial role for setting the SFHs at the low mass end but higher resolution simulations are necessary to confirm this due to our current resolution limits. In the absence of SN the AGN feedback prescription becomes more aggressive causing a larger diversity of SFHs and affecting objects at a broad mass range ($\log(M_{\star}/M_{\odot}) \sim 8.5-11.5$). The later shows the interplay between the two feedback prescriptions.}
\label{fig:sSFRFscattera}
\end{figure}

\begin{figure}
\centering
\includegraphics[scale=0.45]{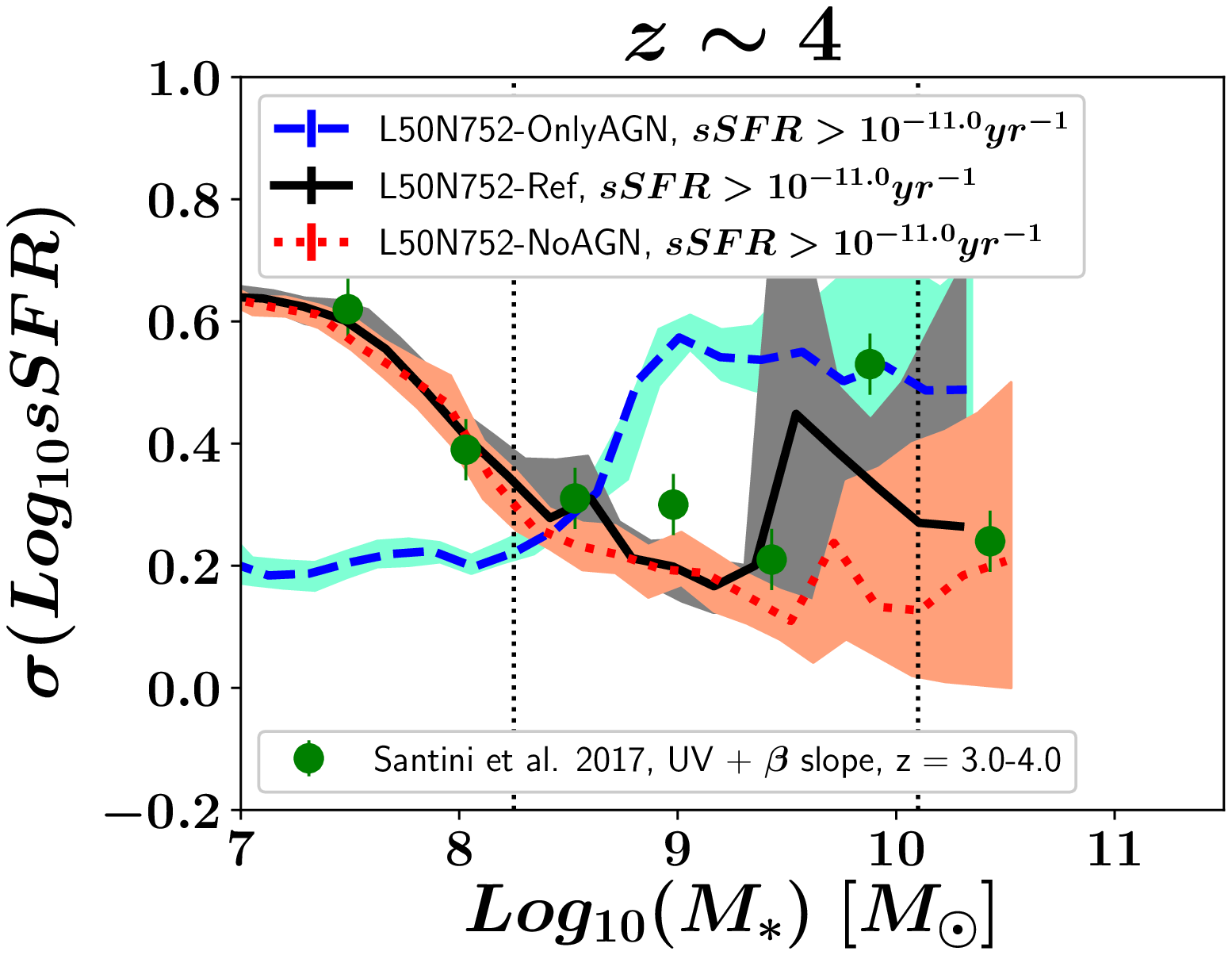}  \hspace{-1.0em}  \vspace{0.5em}
\includegraphics[scale=0.45]{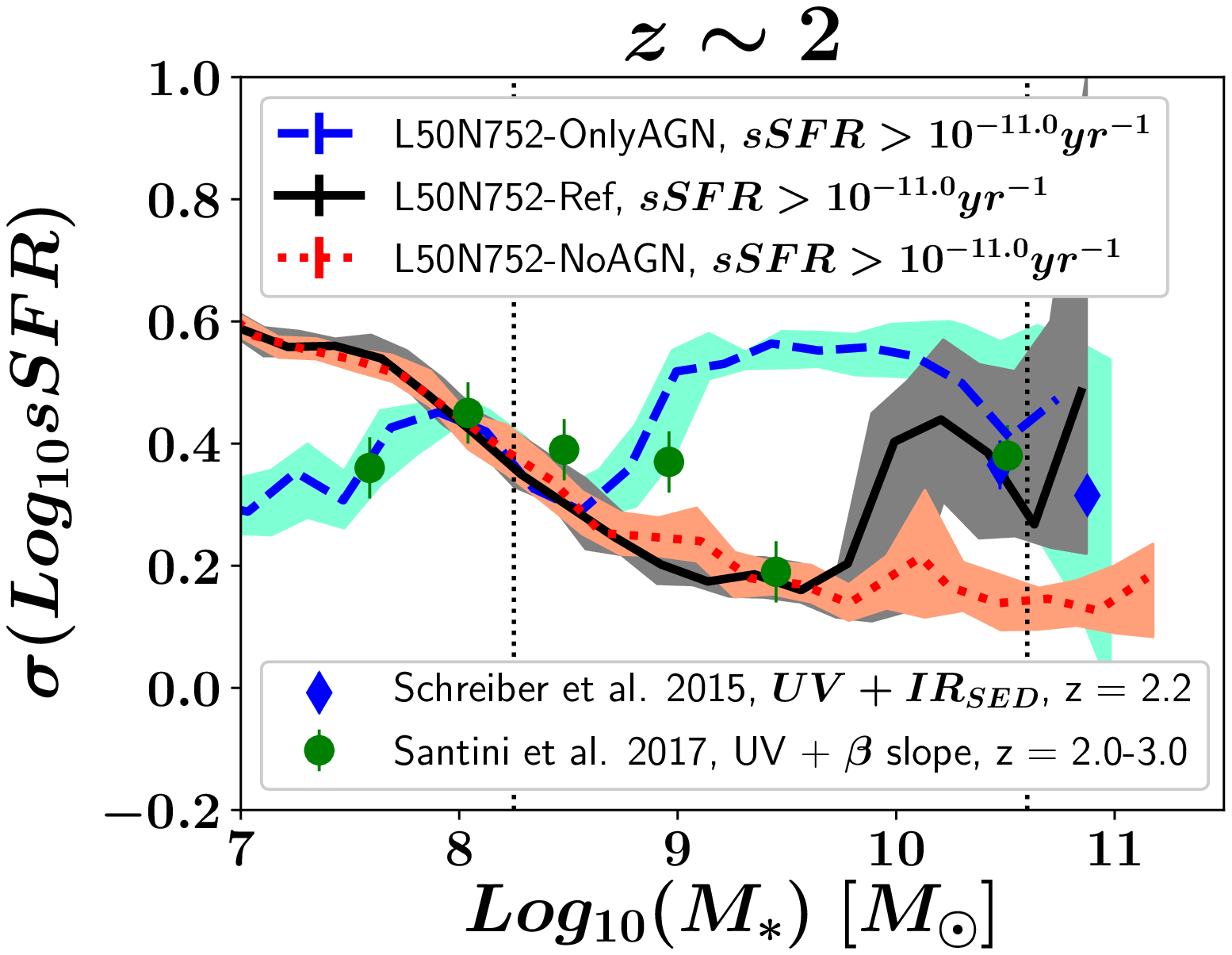}  \hspace{-1.0em}  \vspace{0.5em}
\includegraphics[scale=0.45]{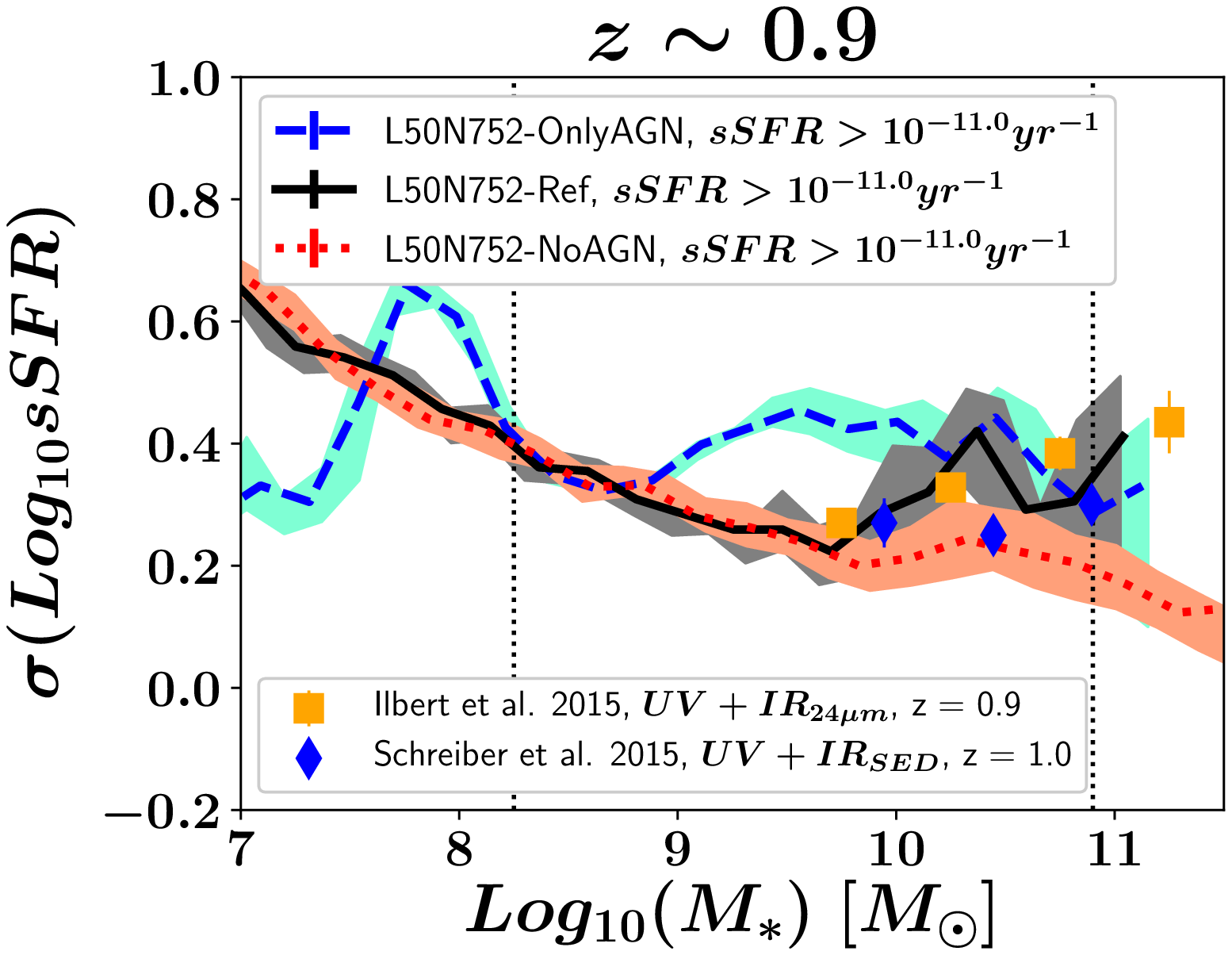}  \hspace{-1.0em}  \vspace{0.5em}
\includegraphics[scale=0.45]{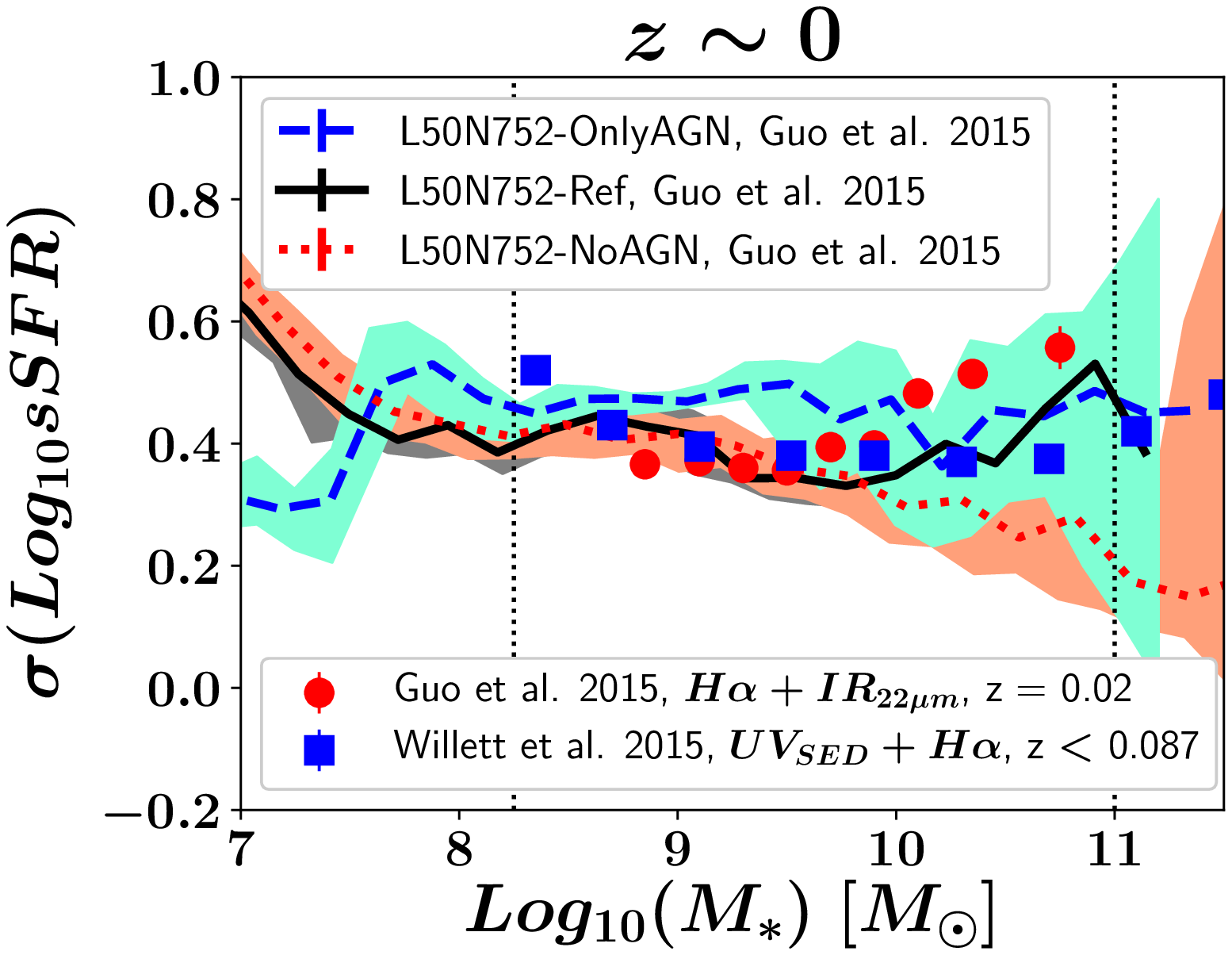}
\caption{The evolution of the  $\sigma_{\text{sSFR, MS, Moderate}}$-$M_{\star}$ relation at $z \sim 0-4$ for simulations which employ different feedback prescriptions. For a simulation without AGN feedback (L50N752-NoAGN, red dotted line) the scatter decreases with stellar mass at all redshifts. Like in Fig. \ref{fig:sSFRFscattera}, where the $\sigma_{\text{sSFR}}$-$M_{\star}$ is presented, we see that the AGN feedback prescription plays a crucial role for determining the scatter of the relation, especially at the high- mass end.}
\label{fig:sSFRFscatteramain}
\end{figure}

\begin{figure}
\centering
\includegraphics[scale=0.45]{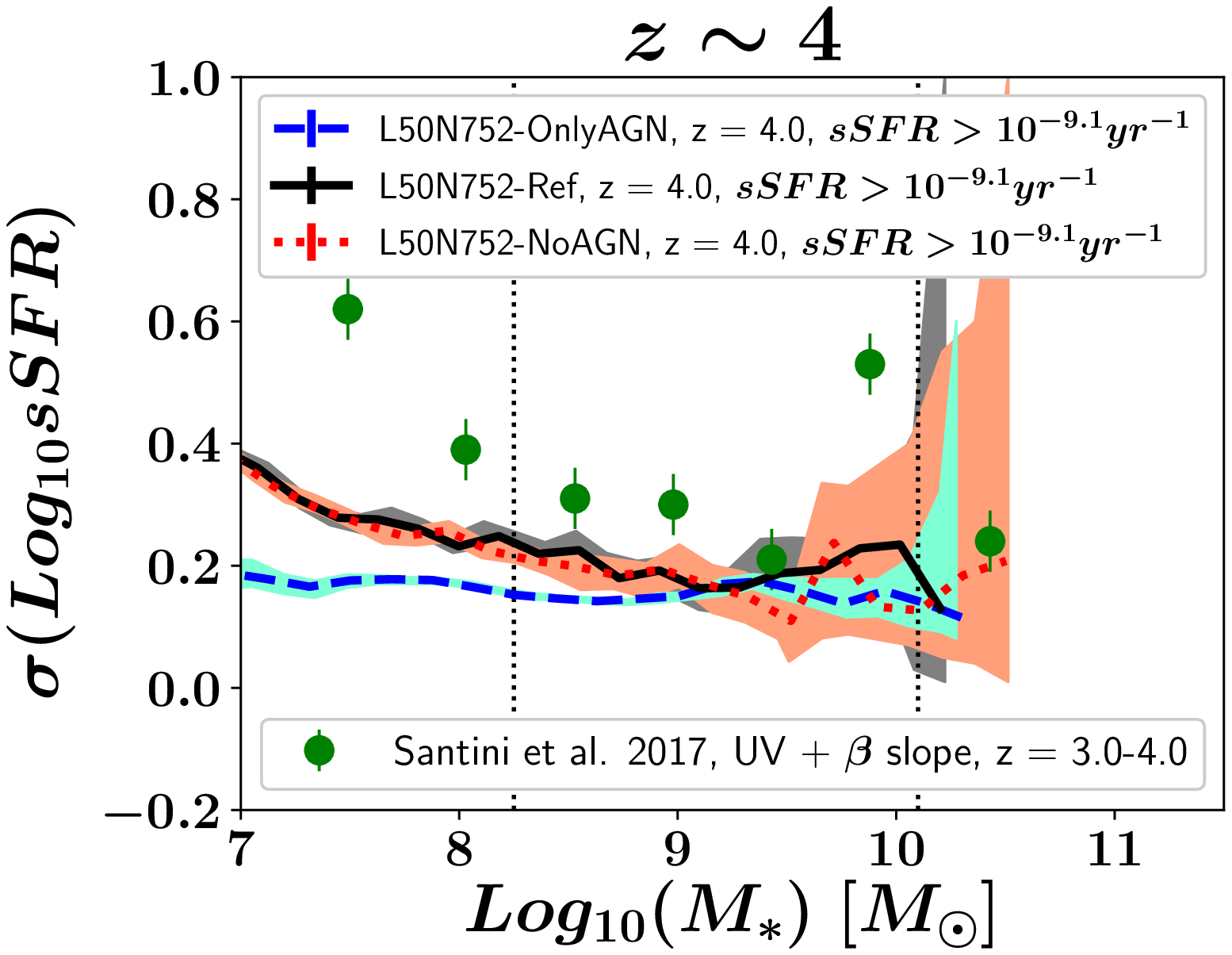}  \hspace{-0.2em} \hspace{-2.5em} \vspace{1em}
\includegraphics[scale=0.45]{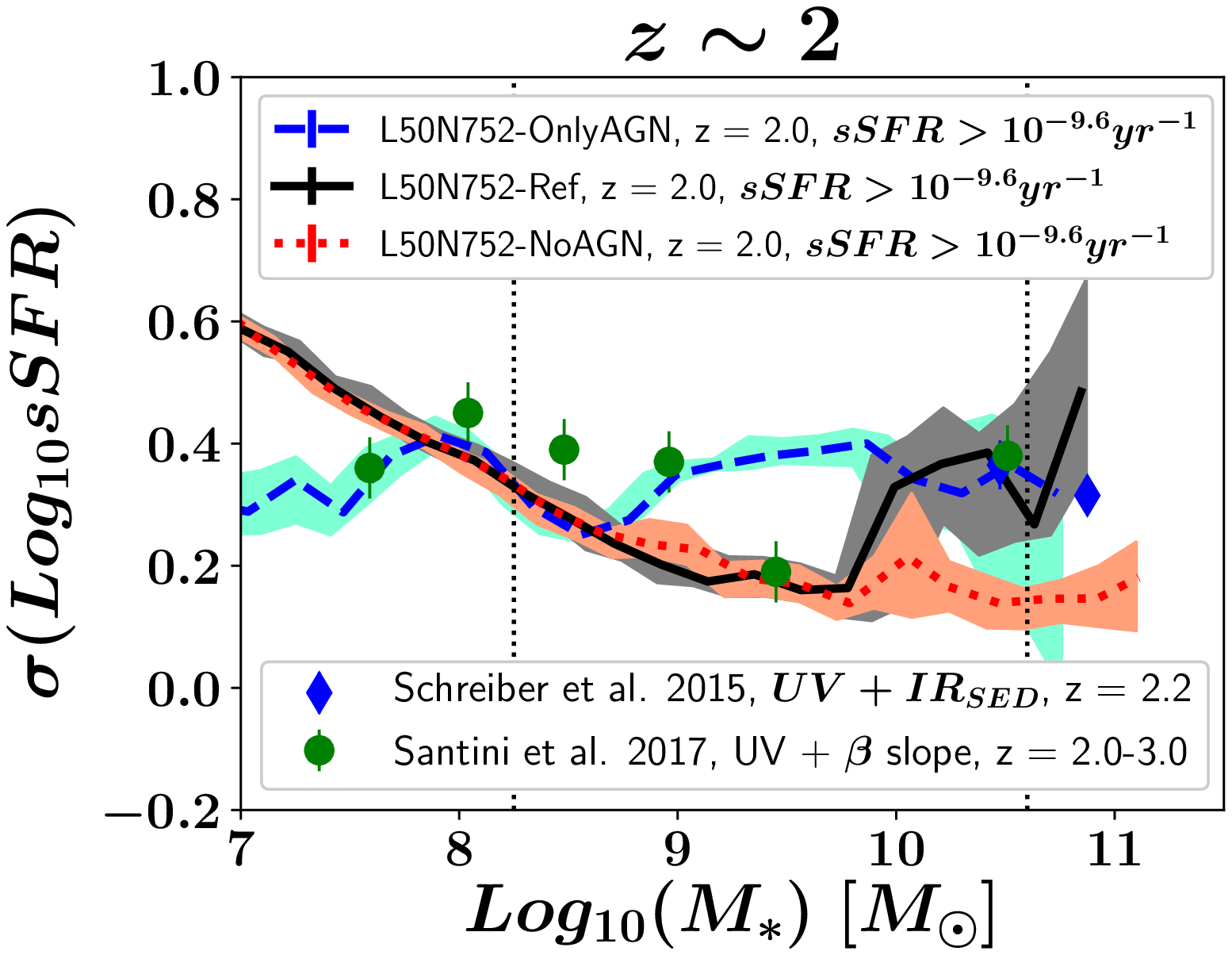}  \hspace{-0.2em} \hspace{-2.5em} \vspace{1em}
\includegraphics[scale=0.45]{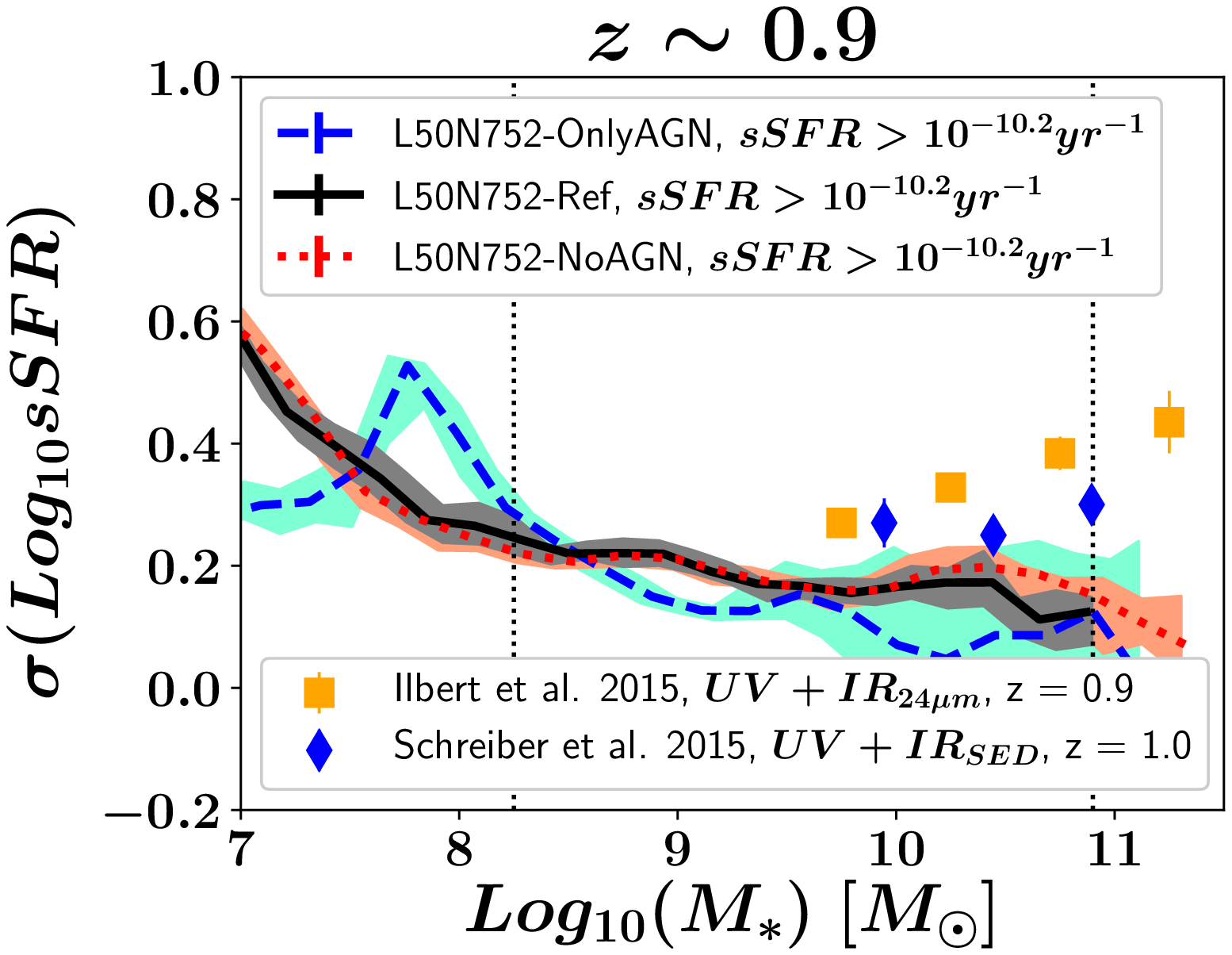}  \hspace{-0.2em} \hspace{-2.5em} \vspace{1em}
\includegraphics[scale=0.45]{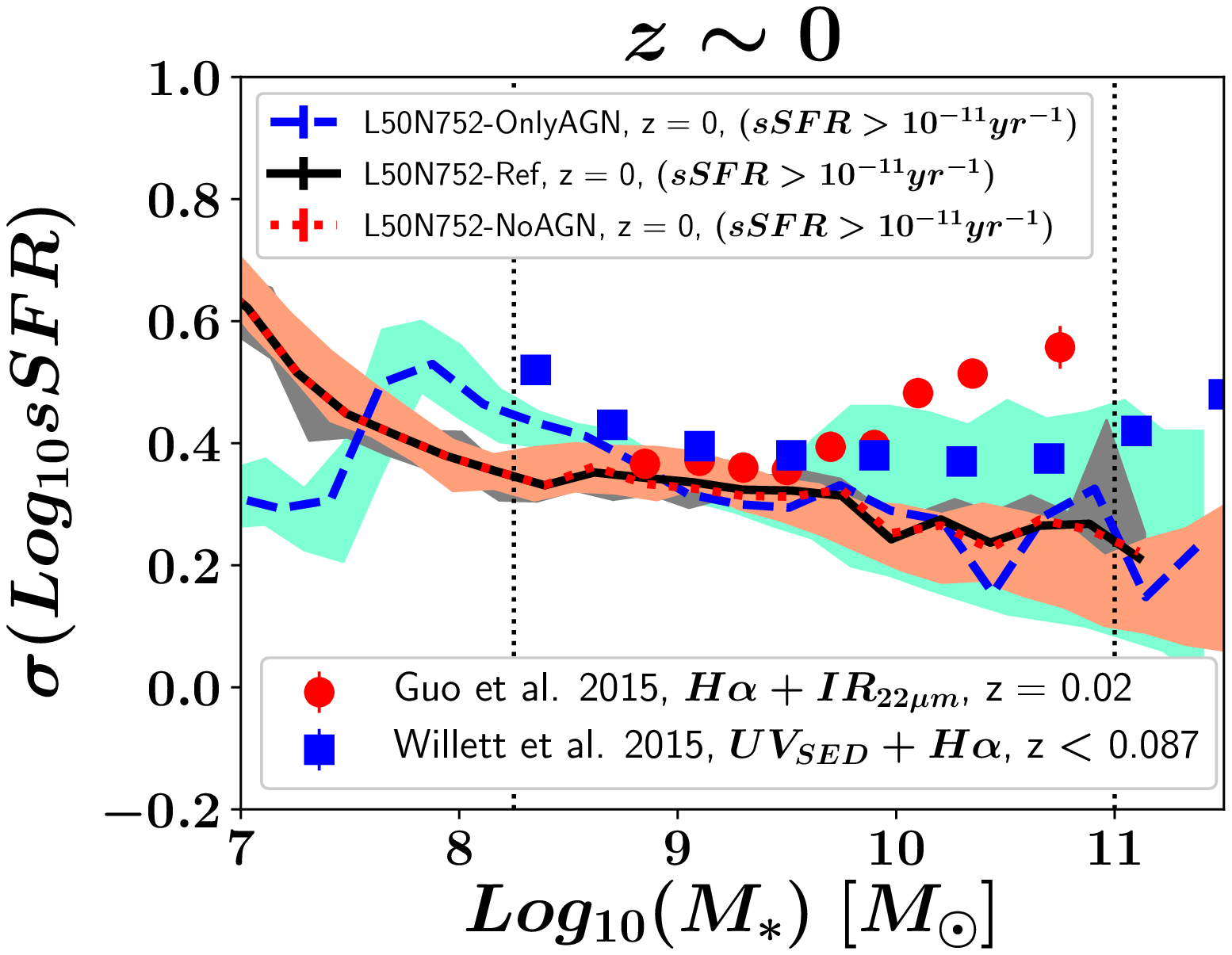}
\caption{The evolution of the  $\sigma_{\text{sSFR, MS}}$-$M_{\star}$ relation at $z \sim 0-4$ for simulations that employ different feedback prescriptions (L50N752-OnlyAGN represented by blue dashed line, L50N752-Ref by black solid line and L50N752-NoAGN by red dashed line). The exclusion of passive objects is severe and objects which are affected by the AGN feedback prescription are not taken into account. Thus, when quenched objects are excluded from the analysis the mechanism does not make its imprint to the  $\sigma_{\text{sSFR, MS}}$-$M_{\star}$ relation, with the difference between the 3 different configurations being small.}
\label{fig:sSFRFscatteramainms}
\end{figure}

At redshift $z \sim 1.4$ (left medium panel of Fig. \ref{fig:sSFRFscatter}) we find that there is an increment of scatter with mass for the EAGLE $\sigma_{\text{sSFR}}$-$M_{\star}$ and $\sigma_{\text{sSFR, MS, Moderate}}$-$M_{\star}$ relations at the high mass end ($\log(M_{\star}/M_{\odot}) \sim 9.5-11.0$) from $\sim 0.2$ dex to $0.45$ and $0.65$ dex, respectively. On the other hand, the $\sigma_{\text{sSFR, MS}}$-$M_{\star}$ relation has an almost constant scatter of $\sim 0.2$ dex with mass at $\log(M_{\star}/M_{\odot}) \sim 10.0-11.0$. The observations for this mass interval \citep{Rodighiero10,Guo2013,Ilbert2015,Schreiber2015} typically lay between the two ``main sequence'' relations $\sigma_{\text{sSFR, MS, Moderate}}$-$M_{\star}$ and $\sigma_{\text{sSFR, MS}}$-$M_{\star}$, something that is possibly related to the uncertainties of removing quiescent objects in the literature \citep{Renzini2015}. The observations of \citet{Ilbert2015} and \citet{Schreiber2015} imply an increasing scatter with mass, while those of \citet{Noeske2007} and \citet{Rodighiero10} a constant ($\sim 0.3$ dex). On the other hand, for the low- mass end ($\log(M_{\star}/M_{\odot}) \sim 8-9.5$) there is a decrement with mass according to the EAGLE reference model in agreement with \citet{Santini2017}. Similarly with higher redshifts the  $\sigma_{\text{sSFR}}$-$M_{\star}$ and $\sigma_{\text{sSFR, MS, Moderate}}$-$M_{\star}$ relations have a U-shape form. The same behavior is found for lower redshifts up to $z \sim0.35$, which reflects the fact that both low and high mass galaxies have a larger scatter/diversity of star formation histories than $M^{\star}$ (characteristic mass) objects. For $z \sim 0$ (bottom right panel of Fig. \ref{fig:sSFRFscatter}) the scatter is constant with mass for both the $\sigma_{\text{sSFR}}$-$M_{\star}$ and $\sigma_{\text{sSFR, MS, Moderate}}$-$M_{\star}$ relations for the $\log(M_{\star}/M_{\odot}) \sim 8.5-9.5$ interval at $\sim 0.4$ dex. The scatter increases for the high mass end to $0.9$ dex when both passive and star forming galaxies are included. The increment is more moderate when cuts similar to the ones of \citet{Guo2015} are applied. On the other hand, the scatter decreases with mass for the `` main sequence'' $\sigma_{\text{sSFR, MS}}$-$M_{\star}$ relation from 0.4 dex to 0.2 dex. 

We note that the EAGLE reference model suggests that both $\sigma_{\text{sSFR, MS, Moderate}}$-$M_{\star}$ and $\sigma_{\text{sSFR, MS}}$-$M_{\star}$ relations are evolving with redshift and are not time independent. In Fig. \ref{fig:sSFRFscatterevl} we present the evolution of the $\sigma_{\text{sSFR, MS, Moderate}}$ at $z \sim 0-4$ for the stellar masses of  $\log(M_{\star}/M_{\odot}) \sim 8.5$ (blue solid line), $\log(M_{\star}/M_{\odot}) \sim 9.5$ (black dotted line) and $\log(M_{\star}/M_{\odot}) \sim 10.5$ (red dashed line). The blue circles, black squares and red triangles represent the compilation of observations of \citep{Guo2013,Ilbert2015,Schreiber2015,Willet2015,Santini2017} at $\log(M_{\star}/M_{\odot}) \sim 8.5$, $\log(M_{\star}/M_{\odot}) \sim 9.5$ and $\log(M_{\star}/M_{\odot}) \sim 11.0$, respectively. Both in simulations and observations the $\sigma_{\text{sSFR, MS, Moderate}}$ at $\log(M_{\star}/M_{\odot}) \sim 8.5$ increases steadily from $\sim$ 0.33 dex at $z \sim 4$ to $\sim$ 0.55 dex. For $\log(M_{\star}/M_{\odot}) \sim 9.5$ the scatter remains almost constant at $\sim0.2$ dex for $z \sim 4.0-1.5$ but increases up to 0.35 at $z \sim 0$ for the redshift interval of $z \sim 1.5-0$. Last, the scatter increases from 0.25 dex to 0.45 dex at $\log(M_{\star}/M_{\odot}) \sim 11.0$.

In Fig. \ref{fig:sSFRFscatterevlMain} we present the evolution of the $\sigma_{\text{sSFR, MS}}$ at $z \sim 0-4$ for the stellar masses of  $\log(M_{\star}/M_{\odot}) \sim 8.5$ (blue solid line), $\log(M_{\star}/M_{\odot}) \sim 9.5$ (black dotted line) and $\log(M_{\star}/M_{\odot}) \sim 11.0$ (red dashed line). The scatter increases at $z \sim 0-4$ from $\sim 0.25$ dex to $\sim 0.35$ dex at $\log(M_{\star}/M_{\odot}) \sim 8.5$ and from $\sim 0.2$ dex to $\sim 0.3$ dex at $\log(M_{\star}/M_{\odot}) \sim 9.5$. However, the $\sigma_{\text{sSFR, MS}}$ remains almost constant at $0.25$ dex at $\log(M_{\star}/M_{\odot}) \sim 10.5$.

In conclusion, the {\it $\sigma_{\text{sSFR}}$-$M_{\star}$} relation (represented by the red solid line) has a U-shape form at all redshifts with the scatter typically decreasing at the $\log(M_{\star}/M_{\odot}) \sim 8-9.5$ mass interval and increasing at $\log(M_{\star}/M_{\odot}) \sim 9.5-11.0$. This implies a diversity of star formation histories for both the low and high mass ends. The above is supported by recent observations, while a complementary work using data from the GAMA survey demonstrates as well the U-shape form of the $\sigma_{\text{sSFR}}$-$M_{\star}$ relation at $z \sim 0$ \citep{Davies2019}. We support that the shape found is not an observational effect and can be found as well in cosmological hydrodynamic simulations like EAGLE or semi-analytic models like SHARK \citep{Shark2019,Davies2019}. Galaxies at the low- and high- mass ends have a larger diversity of SFRs than intermediate mass objects, implying that there are multiple pathways for low and high mass galaxies to evolve. In the following sections we will demonstrate which prescriptions drive the U-Shape form of the $\sigma_{\text{sSFR}}$-$M_{\star}$ relation. We note that the scatter increases with redshift and evolves with time (Fig. \ref{fig:sSFRFscatterevl} and Fig. \ref{fig:sSFRFscatterevlMain}).  The above findings are in agreement with the work of \citet{Kurczynski2017}, which suggests a moderate increase in scatter with cosmic time from 0.2 to 0.4 dex across the epoch of peak cosmic star formation. When moderate sSFR cuts are employed in order to define a main sequence \citep[e.g.  $\sigma_{\text{sSFR, MS}}$-$M_{\star}$,  $\sigma_{\text{sSFR}}$-$M_{\star}$,][]{Ilbert2015,Guo2015} the U-shape form of the relation persist at $z > 0.5$ but is less visible at lower redshifts.

\subsection{The effect of the AGN and SN feedback on the  $\sigma_{\text{sSFR}}$-$M_{\star}$ relation.}
\label{specificSFR-MFeedback}

In this subsection we investigate the effect of Active Galactic Nuclei (AGN) feedback and Supernovae (SN) feedback on the  $\sigma_{\text{sSFR}}$-$M_{\star}$ relation. To do so we compare 3 different configurations that have the same resolution and box size but have different subgrid physics recipes. These include:
\begin{itemize}
\item L50N752-Ref, which is a simulation with the same feedback prescriptions and resolution as the EAGLE reference model L100N1504-Ref (dark solid line of Fig. \ref{fig:sSFRFscattera}),
\item L50N752-NoAGN, which has the same Physics and resolution as L50N752-Ref but does not include AGN feedback (dotted red line of Fig. \ref{fig:sSFRFscattera}),
\item L50N752-OnlyAGN, which has the same Physics, boxsize and resolution as L50N752-Ref but does not include SN feedback (blue dashed line of Fig. \ref{fig:sSFRFscattera}).
\end{itemize}

In Fig. \ref{fig:sSFRFscattera} we present the effect of feedback prescriptions on the  $\sigma_{\text{sSFR}}$-$M_{\star}$ relation in the EAGLE simulations.
\citet{Guo2013} and \citet{Guo2015} find an increasing scatter with mass and suggest that halo/stellar-mass dependent processes such as disk instabilities, bar-driven tidal disruption, minor mergers and major mergers/interactions or stellar feedback are important for large objects. However, the comparison between the reference model (L50N752-Ref, black solid line) and the configuration which does not include the AGN feedback prescription (L50N752-NoAGN, red dotted line) suggests that the effect of the AGN feedback mechanism is mostly responsible for increasing the scatter of the relation for high mass objects ($M_{\star} > 10^{9.5}$ $M_{\odot}$). The prescription increases the diversity in star formation histories at the $\log(M_{\star}/M_{\odot}) \sim 9.5-11.5$ regime (Figs \ref{fig:sSFRFscattera} and \ref{fig:sSFRFscatteramain}) by creating a large number of quenched objects at the high mass end. We note that the absence of SN feedback would result in a more aggressive AGN feedback mechanism, which would significantly increase the dispersion for high mass galaxies (blue dashed line). This shows the interplay between the two different prescriptions and the finding is in agreement with the work of \citet{Bower2017}, which suggests that black hole growth is suppressed by stellar feedback. If there are no galactic winds to decrease the accretion rate for a galaxy with a supermassive black hole, then the later will become bigger and its AGN feedback mechanism will affect more severely the sSFR of the object\footnote{In our AGN feedback scheme halos more massive than the $10^{10}$ $M_{\odot}/h$ threshold are seeded with a Supermassive Black Hole (SMBH), thus even relatively small galaxies at the $\log(M_{\star}/M_{\odot}) \sim 8.5-9$ range which reside halos with $> 10^{10}$ $M_{\odot}/h$ are affected, since the absence of the SNe feedback which would be important at this mass interval allows a fast and significant SMBH growth, which otherwise would not be possible.}. The effect of the AGN feedback for the case of the L50N752-OnlyAGN run (in which SNe feedback is absent) is significant at z $\sim 1-4$, an epoch when the SFRs of simulated objects would be influenced significantly from SNe feedback (For more details see Fig. 7 in \citet{Katsianis2017} which describes the effect of SNe feedback on the star formation rate function). In contrast with  $\sigma_{\text{sSFR}}$-$M_{\star}$ and  $\sigma_{\text{sSFR, MS, Moderate}}$-$M_{\star}$ (Figs \ref{fig:sSFRFscattera} and \ref{fig:sSFRFscatteramain}), we find that the $\sigma_{\text{sSFR, MS}}$-$M_{\star}$ relation is not affected by feedback mechanisms (Fig. \ref{fig:sSFRFscatteramainms}). The different fractions of quenched objects between configurations which employ different feedback prescriptions does not affect the comparison between them since the quenched objects, which increase the scatter, would in any simulation be excluded with the strict selection criterion adopted.

\citet{Santini2017} used the Hubble Space Telescope Frontier Fields to study the main sequence and its scatter. In contrast with \citet{Guo2013} and \citet{Ilbert2015} the authors found a decreasing scatter with mass at all redshifts they considered and they suggested that this behavior is a consequence of the smaller number of progenitors of low mass galaxies in a hierarchical  scenario and/or of the efficient stellar feedback  processes  in  low mass halos. Comparing the reference model (L50N752-Ref, black solid line) with the configuration which does not include the SN feedback mechanism (L50N752-OnlyAGN, blue dashed line) we see that the effect of this mechanism is indeed to increase the scatter of the relation with decreasing mass for low mass objects ($M_{\star} < 10^{8.5}$ $M_{\odot}$). In addition, in low mass galaxies discrete gas accretion events trigger bursts of star formation which inject SNe feedback. Since feedback is very efficient in low mass galaxies this largely suppresses star formation until new gas is accreted \citep[e.g.]{Faucher2018}. However, the finding is below the resolution limit of 100 particles and higher resolution simulations are required to investigate this in the future (Figs \ref{fig:sSFRFscattera} and \ref{fig:sSFRFscatteramain}).

In conclusion, AGN and SN feedback are playing a major role in producing the U-shape to the  $\sigma_{\text{sSFR}}$-$M_{\star}$ and $\sigma_{\text{sSFR, MS, Moderate}}$-$M_{\star}$ relations described in subsection \ref{specificSFR-M} and drive the evolution of the scatter. Both prescriptions give a range of SFHs both at the low- and high-mass ends with AGN feedback increasing the scatter mostly at the $\log(M_{\star}/M_{\odot}) \sim 9.5-11.5$ interval.

\begin{figure}
\centering
\includegraphics[scale=0.45]{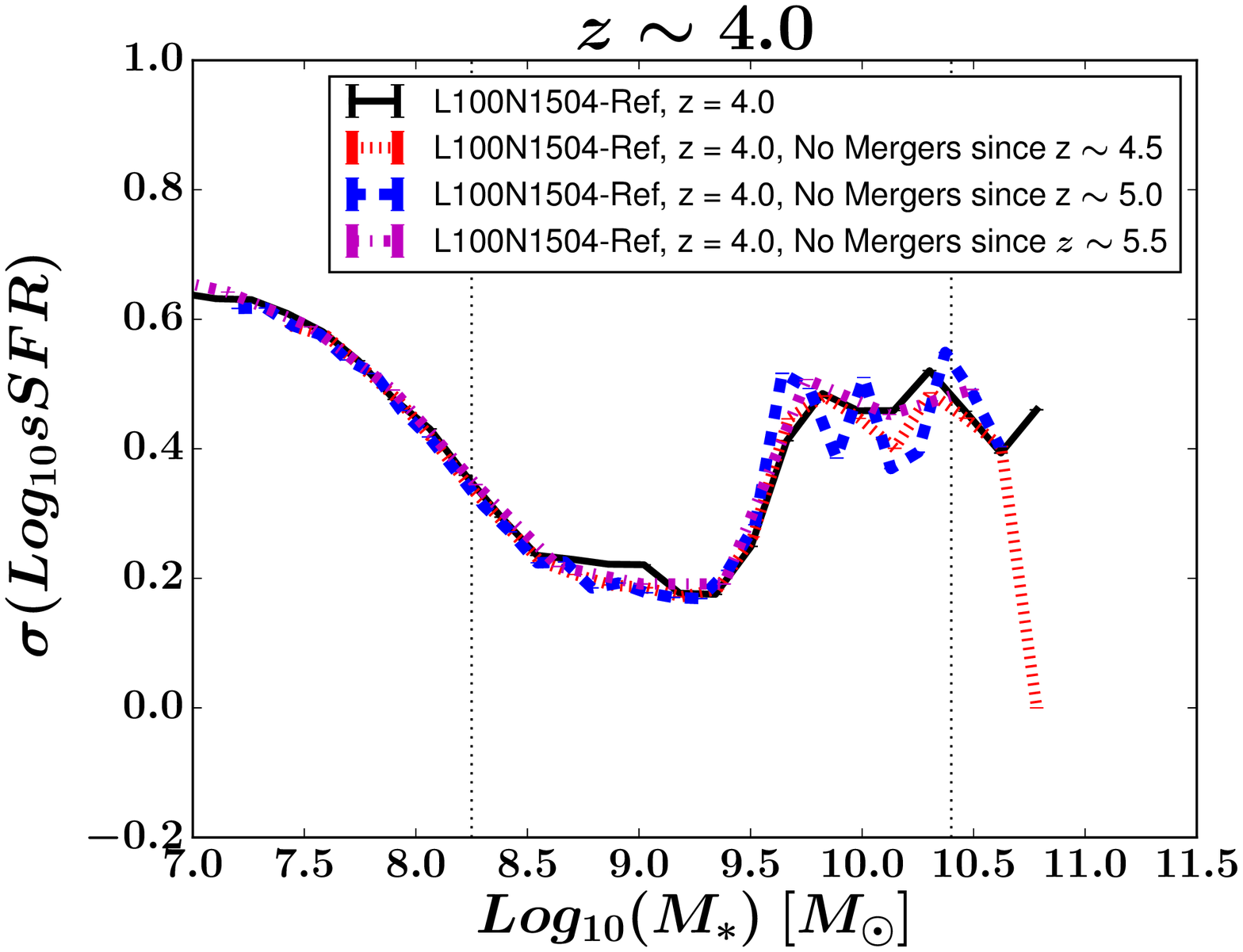} \hspace{-0.2em} \hspace{-2.5em} \vspace{1em}
\includegraphics[scale=0.45]{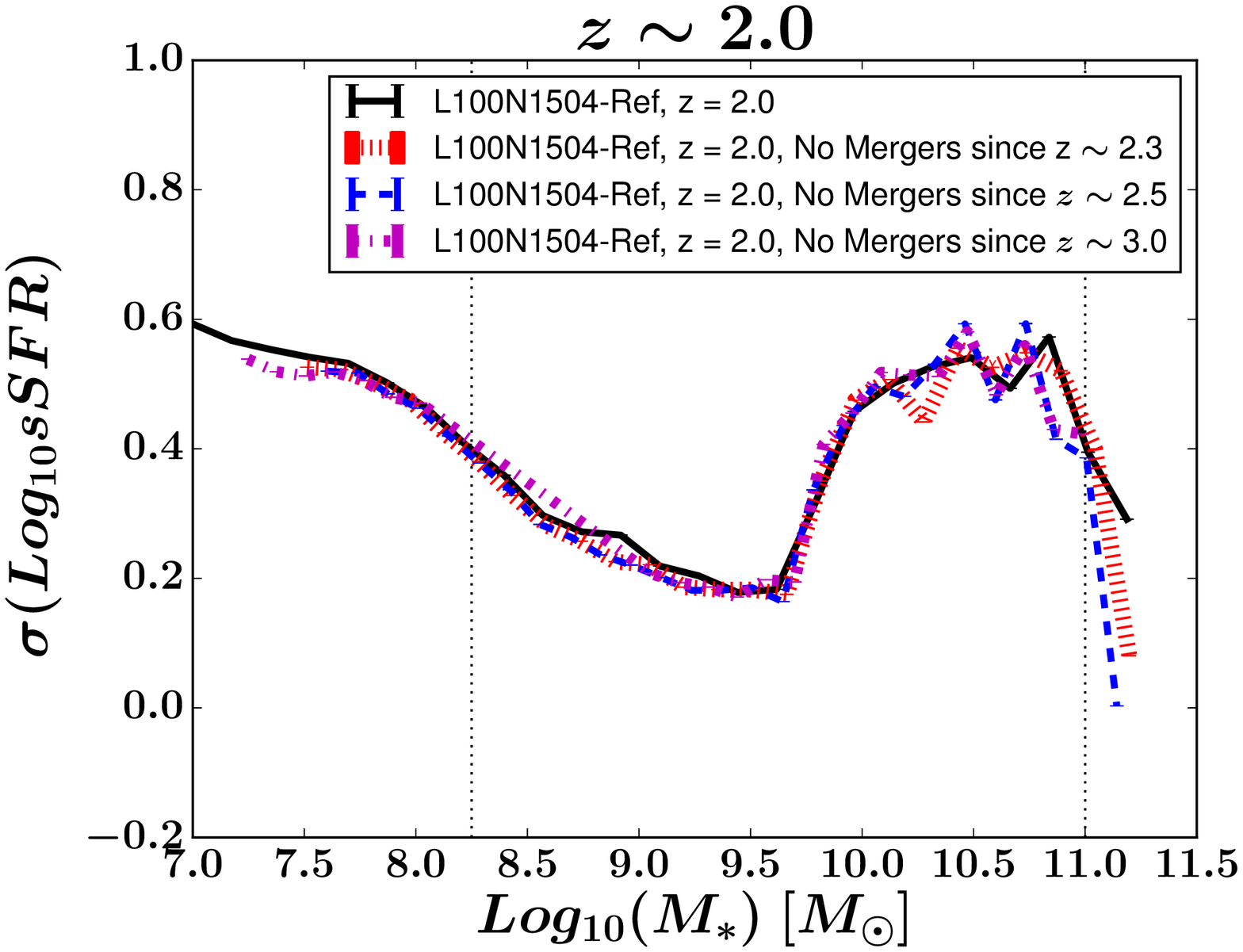} \hspace{-0.2em} \hspace{-2.5em} \vspace{1em}
\includegraphics[scale=0.45]{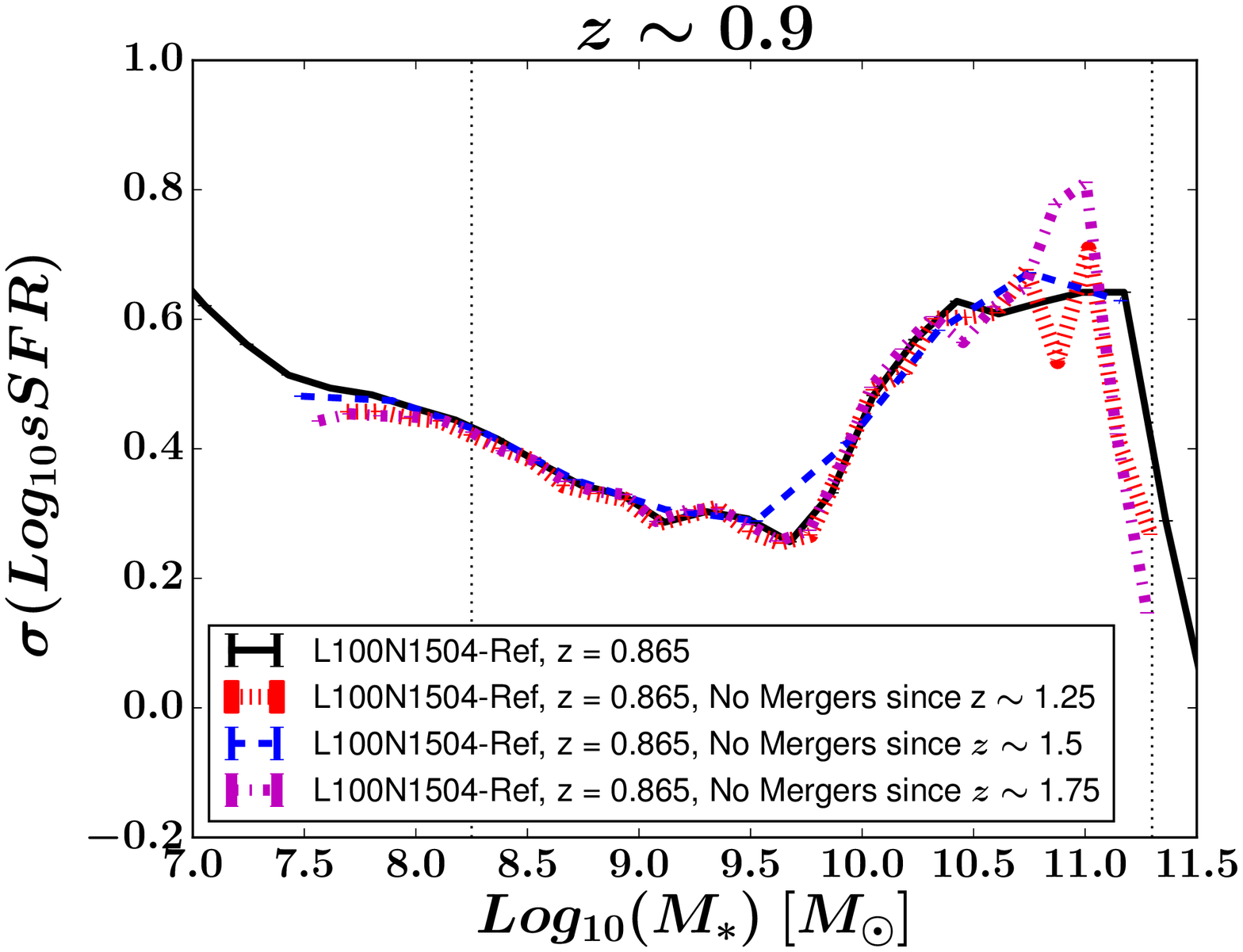} \hspace{-0.2em} \hspace{-2.5em} 
\includegraphics[scale=0.45]{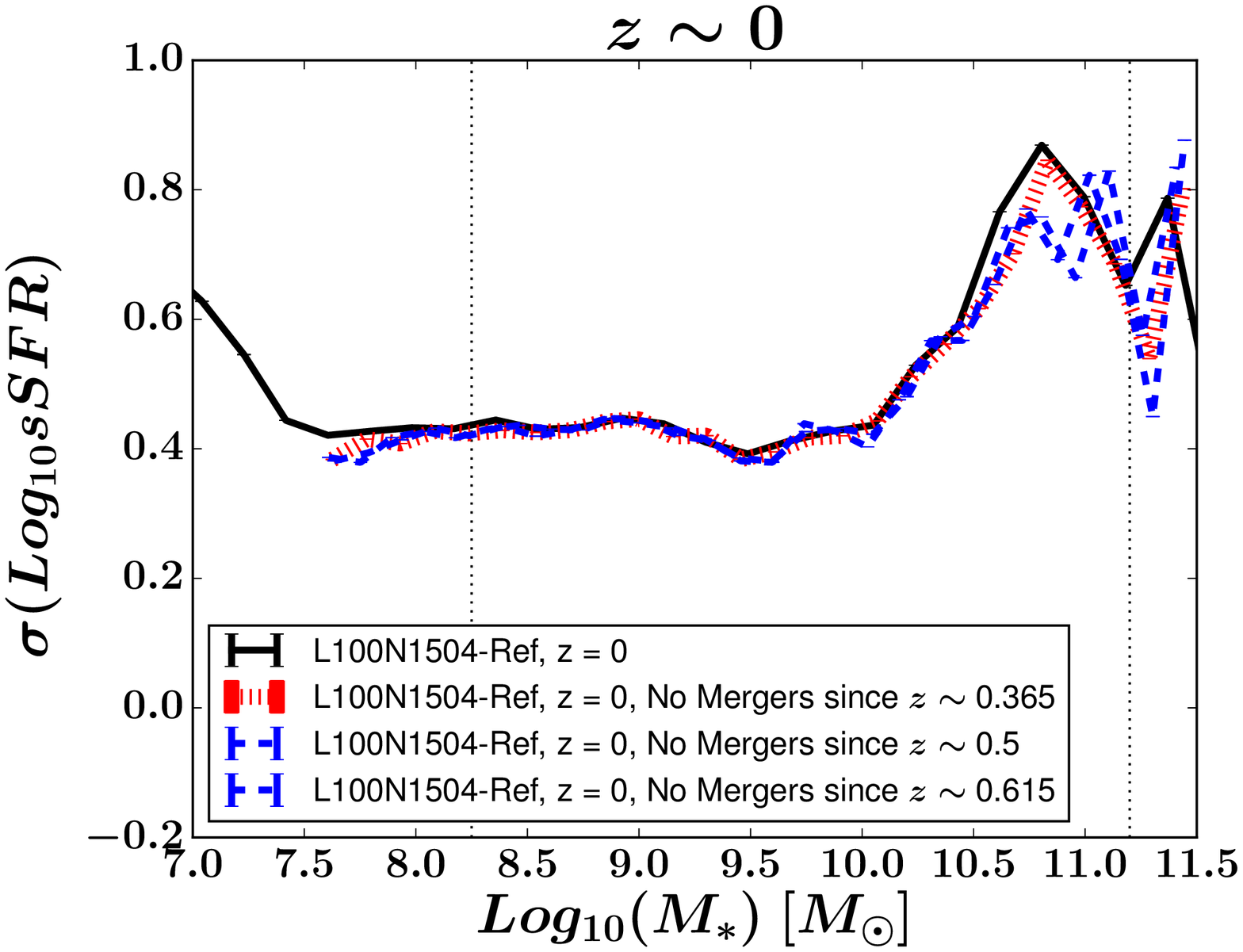}
\caption{The effect of mergers on the  $\sigma_{\text{sSFR}}$-$M_{\star}$ relation at $z \sim 0-4$. The red, blue and magenta lines in each panel represent the results when ongoing mergers and objects that have experienced merging at previous redshifts are excluded. The shape of the  $\sigma_{\text{sSFR}}$-$M_{\star}$ relation is not significantly changed by the presence or absence of these objects.}
\label{fig:sSFRFmergers}
\end{figure}

\subsection{The effect of excluding mergers on the  $\sigma_{\text{sSFR}}$-$M_{\star}$ relation.}
\label{specificSFR-Mmergers}

\citet{Guo2015} pointed out that in massive galaxies interactions like minor and major mergers can induce starbursts followed by strong stellar feedback that can contribute significantly to the spread of sSFRs. In contrast, according to the authors, lower-mass galaxies are supposed to be less affected by the above and thus should have a smaller dispersion of SFHs leading to a scatter that increases with mass for the  $\sigma_{\text{sSFR}}$-$M_{\star}$ relation. On the other hand, \citet{Peng2010} suggest that interacting/merging low-mass satellite galaxies are sensitive to environmental quenching and this could input a significant dispersion to the sSFRs at the low-mass end of the distribution. \citet{Orellana2017} reported that interactions between galaxies can affect the scatter for a range of masses. As \citet{Guo2015} suggest the effect of the above mechanisms to the sSFR dispersion are difficult to examine.

\citet{Qu2017} and \citet{Lagos2018} studied the impact of mergers on mass assembly and angular momentum on the EAGLE galaxies. The authors found that the reference model is able to reproduce the observed merger rates and merger fractions of galaxies at various redshifts \citep{Conselice2003,Kartaltepe2007,Ryan2008,Lotz2008,Conselice2009,deRavel2009,Williams2011,LopezSanjuan,Bluck2012,Stott2013,Robotham2014,Man2016}. Thus EAGLE can be used to study the effect of mergers on the  $\sigma_{\text{sSFR}}$-$M_{\star}$ relation. We identify mergers using the merger trees available in the EAGLE database. These merger trees were created using the D-Trees algorithm of \citet{Jiang2014}. \citet{Qu2017} described how this algorithm was adapted to work with EAGLE outputs. We consider that a merger (major or minor) occurs when the stellar mass ratio between the two merging systems, $\mu = M_2/M_1$ is above 0.1 \citep{Crain2015}. The separation criterion, $R_{merge}$, is defined as $R_{merge} = 5 \times R_{1/2}$, where $R_{1/2}$ is the half-stellar mass radius of the primary galaxy \citep{Qu2017}. The above selection method to identify  mergers and to separate into major or minor mergers is widely assumed in the EAGLE literature \citep{Jiang2014,Crain2015,Qu2017,Lagos2018}. The fraction of mergers in the reference model at $z \sim 0$ increases towards higher masses \citep{Lagos2018}. Thus, it would be expected that the mechanism  affects mostly the SFHs of high mass objects \citep{Guo2015}.

In Fig. \ref{fig:sSFRFmergers} we present the effect of mergers on the  $\sigma_{\text{sSFR}}$-$M_{\star}$ relation in the EAGLE simulations at $z \sim 0-4$. In the bottom panel the black line represents the reference model at $z \sim 0$, in which all galaxies are considered. The red dotted line represents the same but when ongoing mergers and objects that have experienced merging from $z \sim 0.365$ are not included in the analysis. The blue dashed line describes the results when objects that have experienced merging from $z \sim 0.5$ are excluded. The magenta represents the same when galaxies that have experienced merging from $z \sim 0.615$ are excluded. The above analysis allows us to quantify the effect of ongoing (at $z = 0$), ongoing + recent ($z \sim 0-0.35$) and ongoing + recent + past ($z \sim 0-0.65$) mergers to the  $\sigma_{\text{sSFR}}$-$M_{\star}$ relation and has been done similarly for $z \sim 0, \, 0.9, \, 2, \, 4 $. We see that according to the reference model, mergers do not induce a significant dispersion in the star formation histories of galaxies. The above findings can be seen at all redshifts considered. This implies that recent mergers, despite their importance for galaxy formation and evolution, do not impart a significant scatter on the $\sigma_{\text{sSFR}}$-$M_{\star}$ relation. 

\section{The effect of SFR and stellar mass diagnostics on the  $\sigma_{\text{sSFR}}$-$M_{\star}$ relation.}
\label{specificSFR-Mdiagnostics}

To obtain the intrinsic properties of galaxies, observers have to rely on models for the observed light. Stellar masses are typically calculated via the Spectral Energy Density (SED) fitting technique, while for the case of SFRs different authors employ different methods [e.g. Conversion of IR+UV luminosities to SFRs \citep{Arnouts2013,Whitaker2014}, SED fitting \citep{Bruzualch03}, conversion of UV, H$\alpha$ and IR luminosities \citep{Katsianis2017}]. However, there is an increasing number of reports that different techniques give different results, most likely due to systematic effects affecting the derived properties \citep{Bauer11,Utomo2014,Fumagalli2014,Katsianis2015,Davies2016,Davies2017}. \citet{Boquien2014} argued that SFRs obtained from SED modeling, which take into account only FUV and U bands are overestimated. \citet{Hayward2014} noted that the SFRs obtained from IR luminosities \citep[e.g. ][]{Noeske2007,Daddi2007} can be artificially high. \citet{Ilbert2015} compared SFRs derived from SED and UV+IR, and find a tension reaching 0.25 dex. \citet{Guo2015} suggested that sSFR based on mid-IR emission may be significantly overestimated \citep{Salim2009,Chang2015,Katsianis2016,Katsianis2017}. All the above uncertainties on the determination of intrinsic properties could possibly affect the observed/derived  $\sigma_{\text{sSFR}}$-$M_{\star}$ relation.

\citet{Camps2016}, \citet{Trayford2017} and \citet{Camps2018} presented a procedure to post-process the EAGLE galaxies and produce mock observations that describe how galaxies appear in various bands (e.g. GALEX-FUV, MIPS${\rm _{160}}$, SPIRE${\rm _{500}}$). The authors did so by performing a full 3D radiative transfer simulation to the EAGLE galaxies using the SKIRT code \citep{Baes2003,Baes2011,Marko2012,Cb2015,Peest2017,Marko2016,Marko2017,Behrens2018}. In this section we use the artificial SEDs present in \citet{Camps2018}, in order to study how typical SFR/${\rm M_{\star}}$ diagnostics affect the $\sigma_{\text{sSFR}}$-$M_{\star}$ relation. We stress that the EAGLE objects that were post-processed by SKIRT were galaxies with stellar masses $M_{\star} > 10^{8.5} M_{\odot}$, above the resolution limit of 100 gas particles, and sufficient dust content. We use the Fitting and Assessment of Synthetic Templates (FAST) code \citep{Kriek2009} to fit the mock SEDs to identify the SFRs and stellar masses of the EAGLE+SKIRT objects like in a range of observational studies \citep{Gonzalez2012,Gonzalez2014,Botticella2017,Soto2017,Aird2017}. Doing so enables us to evaluate the effect of different SFR/stellar mass diagnostics on the derived $\sigma_{\text{sSFR}}$-$M_{\star}$ relation and thus isolating the systematic effect on the $\sigma_{\text{sSFR}}$. We assume an exponentially declining SFH [${\rm SFR \sim exp(-t/tau)}$]{\footnote{We note that this parameterization despite the fact that is commonly used in the literature could misinterpret old stellar light for an exponentially increasing contribution originating from a younger stellar population. This can underestimate by a factor of two both SFRs and stellar masses. In addition, exponentially  SFHs may not be representative in describing star forming galaxies \citep{Ciesla2017,Iyer2017,Carnall2018,Leja2018}.}}  \citep{Longhetti2009,Micha2012,Botticella2012,Fumagalli2016,Blancato2017,Abdurrouf2019}, a Chabrier IMF \citep{chabrier03}, a \citet{Calzetti2000} dust attenuation law \citep{Mitchel2013,Sklias2014,Cullen2018,Mclure2018} and a metallicity 0.02 $Z_{\odot}$ \citep{Ono2010,Greisel2013,Chan2016,Mclure2018b}. The above data are labelled in this work as L100N1504-Ref+SKIRT.

\begin{figure}
\centering
\includegraphics[scale = 0.45]{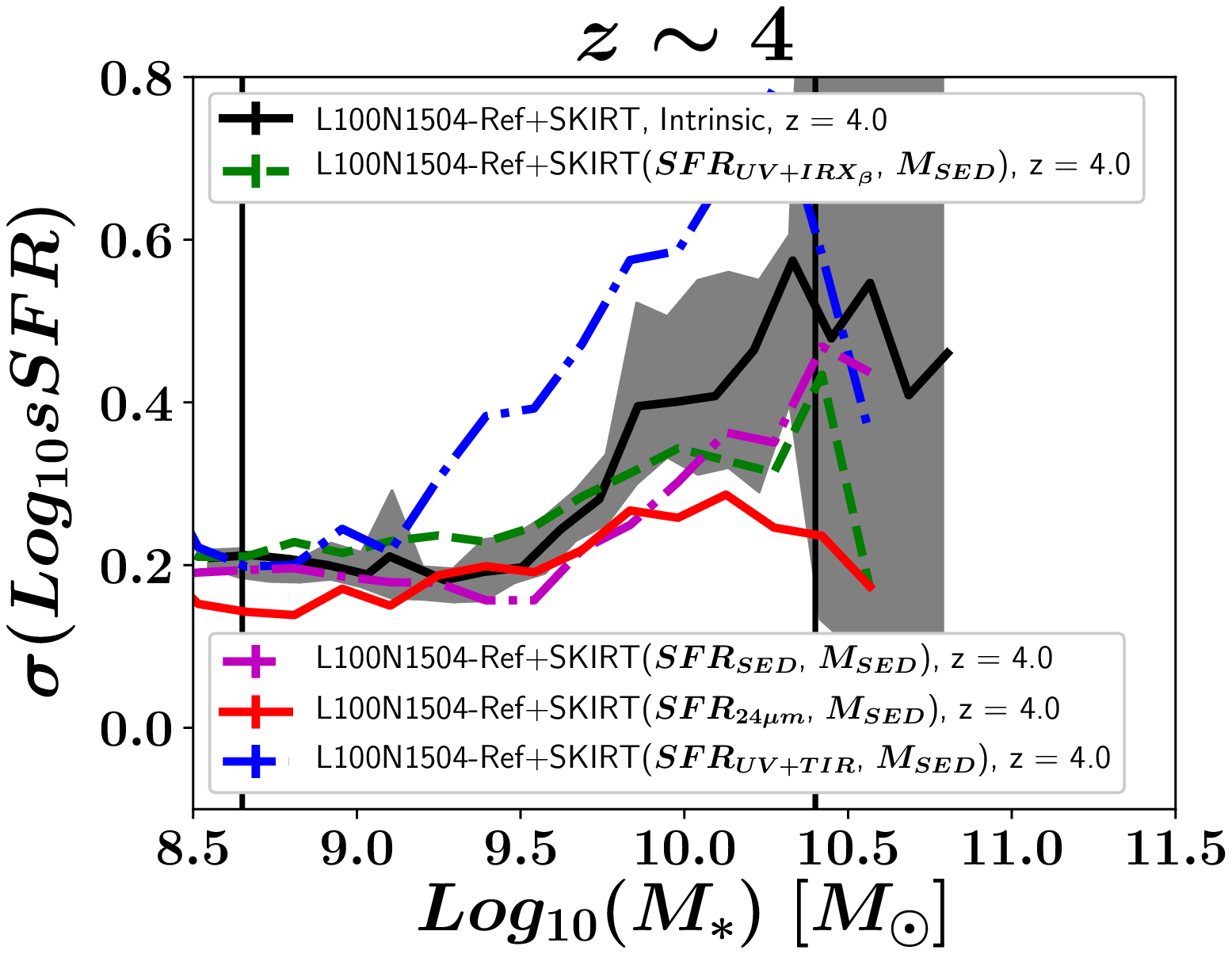} \hspace{-0.5em} \hspace{-5.5em} \vspace{-3.3em}
\includegraphics[scale = 0.45]{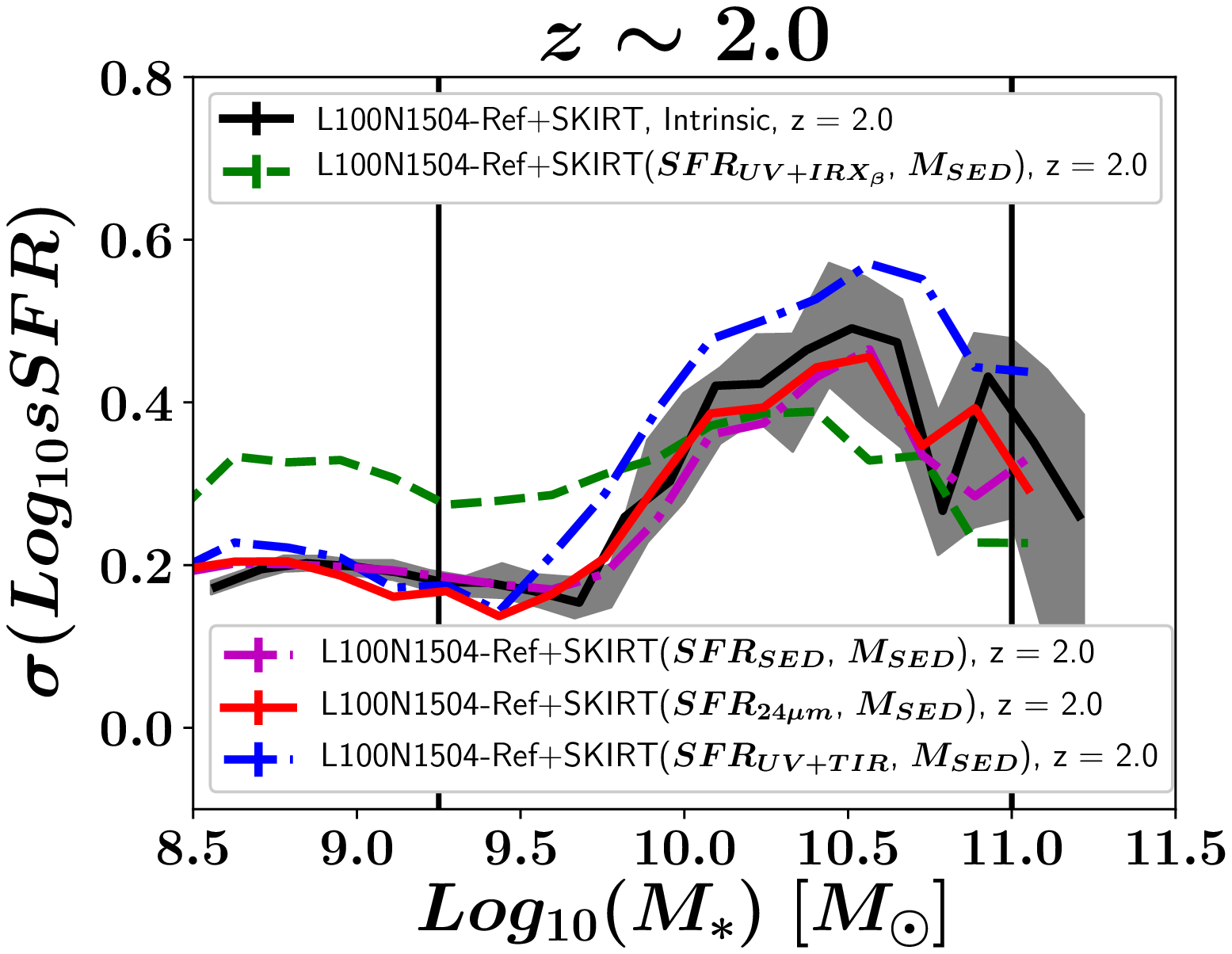} \hspace{-0.5em} \hspace{-5.5em} \vspace{-1em}
\includegraphics[scale = 0.45]{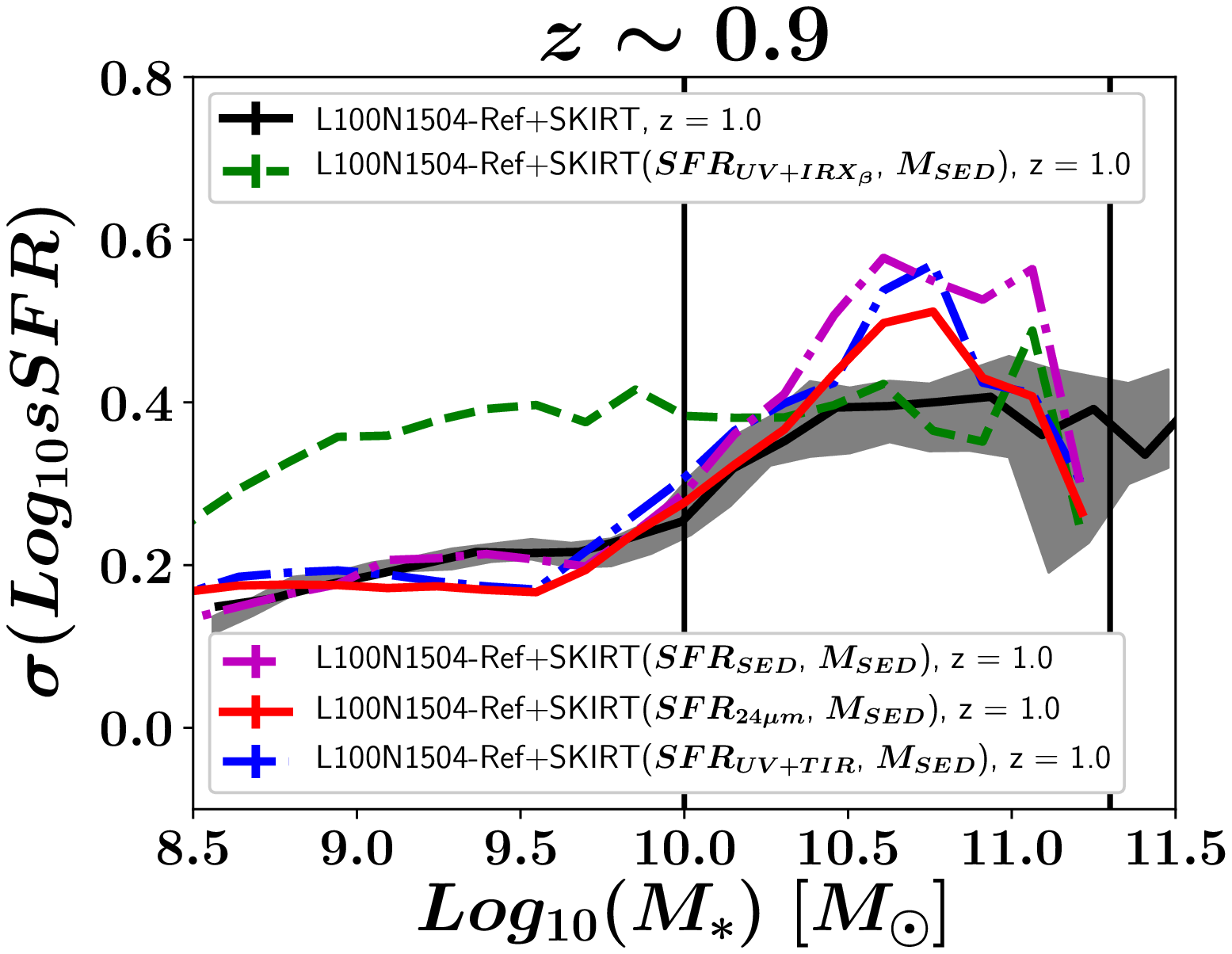} \hspace{-0.5em} \hspace{-5.5em} \vspace{1em}
\includegraphics[scale = 0.45]{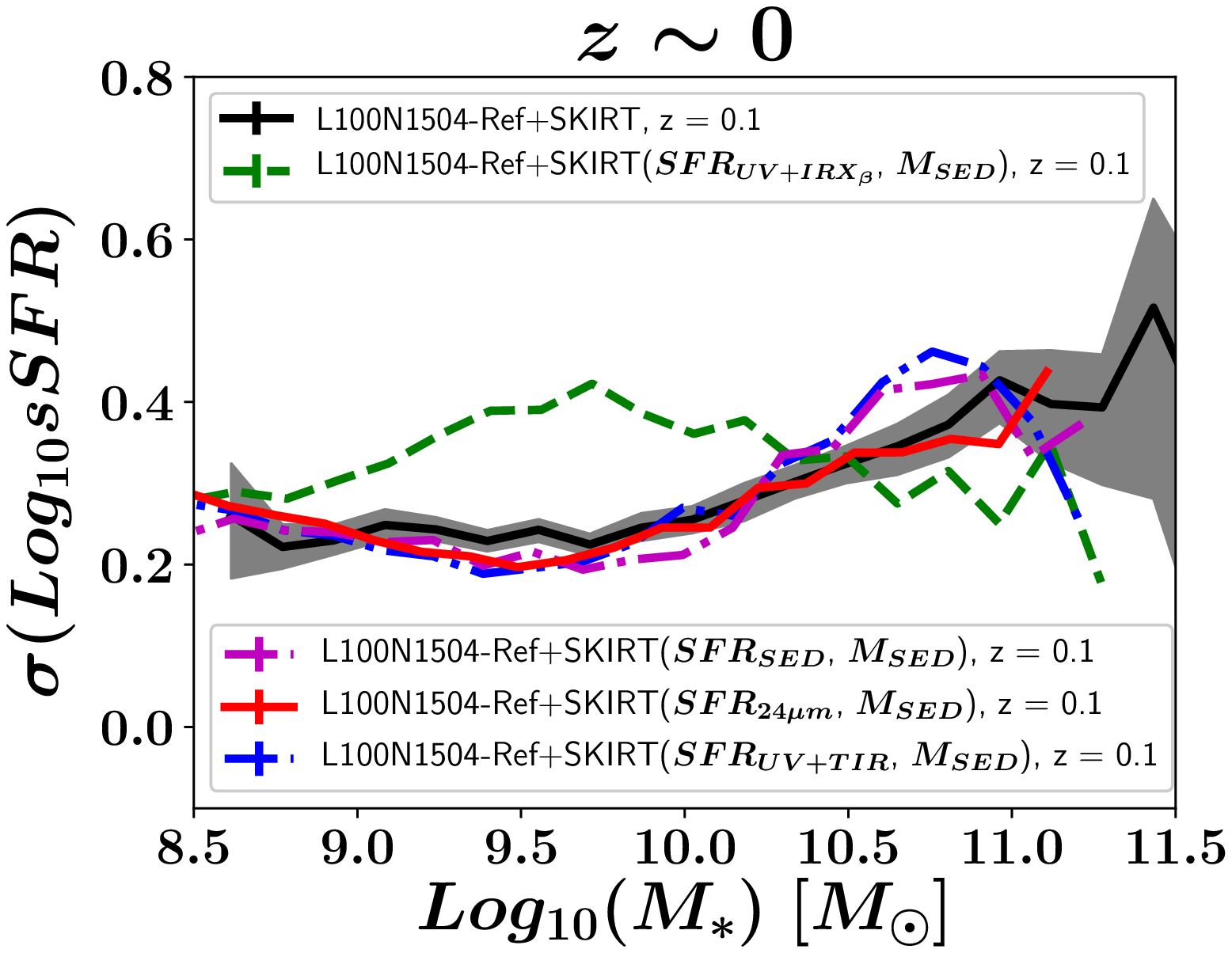}
\caption{The evolution of the  $\sigma_{\text{sSFR}}$-$M_{\star}$ relation for the galaxies of the EAGLE+SKIRT mock survey at $z \sim 0-4$. The black solid lines represents the relation if the stellar masses and SFRs are taken from the intrinsic EAGLE properties. The green dashed line represents the relation if SFRs are obtained using the UV luminosities and the IRX-$\beta$ relation \citep{meurer1999,bouwens2012,Kennicutt2012}. The magenta dotted line describes the same when SFRs and stellar masses are both inferred using the SED fitting technique. The red solid line is the  ${\rm \sigma_{\text{sSFR}}}$-${\rm M_{\star}}$ relation retrieved when stellar masses are calculated via SED fitting and SFRs by combining the Infrared luminosity estimated from the $24 \mu m$ luminosity \citep{Kennicutt2012,DaleHelou2002}. The blue dotted line is the  $\sigma_{\text{sSFR}}$-$M_{\star}$ relation obtained if stellar masses are calculated via SED fitting and SFRs by combining the UV and total Infrared luminosities \citep{DaleHelou2002,Kennicutt2012}. The mock survey suggests that the relation is affected the by ${\rm SFR/M_{\star}}$ diagnostics.}
\label{fig:indicators}
\end{figure}

The black line of Fig. \ref{fig:indicators} represents the intrinsic scatter of the sSFRs of the EAGLE+SKIRT objects (intrinsic SFRs and stellar masses). The vertical lines define the mass interval at which the above data fully represent the total EAGLE sample. The shaded grey region represents the 95 \% bootstrap confidence interval for 5000 re-samples for the $\sigma_{\text{sSFR}}$-$M_{\star}$ relation. For clarity we present the above only for the reference model. The main results are the following:

\begin{itemize}
\item The magenta dashed line describes the  $\sigma_{\text{sSFR}}$-$M_{\star}$ relation when SFRs and stellar masses are both inferred using the SED fitting technique from the mock survey.  The method is used in a range of observational studies to obtain the ${\rm SFR-M_{\star}}$ relation and its scatter \citep{deBarros,Salmon2015}. According to the EAGLE+SKIRT data we see that the shape remain relatively unchanged with respect to the intrinsic relation (black solid line), but the scatter is slightly underestimated ($\sim$ 0.5 dex) at the high mass end at $z \sim4$ but overestimated by $\sim 0.15 $ dex at the mass interval of $\log(M_{\star}/M_{\odot}) \sim 10.0-11.0$. The above gives the false impression that the scatter increases more sharply with mass.

\item The dark green dashed line represents the  $\sigma_{\text{sSFR}}$-$M_{\star}$ relation when SFRs are obtained using the FUV luminosity \citep{Kennicutt2012} and the IRX-$\beta$ relation \citep{meurer1999,bouwens2012,Katsianis2016} while the stellar masses are calculated through the SED fitting technique. This combination to obtain properties is widely used in the literature to estimate the ${\rm SFR-M_{\star}}$ relation and its scatter \citep{Santini2017}. We see that this method implies an artificially higher $\sigma_{\text{sSFR}}$ (with respect the intrinsic-black solid line) at the low mass end. This gives an artificial mass independent  ${\rm \sigma_{\text{sSFR}}}$-${\rm M_{\star}}$ relation with a scatter of $\sim 0.35$ dex, which is not evolving significantly from $z \sim2$ to $z\sim0$.

\item  The red solid line is the  ${\rm \sigma_{\text{sSFR}}}$-${\rm M_{\star}}$ relation retrieved when stellar masses are calculated via SED fitting and SFRs by combining the FUV and Infrared luminosity estimated from the $24 \mu m$ luminosity \citep{Kennicutt2012,DaleHelou2002}. Combining IR and UV luminosities to obtain SFRs in observations is a clasic method in the literature \citep{Daddi2007b,Santini2009,Heinis2014} to obtain the ${\rm SFR-M_{\star}}$ relation. At $z \sim 4$ the scatter is underestimated with respect the reference black line by $0.2$ dex at high masses and imply a dispersion with a constant scatter around 0.15 dex. At lower redshifts there is a good agreement (within 0.05 dex) from the reference intrinsic black line.

\item The blue dotted line describes the  $\sigma_{\text{sSFR}}$-$M_{\star}$ relation when SFRs are derived from UV$+$TIR and stellar masses from the SED fitting technique. According to the EAGLE+SKIRT data this technique agrees well with the intrinsic relation, except redshift 4 where the derived relation is mass independent with a scatter of 0.2 dex. Similarly with the magenta line, which represents the results from SED fitting) the scatter is overestimated with respect the reference black line by $0.15$ dex for high mass objects at the mass interval of $\log(M_{\star}/M_{\odot}) \sim 10.0-11.0$ for $z \sim 0.9$.

\end{itemize}
  
In conclusion, according to the EAGLE+SKIRT data, the inferred shape and normalization of the  ${\rm \sigma_{\text{sSFR}}}$-${\rm M_{\star}}$ relation can be affected by the methodology used to derive SFRs and stellar masses in observations. This can affect the conclusions about its shape and it is important for future observations to investigate this further \citep{Davies2019}. However, we note that having access to IR data and deriving SFRs and stellar masses from SED fitting or combined UV+IR luminosities typically give a ${\rm \sigma_{\text{sSFR}}}$-${\rm M_{\star}}$ relation close to the intrinsic simulated relation and can probe successfully the shape of the relation for $\log(M_{\star}/M_{\odot}) \sim 9.0-11.0$.


\section{Conclusion and discussion}
\label{ConDis}

The  ${\rm \sigma_{\text{sSFR}}}$-${\rm M_{\star}}$ relation reflects the diversity of star formation histories for galaxies at different masses. However, it is difficult to decipher the true shape of the relation, the intrinsic value of the scatter and which mechanisms important for galaxy evolution govern it, solely by relying on observations. In this paper we presented the evolution of the  intrinsic ${\rm \sigma_{\text{sSFR}}}$-${\rm M_{\star}}$ relation employing the EAGLE suite of cosmological simulations and a compilation of multiwavelength observations at various redshifts. We deem the EAGLE suite appropriate for this study as it is able to reproduce the observed star formation rate and stellar mass functions \citep{Furlong2014,Katsianis2017} for a wide range of SFRs, stellar masses and redshifts. The investigation is not limited by the shortcomings encountered by galaxy surveys and address a range of redshifts and mass intervals. Our main conclusions are summarized as follows: 

\begin{itemize}
  
\item In agreement with recent observational studies \citep{Guo2013,Ilbert2015,Willet2015,Santini2017} the EAGLE reference model suggests that the  ${\rm \sigma_{\text{sSFR}}}$-${\rm M_{\star}}$ relation is evolving with redshift and the dispersion is mass dependent (Section \ref{specificSFR-M}). This is in contrast with the widely accepted notion that the dispersion is mass/redshift independent with a constant scatter ${\rm \sigma_{\text{sSFR}} \sim 0.2-0.3}$ \citep{Noeske2007,Elbaz2007,Rodighiero10,Whitaker2012}. We find that the  ${\rm \sigma_{\text{sSFR}}}$-${\rm M_{\star}}$ relation has a U-shape form with the scatter increasing both at the high and low mass ends. Any interpretations of an increasing \citep{Guo2013,Ilbert2015} or decreasing dispersion \citep{Santini2017} with mass may be misguided, since they usually focus on limited mass intervals (Subsection \ref{specificSFR-MFeedback}). The finding about the U-shape form of the relation is supported by results relying on the GAMA survey \citep{Davies2019} at $z \sim 0$.
  
\item AGN and SN feedback are driving the shape and evolution of the  ${\rm \sigma_{\text{sSFR}}}$-${\rm M_{\star}}$ relation in the simulations (Subsection \ref{specificSFR-MFeedback}) . Both mechanisms cause a diversity of star formation histories for low mass (SN feedback) and high mass galaxies (AGN feedback).

\item Mergers do not play a major role on the shape of the  ${\rm \sigma_{\text{sSFR}}}$-${\rm M_{\star}}$ relation (Subsection \ref{specificSFR-Mmergers}).

\item We employ the EAGLE/SKIRT mock data to investigate how different ${\rm SFR/M_{\star}}$ diagnostics affect the  ${\rm \sigma_{\text{sSFR}}}$-${\rm M_{\star}}$ relation. The shape of the relation remains relatively unchanged if both the SFRs and stellar masses are inferred through SED fitting, or combined UV+IR data. However, SFRs that rely solely on UV data and the IRX-${\rm \beta}$ relation for dust corrections imply a constant scatter with stellar mass with almost no redshift evolution. Methodology used to derive SFRs and stellar masses  can affect the inferred  ${\rm \sigma_{\text{sSFR}}}$-${\rm M_{\star}}$ relation in observations and thus compromising the robustness of conclusions about its shape and normalization. 

\end{itemize}  
  

\acknowledgments
\section*{Acknowledgments}

We would like to thank the anonymous referee for their suggestions and comments which improved significantly our manuscript. In addition, we would like to thank Jorryt Matthee for discussions and suggestions. This work used the DiRAC Data Centric system at  Durham  University,  operated  by  the  Institute for  Computational  Cosmology  on  behalf  of  the  STFC  DiRAC  HPC  Facility  (www.dirac.ac.uk). A.K has been supported by the {\it Tsung-Dao Lee Institute Fellowship}, {\it Shanghai Jiao Tong University} and {\it CONICYT/FONDECYT fellowship, project number: 3160049}. G.B. is supported by {\it CONICYT/FONDECYT, Programa de Iniciacion, Folio 11150220}. V.G. was supported by {\it CONICYT/FONDECYT iniciation grant number 11160832}. X.Z.Z. thanks supports from the National Key Research and Development Program of China (2017YFA0402703), NSFC grant (11773076) and the Chinese Academy of Sciences (CAS) through a grant to the CAS South America Center for Astronomy (CASSACA) in Santiago, Chile. AK would like to thank his family and especially George Katsianis, Aggeliki Spyropoulou John Katsianis and Nefeli Siouti for emotional support. He would also like to thank Sophia Savvidou for her IT assistance.

\appendix
\section{The evolution of the specific star formation rate function.}
\label{SFRFEAGLE}

\begin{figure}
\centering
\includegraphics[scale=0.40]{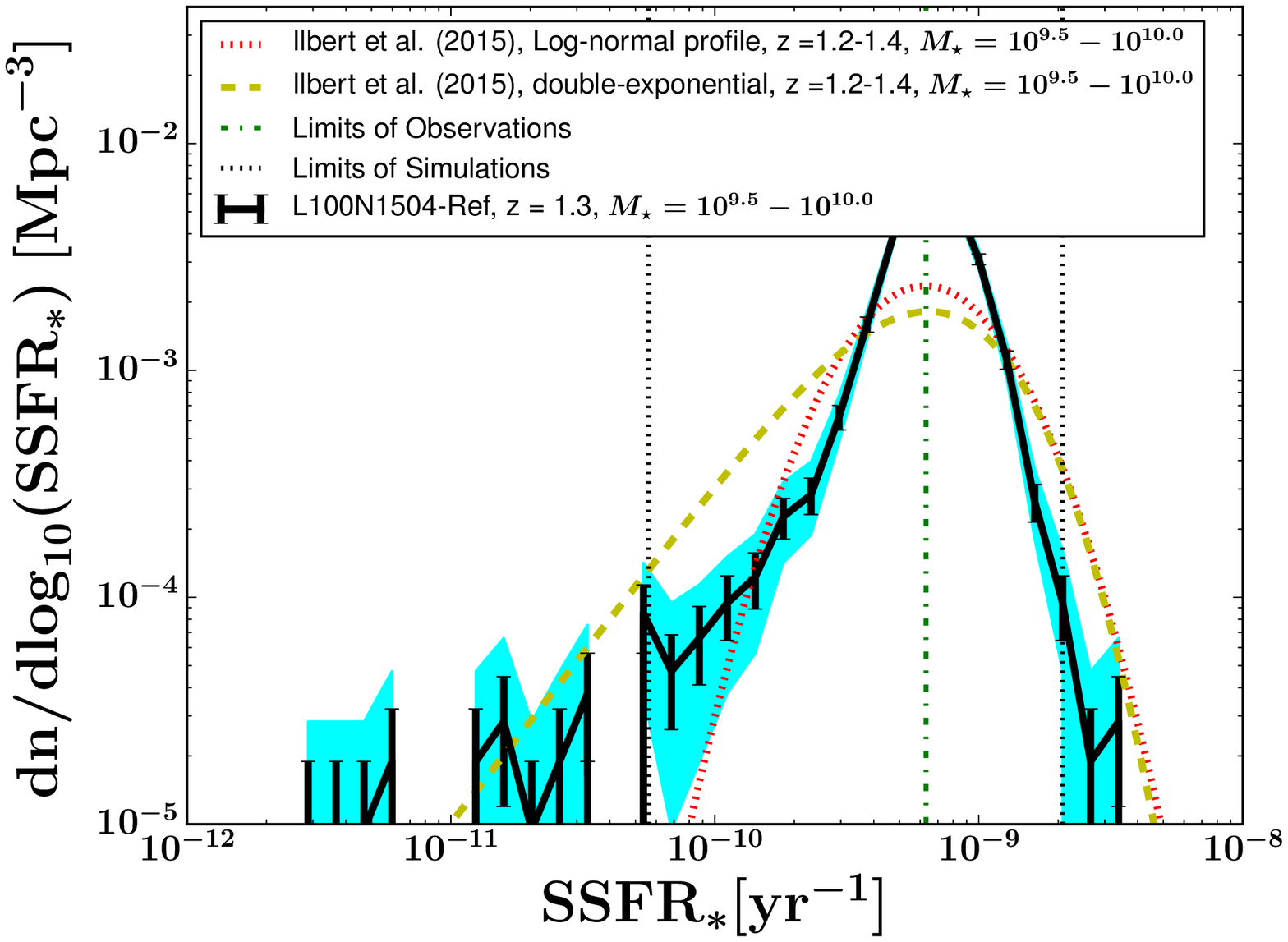} \hspace{-3.5em} \vspace{-1.0em}%
\includegraphics[scale=0.40]{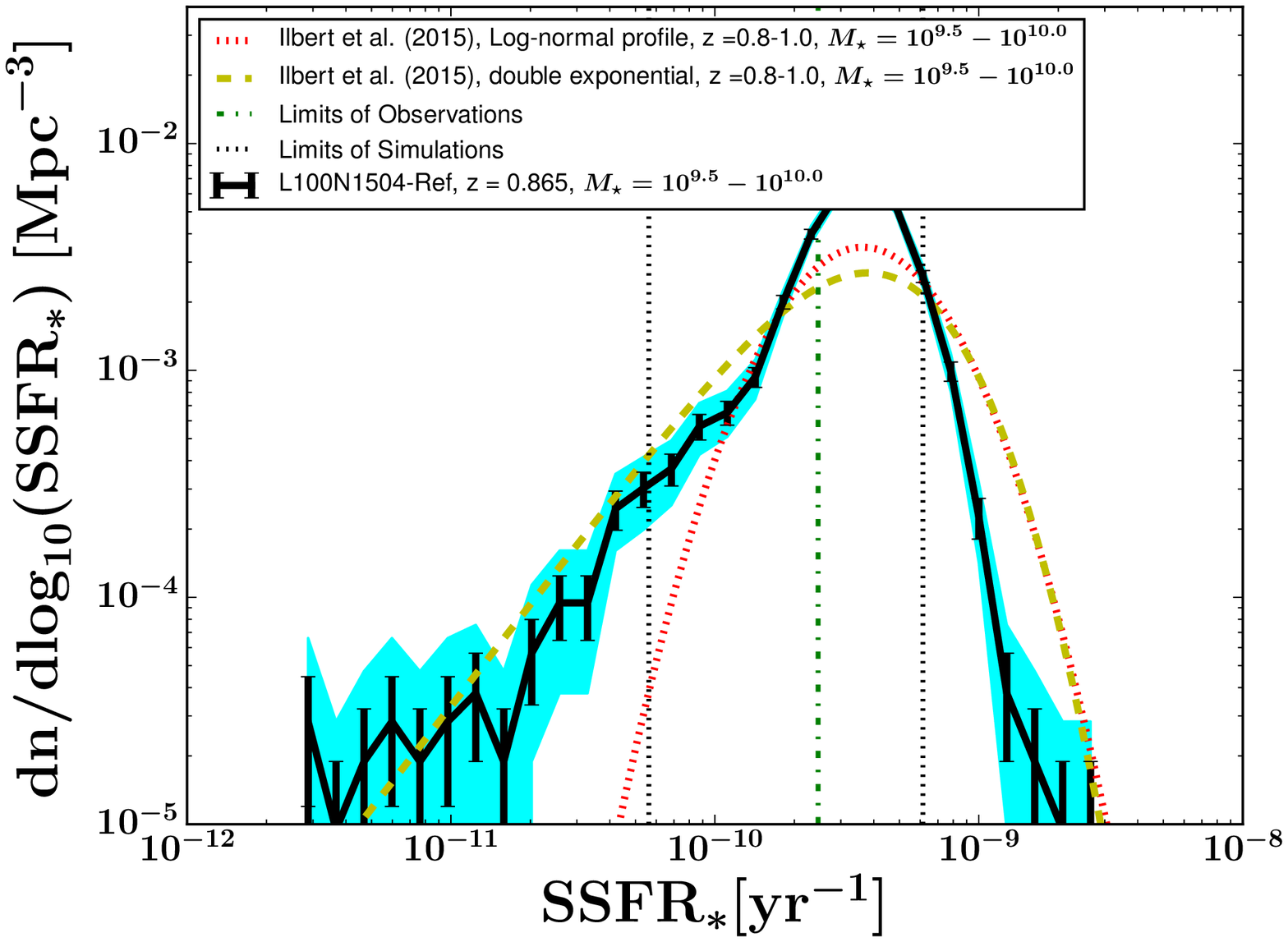} \hspace{-3.5em} \vspace{-1.0em}%
\includegraphics[scale=0.40]{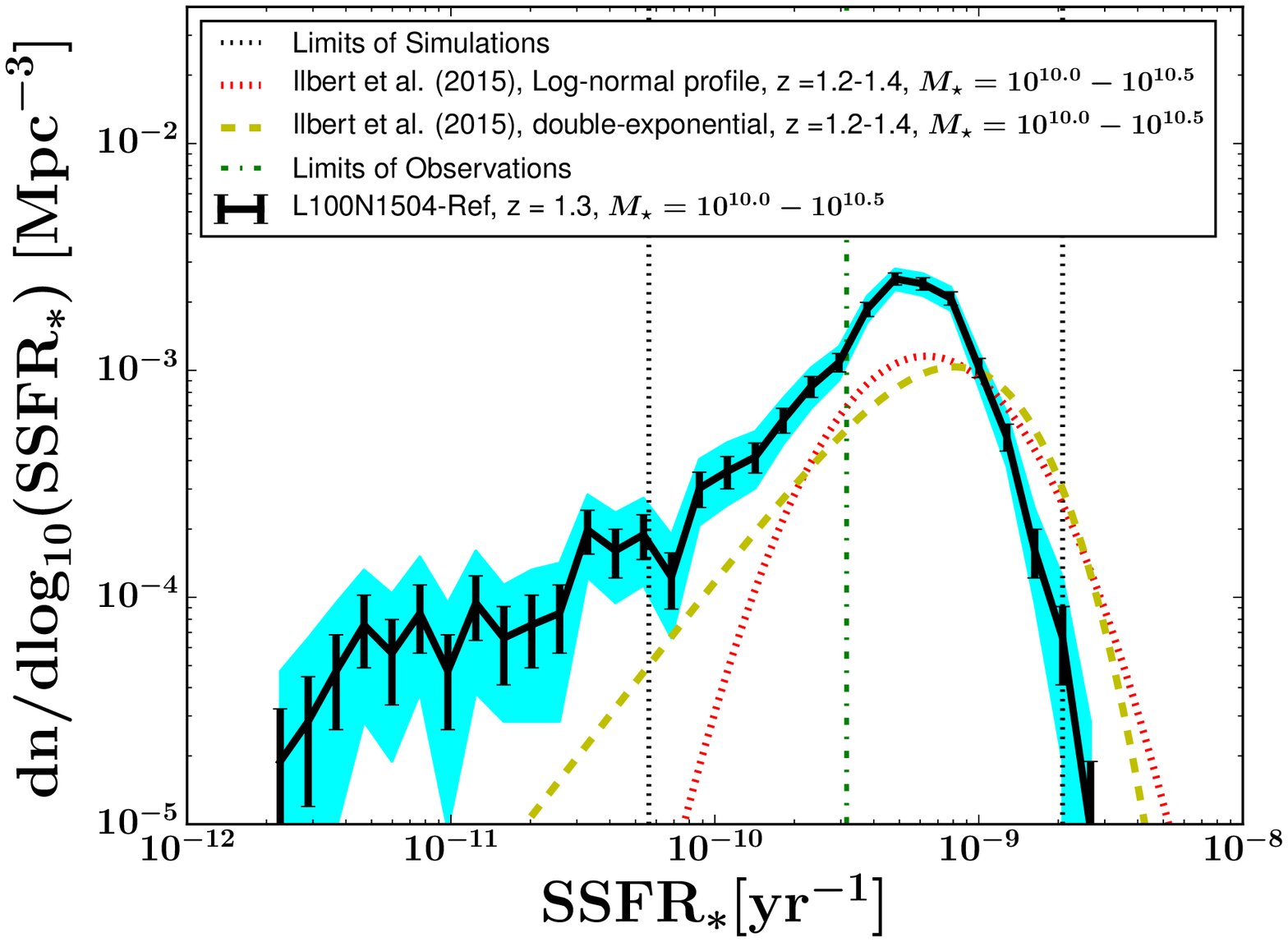} \hspace{-3.5em}%
\includegraphics[scale=0.40]{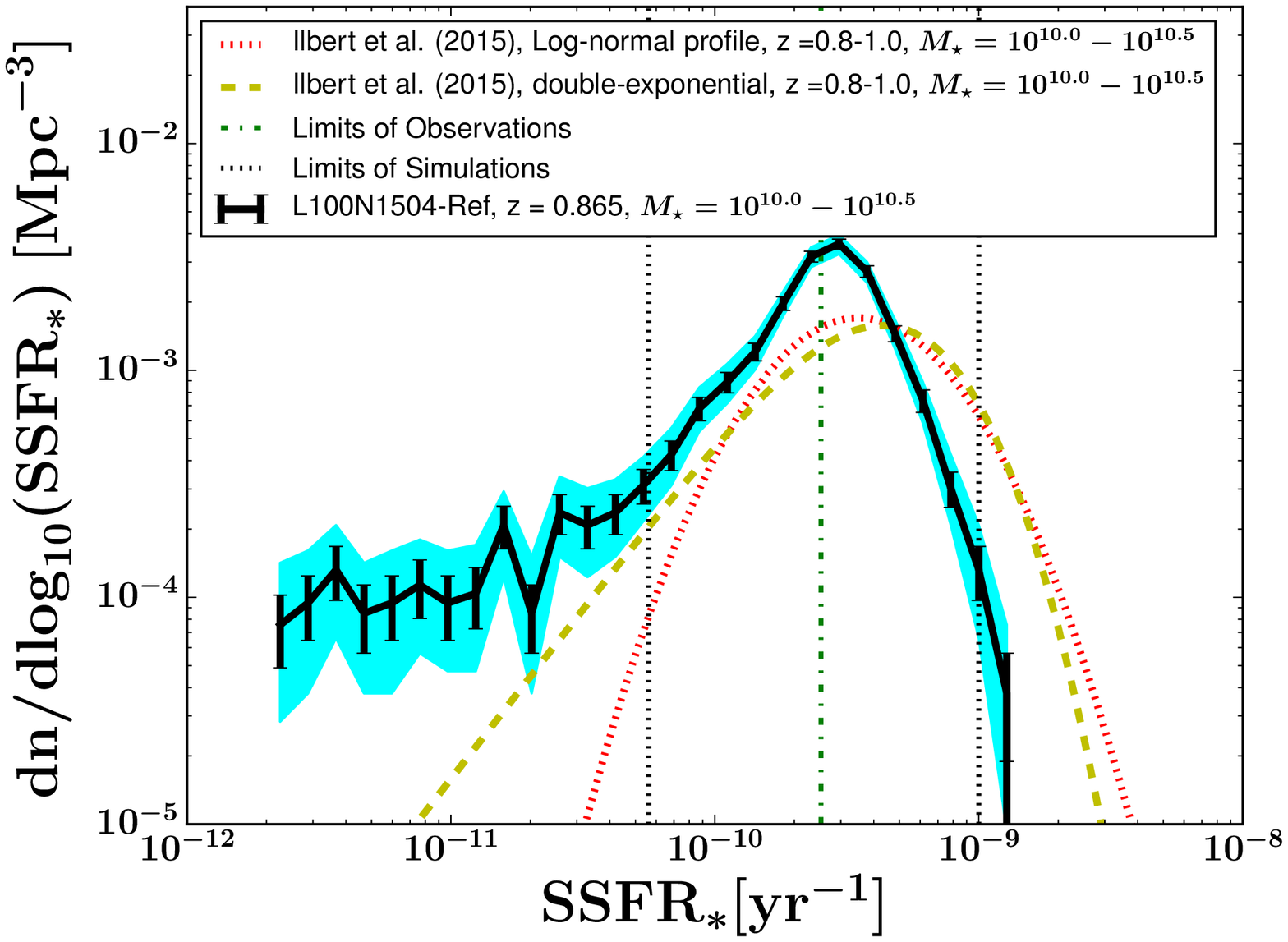} \hspace{-3.5em} \vspace{-1.0em}%
\includegraphics[scale=0.40]{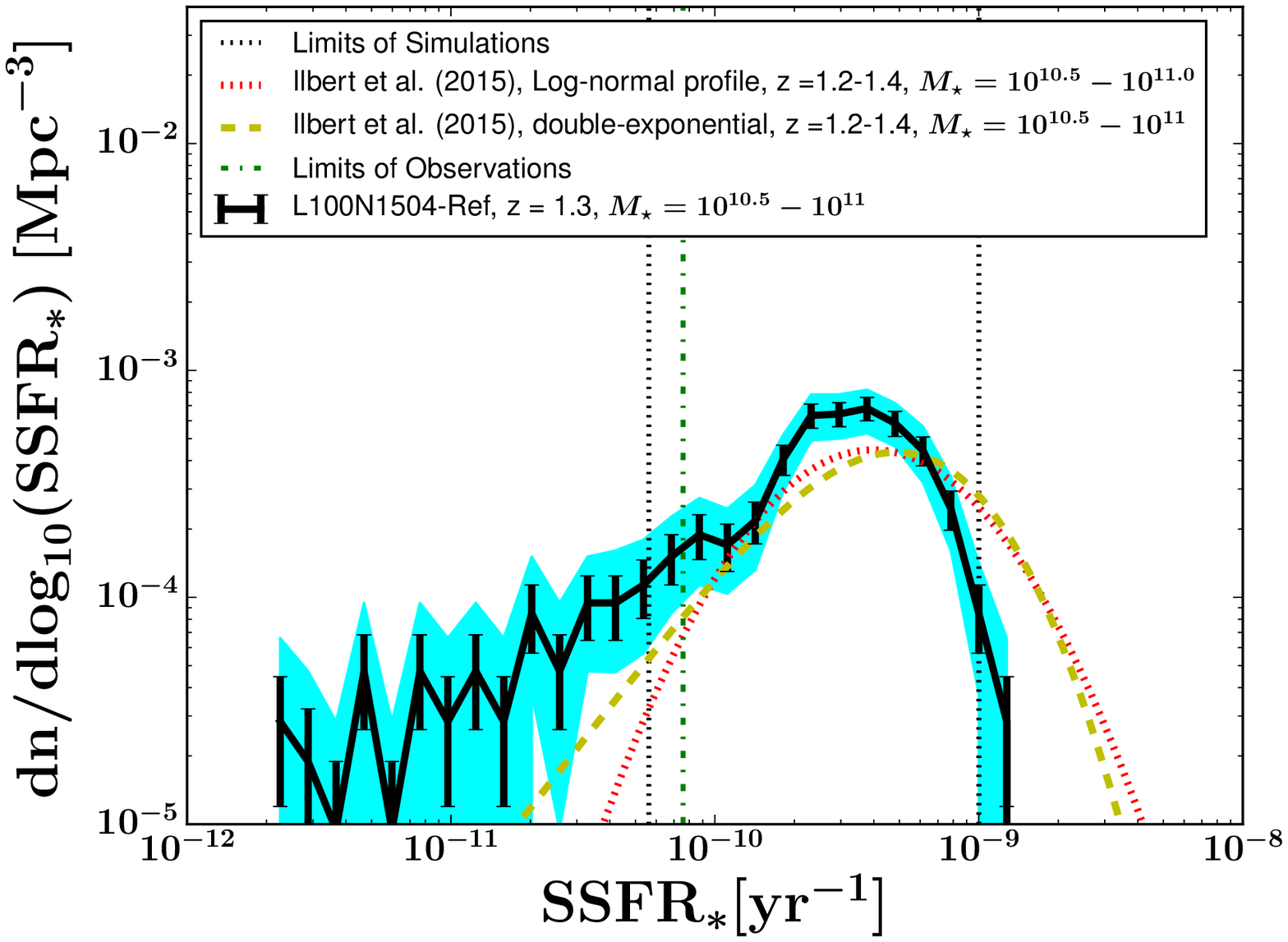} \hspace{-3.5em}%
\includegraphics[scale=0.40]{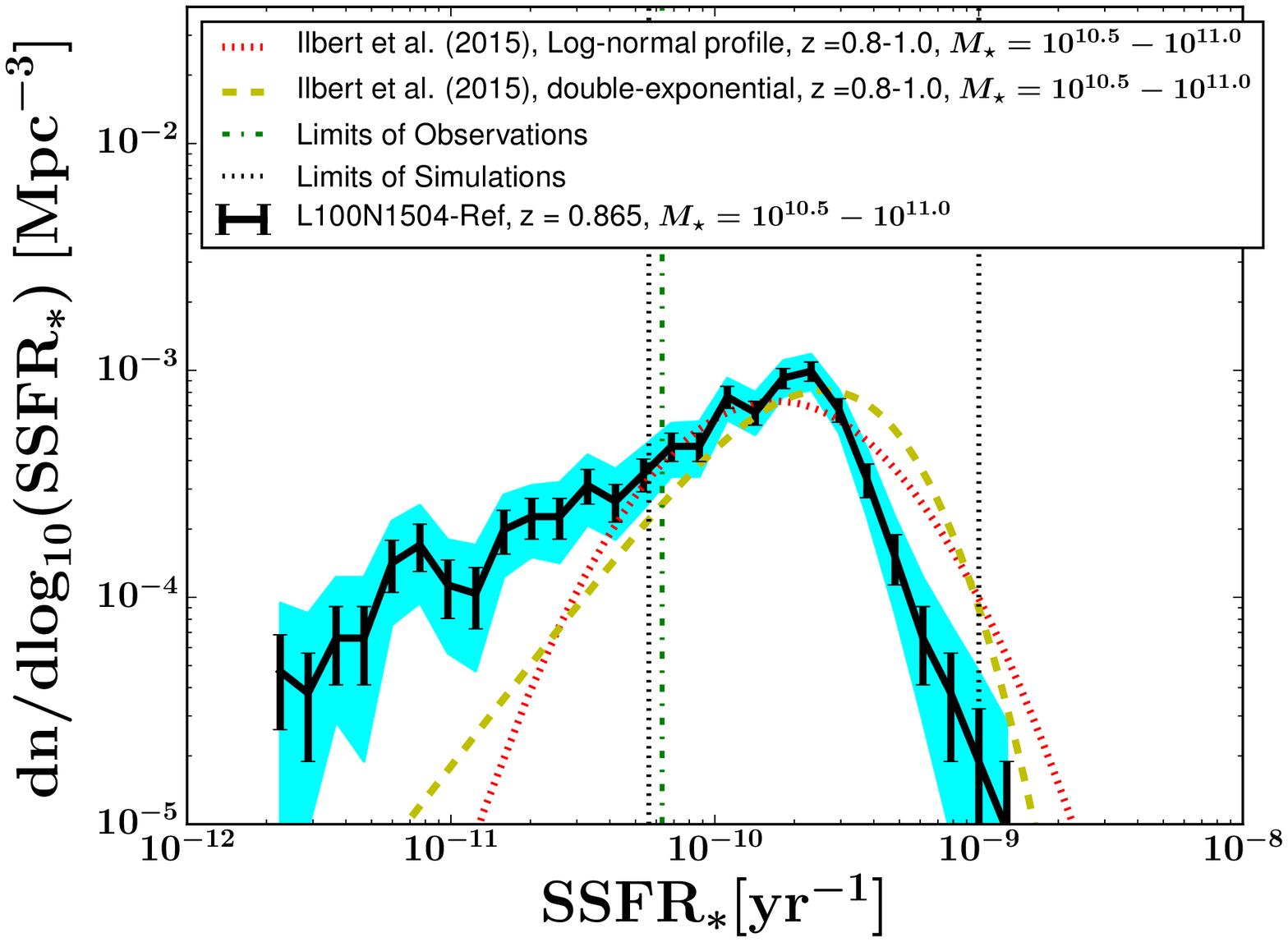}
\caption{The simulated and observed specific star formation rate functions at ${\rm 1.2 < z < 1.4}$ (left panels) and ${\rm 0.8 < z < 1.0}$ (right panels) per stellar mass bin of ${\rm  9.5 < \log(M_{\star}/M_{\odot}}) < 10.0$ (top panels), $ 10.0 < {\rm \log(M_{\star}/M_{\odot})} < 10.5$ (medium panels) and $ 10.5 <{\rm  \log(M_{\star}/M_{\odot})} < 11.0$ (bottom panels). The black solid line corresponds to the EAGLE sSFRF while the orange dashed line represents the best-fit of the sSFR(UV+IR) function with a double-exponential profile from \citet{Ilbert2015}. The dotted line represents a log-normal fit to the data of \citet{Ilbert2015}. The dark green vertical line represents the limits of the results. We note that the observed distributions are shifted by 0.2 dex in order to account for the differences with simulations reported in \citet{Furlong2014} and \citet{McAlpine2017}. The cyan area represents the 95 $\%$ bootstrap confidence interval for 1000 re-samples of the EAGLE sSFRs, while the black errorbars represent the 1 sigma poissonian errors.}
\label{fig:sSFRFhigh}
\end{figure}

\begin{figure}
  \centering
\includegraphics[scale=0.40]{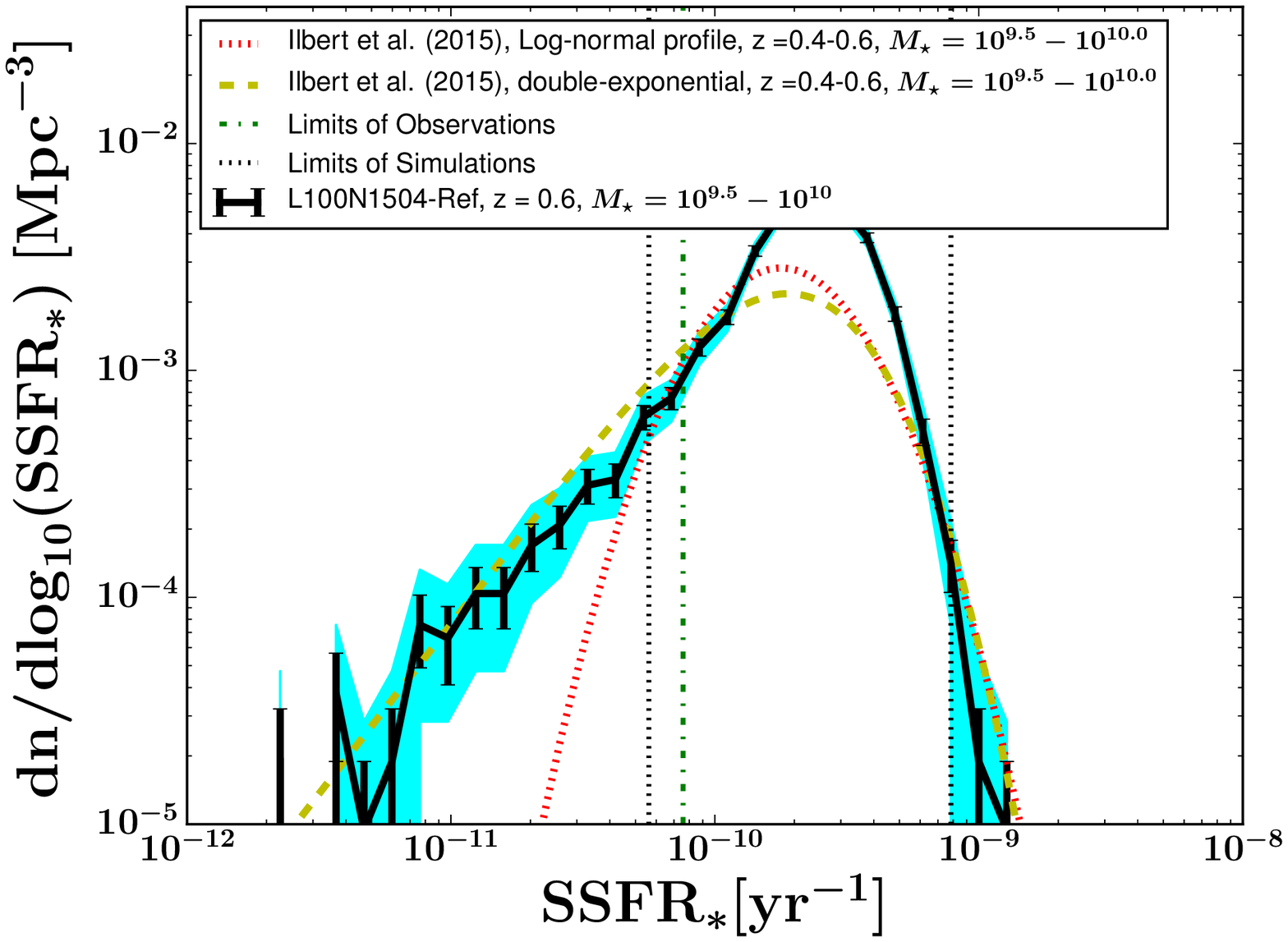} \hspace{-3.5em} \vspace{-1.0em}%
\includegraphics[scale=0.40]{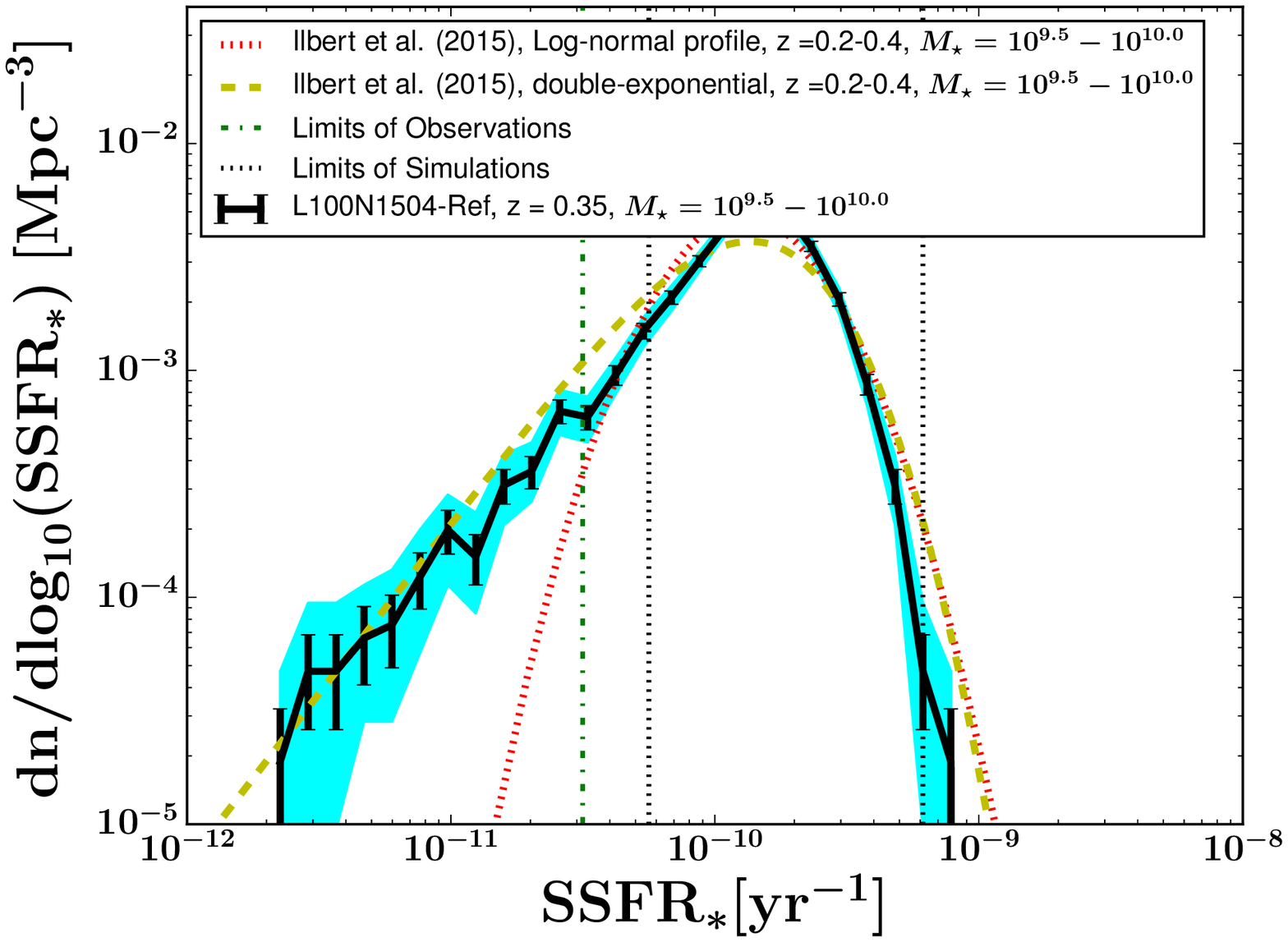} \hspace{-3.5em} \vspace{-0.9em}%
\includegraphics[scale=0.40]{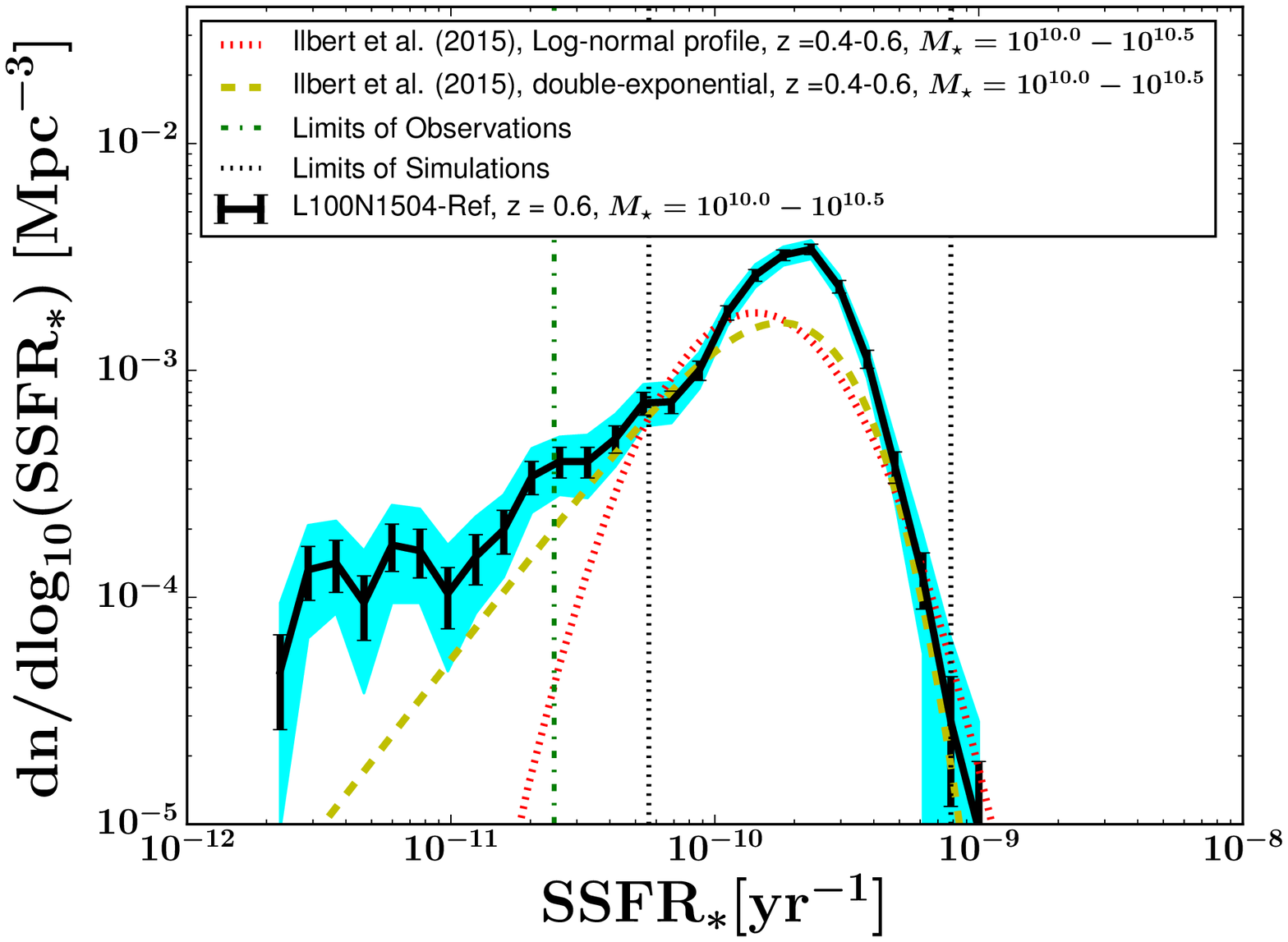} \hspace{-3.5em}
\includegraphics[scale=0.40]{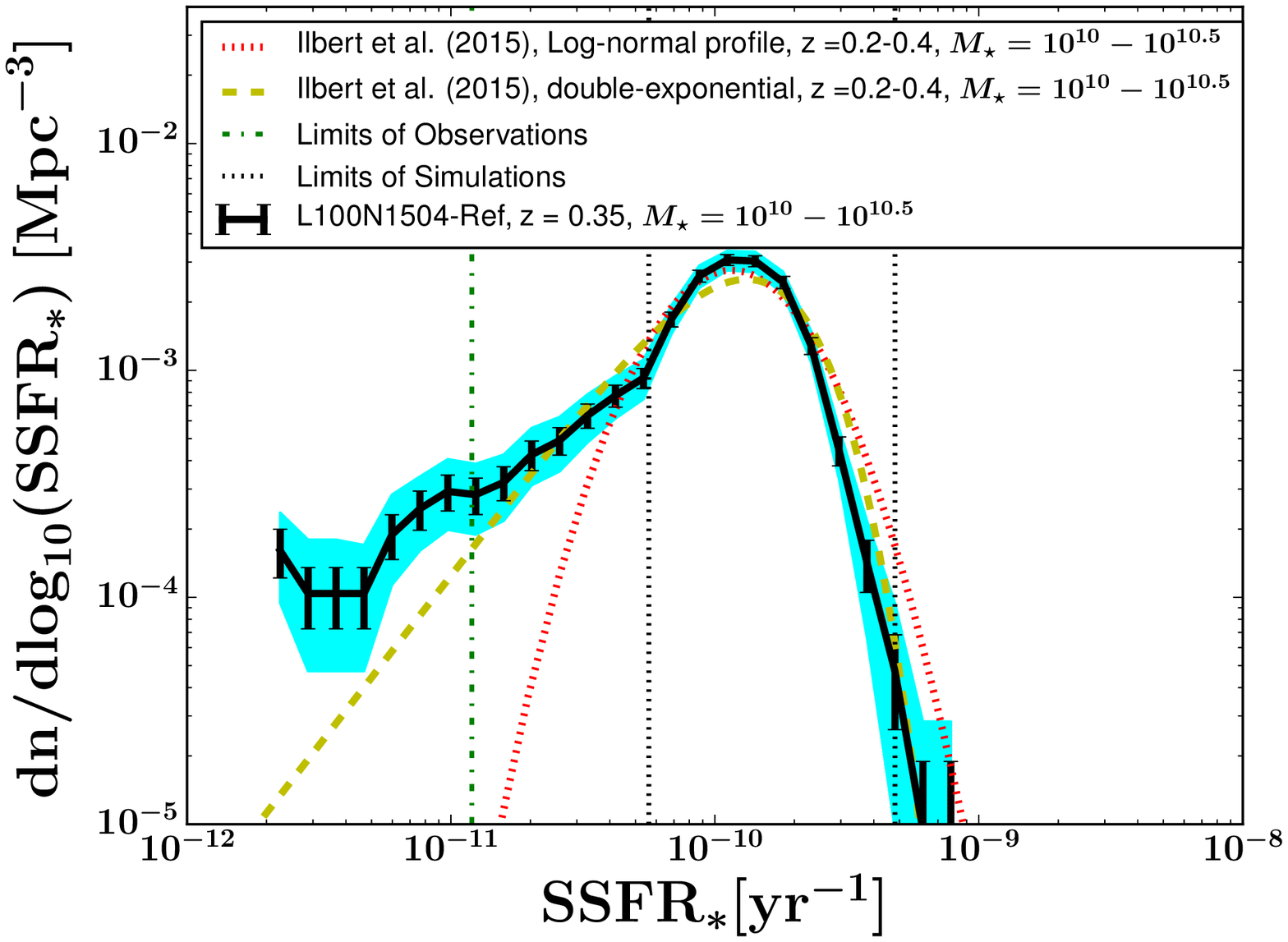} \hspace{-3.5em} \vspace{-0.9em}%
\includegraphics[scale=0.40]{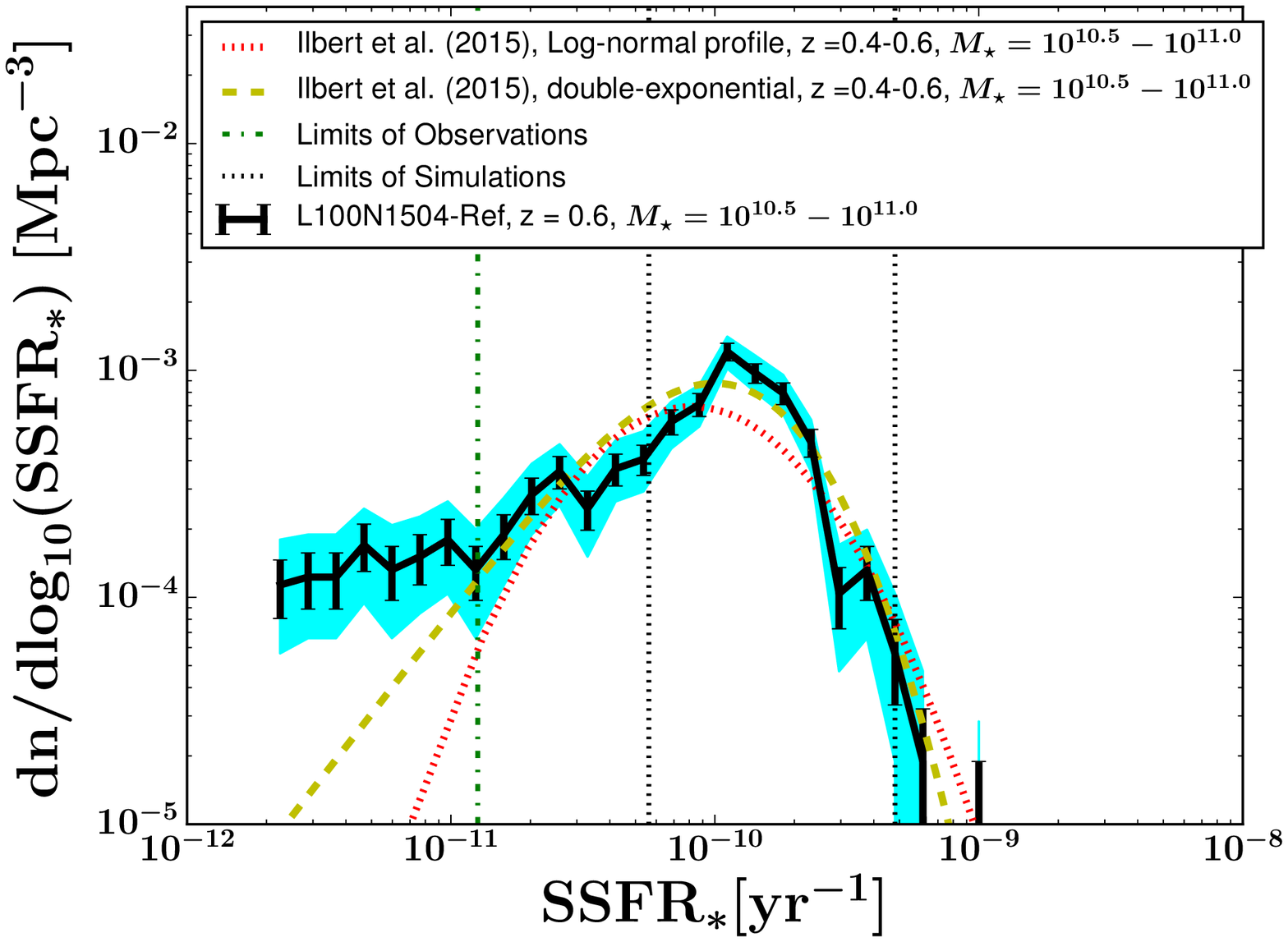} \hspace{-3.5em}
\includegraphics[scale=0.40]{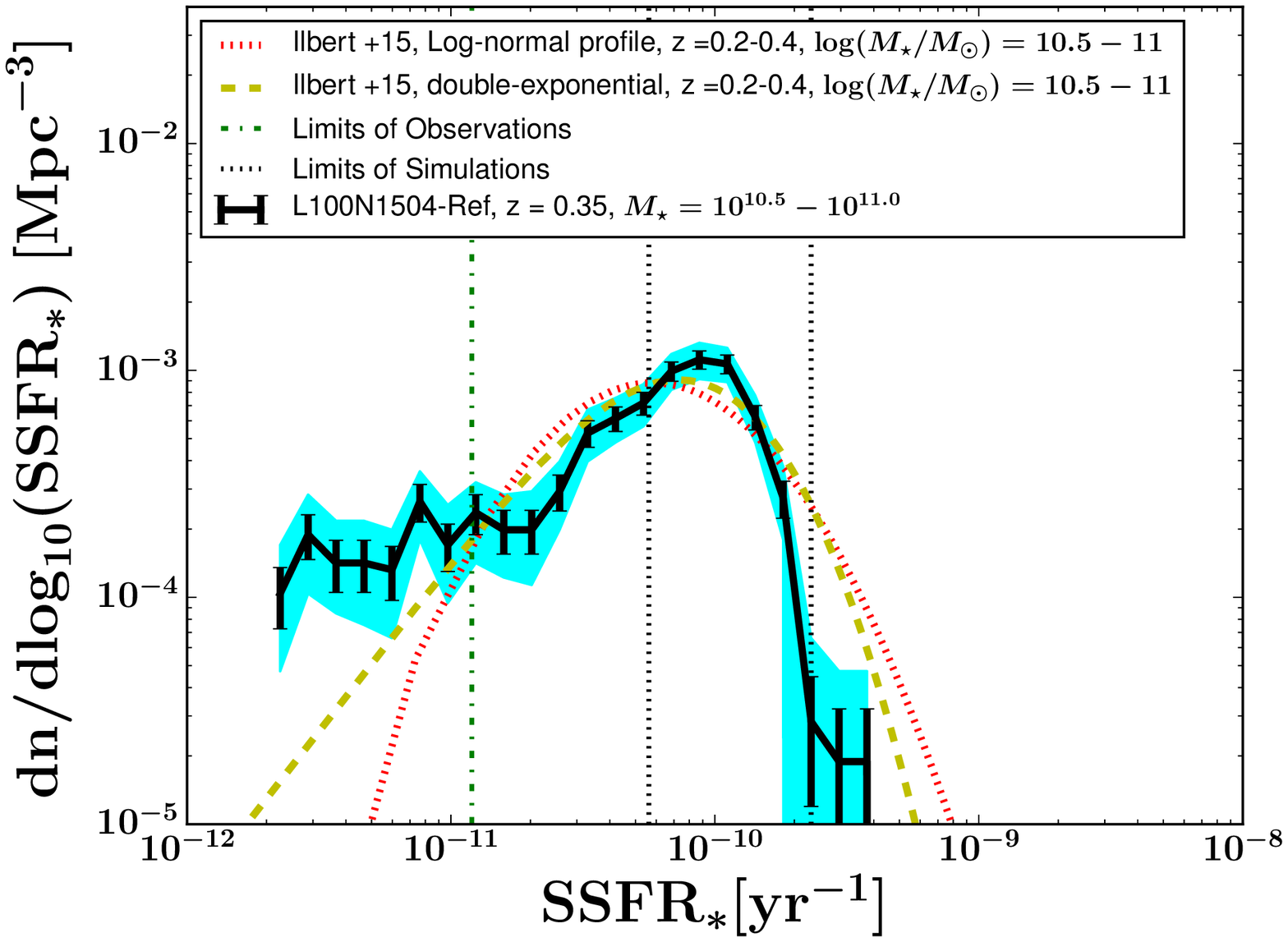}
\caption{The simulated and observed specific star formation rate functions at $0.4 < z < 0.6$ (left panels) and $0.2 < z < 0.4$ (right panels) per stellar mass bin of $ 9.5 < {\rm \log(M_{\star}/M_{\odot})} < 10.0$ (top panels), $ 10.0 < {\rm \log(M_{\star}/M_{\odot})} < 10.5$ (medium panels) and $ 10.5 < {\rm \log(M_{\star}/M_{\odot})} < 11.0$ (bottom panels). The black solid line corresponds to the EAGLE sSFRF while the orange dashed line represents the best-fit of the sSFR(UV+IR) function with a double-exponential profile from \citet{Ilbert2015}. The dotted line represents a log-normal fit to the data of \citet{Ilbert2015}. The dark green vertical line represents the limits of the observations. We note that the observed distributions are shifted by 0.2 dex in order to account for the differences with simulations reported in \citet{Furlong2014} and \citet{McAlpine2017}. The cyan area represents the 95 $\%$ bootstrap confidence interval for 1000 re-samples of the EAGLE sSFRs, while the black errorbars represent the 1 sigma poissonian errors.}
\label{fig:33}
\end{figure}

In this appendix, we base our analysis of the dispersion of the sSFRs at different mass intervals on their distribution/histogram, namely the specific Star Formation Rate Function (sSFRF), following \citet{Ilbert2015}, because studies of the dispersion that rely solely on 2d scatter plots (i.e. displays of the location of the individual sources in the plane) are not able to provide a quantitative information of how galaxies are distributed around the mean sSFR and cannot account for galaxies that could be under-sampled or missed by selection effects. In section \ref{specificSFR-M} we present the evolution of the  ${\rm \sigma_{\text{sSFR}}}$-${\rm M_{\star}}$ relation at ${\rm z \sim0-4}$ in order to visualize the scatter across galaxies, its shape and its evolution. We present the distribution of the sSFR of the EAGLE reference model (L100N1504-Ref) and compare it with observations \citep{Ilbert2015} in Fig. \ref{fig:sSFRFhigh} (${\rm z \sim0.8-1.4}$) and Fig. \ref{fig:33} (${\rm z \sim0.2-0.6}$). The EAGLE SFRs are reported to be 0.2 dex lower than observations but are able to replicate the observed evolution and shape of the Cosmic Star Formation Rate Density \citep{Furlong2014,Katsianis2017}, evolution of star formation rate function \citep{Katsianis2017} and SFR-BHAR \citep{McAlpine2017}. Following, \citet{McAlpine2017} we decrease the observed sSFRs by 1.58 in order to focus solely on the scatter of the distribution and its shape. We note that this discrepancy may have its roots to the methodologies used by observers to obtain the intrinsic SFRs and stellar masses \citep{Katsianis2016}.

In the top panels of Fig. \ref{fig:sSFRFhigh} and Fig. \ref{fig:33} we present the comparison between the simulated and observed data. The black solid line corresponds to the results from the EAGLE reference model, while the orange dashed line (double-exponentional) and red dotted lines (log-normal profile) represent the fits for the observed distribution. The green vertical line mark the limits of the observations. The cyan area represents the 95 $\%$ bootstrap confidence interval for 1000 re-samples of the EAGLE SFRs, while the black errorbars represent the 1 sigma poissonian errors. We can see that the agreement between the simulated and observed sSFR functions is typically good at low mass objects but breaks down for galaxies more massive than ${\rm \log(M_{\star}/M_{\odot}) > 10}$, at ${\rm z > 0.8}$ with the L100N1504-Ref distribution being shifted to lower sSFR and having larger peak values in comparison with the observations. \citet{Ilbert2015} were not able to directly constrain the full shape of the sSFR function, despite the fact that they combined both GOODS and COSMOS data. In most of the redshift and mass bins, the sSFR function is incomplete below the peak in sSFR. The authors tried to discriminate between a log-normal and a double-exponential profile but were not able to sample sufficiently low sSFR to see any advantage of using either one or the other parametrization. They argued that the fit with a double-exponential function is more suitable than the log-normal function at ${\rm z \sim 0}$. The EAGLE reference model points to the direction that the sSFRF at the different mass bins and redshifts follows a double-exponential function. However, for higher redshifts the simulated distributions are slightly flatter than the double-exponential fits of the observations. A double-exponential profile, that is not commonly used to describe the sSFR distribution \citep{Ilbert2015}, allows a significant density of star-forming galaxies with a low sSFR and the confirmation of this shape from future observations is important.

\bibliographystyle{apj}	
\bibliography{Katsianis_mnrasRevScatter}



\end{document}